\documentclass[journal,10pt, twocolumn]{IEEEtran}
\usepackage[T1]{fontenc}
\usepackage{url}
\usepackage{color}
\usepackage{subcaption} 
\usepackage{amsmath,amssymb,amsfonts}
\usepackage{algorithmic}
\usepackage{graphicx}
\usepackage{textcomp}
\usepackage{booktabs}
\usepackage{cite}
\usepackage{caption}
\usepackage{graphicx}
\usepackage{mathtools}
\usepackage{multirow}
\usepackage{multicol}
\usepackage{xcolor}
\usepackage[ruled,vlined]{algorithm2e}
\usepackage{float}
\usepackage{graphicx}
\usepackage{subcaption}
\usepackage{gensymb}
\usepackage{enumitem}
\usepackage{epstopdf}
\usepackage{amsmath}
\usepackage{rotating}
\usepackage[edges]{forest}
\usepackage{tikz}
\usetikzlibrary{trees}
\usetikzlibrary{shadows}
\usepackage{fixltx2e}
\usepackage{tabularx}

\begin{document}

\title{An In-Depth Survey on Virtualization Technologies in 6G Integrated Terrestrial and Non-Terrestrial Networks}

	\author{{Sahar Ammar, Chun Pong Lau, and Basem Shihada} 

 \thanks{ The authors are with CEMSE Division,  King Abdullah University of Science and Technology (KAUST),  Thuwal, Makkah Province, Saudi Arabia. E-mail: \{sahar.ammar, lau.pong, basem.shihada\}@kaust.edu.sa}

	} 
	
	\maketitle
	\begin{abstract}
6G networks are envisioned to deliver a large diversity of applications and meet stringent quality of service (QoS) requirements. Hence, integrated terrestrial and non-terrestrial networks (TN-NTNs) are anticipated to be key enabling technologies. However, the TN-NTNs integration faces a number of challenges that could be addressed through network virtualization technologies such as Software-Defined Networking (SDN), Network Function Virtualization (NFV) and network slicing. In this survey, we provide a comprehensive review on the adaptation of these networking paradigms in 6G networks. We begin with a brief overview on NTNs and virtualization techniques. Then, we highlight the integral role of Artificial Intelligence in improving network virtualization by summarizing major research areas where AI models are applied. Building on this foundation, the survey identifies the main issues arising from the adaptation of SDN, NFV, and network slicing in integrated TN-NTNs, and proposes a taxonomy of integrated TN-NTNs virtualization offering a thorough review of relevant contributions. The taxonomy is built on a four-level classification indicating for each study the level of TN-NTNs integration, the used virtualization technology, the addressed problem, the type of the study and the proposed solution, which can be based on conventional or AI-enabled methods. Moreover, we present a summary on the simulation tools commonly used in the testing and validation of such networks. Finally, we discuss open issues and give insights on future research directions for the advancement of integrated TN-NTNs virtualization in the 6G era.
    
\end{abstract}


\section{Introduction}
Since the introduction of the first generation (1G) in the 1980s, cellular networks have been evolving rapidly with the development of a new generation roughly every ten years. First, the analog mobile networks provided voice communications in the 1G era. Then, digitalization was introduced in 2G allowing not only voice calling but also data services, including short  messaging service (SMS). In the beginning of the second millennia, the third generation (3G) emerged to offer new data services such as video calling and internet access using multiple access techniques based on Code-division multiple access (CDMA). Around ten years later, 4G revolutionized the daily lives of people around the world with the proliferation of smart devices and the development of mobile-oriented applications and social media. To achieve high data transmission rates, multiple technologies were employed in 4G Long-Term Evolution (LTE) networks, including multiple input and multiple output (MIMO) antennas and orthogonal frequency division multiplexing (OFDM) \cite{Dang2020}. Further technological innovations continue to emerge with the deployment of the current generation (5G) and the advancement of research towards the next generation (6G).              

Unlike, the previous generations (1G to 4G) which aimed to connect people through mobile broadband communications, 5G and beyond focus on connecting devices with the rise of Internet of Things (IoT) \cite{tong_zhu_2021}. Technologies including network densification, massive MIMO and network slicing are employed in 5G to cope with the increasing number of connected devices and support new services. 5G networks enabled a variety of applications such as high-definition video streaming, virtual reality (VR) and augmented reality (AR) applications, smart cities, remote healthcare and industrial and vehicular automation. 
In the IMT-2020 vision, the ITU-R identified three categories of 5G use cases, namely enhanced mobile broadband (eMBB), ultra-reliable low-latency communications (URLLC), and Massive machine-type communications (mMTC) \cite{series2015imt}. 

While 5G networks are being deployed and commercialized, researchers are shifting their focus to the next generation of communication networks. 6G networks are envisioned to be human-centric, rather than machine-, application- or data-centric \cite{Dang2020} and to have enhanced capabilities compared to 5G networks. In terms of key performance indicators (KPIs), higher data rates, lower latency, increased reliability and security, and massive connectivity are expected. For example, peak data rates of 1 $Tb/s$ are anticipated in 6G, compared to a few tens of $Gb/s$ for 5G. A reduced latency is also envisaged going from 1 $ms$ in 5G to 10–100 $\micro s$ in 6G. Additionally, reliability is estimated to reach seven nines (99.99999 $\%$) in 6G compared to five nines (99.999 $\%$) for 5G, while security is expected to attain very high levels in next generation networks \cite{wu20216g}. Furthermore, other key capabilities of 6G are envisioned by Huawei's researchers in their book "6G: The Next Horizon" \cite{tong_zhu_2021}, including better spectral, energy and cost efficiency, very high localization and sensing accuracy, and higher intelligence. Regarding application scenarios, 6G networks are envisaged to support five categories \cite{Dang2020}: 
\begin{enumerate}
    \item Enhanced Mobile Broadband Plus (eMBB-Plus) is the improved version of eMBB in 5G, enabling traditional mobile communications with higher data rates, extended coverage and enhanced security, and privacy.  
    \item Big Communications (BigCom) aims to connect the unconnected by providing basic connectivity services to users in remote areas.
    \item Secure Ultra-Reliable Low-Latency Communications (SURLLC) is an enhanced version of unified URLLC and mMTC from 5G, it offers higher reliability, lower latency and increased security levels.
    \item Three-Dimensional Integrated Communications (3D-InteCom) incorporates the application scenarios where the network operation and optimization consider three-dimensional communications taking into account the heights of aerial, space, and underwater nodes. 
    \item Unconventional Data Communications (UCDC) encompasses applications that do not fit into any of the other categories, including new communication paradigms such as holographic and tactile communications.
\end{enumerate}

In order to support the large variety of applications and satisfy the target KPIs of 6G networks, six categories of key enabling technologies are discussed in \cite{tong_zhu_2021}:
\begin{enumerate}
    \item New spectrum: Frequencies in the mmWave, THz, and optical bands are necessary to serve applications requiring ultra-high data rates, such as AR/VR and holographic applications.
    \item Joint sensing and communication (JSAC): Using mmWave and THz frequencies, sensing features can be incorporated into communication systems enabling higher accuracy and resolution. While communication signals can be exploited to perform object detection and activity recognition, sensing services can improve the communication performance through beam alignment and channel prediction.
    \item Artificial intelligence (AI) technologies: The role of AI in 6G networks can be examined from two perspectives; networking for AI where 6G networks will be designed and optimized to accommodate AI applications, and AI for networking where AI techniques are employed to optimize the network's operation and management \cite{B.3.1}.
    \item Integrated terrestrial and non-terrestrial networks (NTNs): Users in remote and rural areas remain unconnected or under-connected, leading to the digital divide. Thus, the integration of NTN with terrestrial networks advocates for cost-effective, seamless global connectivity.
    \item Native trustworthiness: As 6G is expected to be human-centric, network security and data privacy are critical features. Secure communications can be offered by blockchain and quantum technologies.
    \item Green communications and sustainable networking: With the growing number of network equipment's and connected devices, energy efficiency has become a critical aspect in the design of 6G networks. This can be realized through energy-aware protocols and energy harvesting methodologies.
\end{enumerate}

In this work, we focus on the integration of terrestrial and non-terrestrial networks (TN-NTN) in the 6G era, specifically with respect to the aspects of network virtualization. In fact, the integration of non-terrestrial platforms in 6G networks introduces multiple challenges particularly in terms of network management, network interoperability, and QoS requirements assurance. This is mainly due to the large-scale and heterogeneous network topology, the dynamic environment and limited on-board resources of network nodes such as satellites, high altitude platform stations (HAPS) and unmanned aerial vehicles (UAVs). In this context, network virtualization technologies, including Software-Defined Networking (SDN), Network Function Virtualization (NFV), and Network Slicing (NS), can be adopted to tackle these issues. On the one hand, SDN promotes network programmability and reconfigurability by decoupling the data/control planes and implementing the network control logic in a logically centralized fashion using SDN controllers \cite{SDN2014survey}. On the other hand, NFV improves network flexibility and reduces deployment costs through the separation of the network functions from the underlying hardware on which they are running and the creation of virtual network functions (VNFs) implemented on virtual machines \cite{B.1.9}. Moreover, network slicing provides multi-tenant software-oriented networks and offers optimized solutions for various market scenarios with different performance requirements by enabling multiple virtual customized networks to operate on shared physical infrastructure \cite{B.1.2}. Therefore, employing these networking paradigms in next-generation networks enables a seamless TN-NTN integration, efficient network management, and enhanced network performance.

This survey offers a comprehensive review on the application of network virtualization approaches in 6G integrated TN-NTNs. We consider the three NTNs segments including Satellite-Terrestrial (S-T), Aerial-Terrestrial (A-T) and Satellite-Aerial-Terrestrial (S-A-T), and we cover the three main virtualization technologies, namely SDN, NFV and network slicing. In Table \ref{tab:Survey_Comparison}, we provide a summary on the main related surveys and a comparison in terms of covered topics. Firstly, the surveys in \cite{rinaldi2020non,Liu2018,A.2} provide a global overview on non-terrestrial networks taking into account the three NTNs segments. They present the NTNs integration with terrestrial networks from different aspects including architectures, use cases, network management, as well as performance analysis and network optimization. Secondly, references \cite{B.1.1,B.1.2,B.1.5,B.1.8} present comprehensive reviews on network virtualization and softwarization, particularly the concepts of SDN, NFV and NS while detailing their architectures, key principles, enabling technologies and use cases. Hence, these works focus on either integrated terrestrial and non-terrestrial networks, or on virtualization technologies, independently. Thirdly, the studies presented in \cite{B.2.3.1,B.2.3.2,B.2.3.3} review research efforts combining network virtualization technologies with non-terrestrial networks. While \cite{B.2.3.1} and \cite{B.2.3.3} focus on SDN/NFV and network slicing in 5G, respectively, in the context of UAV networks, the authors of \cite{B.2.3.2} discuss the SDN paradigm in satellite networks (S-T segment). Therefore, although the aforementioned surveys \cite{B.2.3.1,B.2.3.2,B.2.3.3} combine NTNs with virtualization techniques, they either consider only one NTNs segment, namely the A-T segment in \cite{B.2.3.1,B.2.3.3} and the S-T segment in \cite{B.2.3.2}, or cover a specific virtualization technology. The main contributions of this work can be summarized as follows:
\begin{itemize}
    \item We give an overview on non-terrestrial networks and the challenges of their integration in 6G, and a background on network virtualization and its enablers i.e. SDN, NFV and network slicing.
    \item We highlight the role of AI models in network virtualization and summarize the major research areas where AI algorithms are usually employed in SDN, NFV, and network slicing.
    \item We outline the main challenges associated with the adaptation of SDN, NFV, and network slicing technologies in integrated terrestrial and non-terrestrial networks.
    \item We propose a taxonomy of integrated TN-NTNs virtualization where we comprehensively review the relevant contributions and categorize them based on a four-level classification.  
    \item We provide a summary on the simulation and implementation tools utilized in the performance evaluation and testing of virtualization-based networks. 
    \item We identify several open issues and give insights on future research directions.
\end{itemize}

\begin{table*}
\centering
\begin{tabularx}{\textwidth}{ |p{0.05\textwidth}|X|p{0.02\textwidth}p{0.02\textwidth}p{0.02\textwidth}p{0.02\textwidth}p{0.02\textwidth}p{0.02\textwidth}|}

\hline
\multirow{3}{*}{Ref.} &
  \multirow{3}{*}{Summary} &
  \multicolumn{6}{c|}{Covered Topics} \\ \cline{3-8} 
 &
   &
  \multicolumn{3}{c|}{NTNs Segment} &
  \multicolumn{3}{c|}{\begin{tabular}[c]{@{}l@{}}Virtualization\\ Technologies\end{tabular}} \\ \cline{3-8} 
 &
   &
   \multicolumn{1}{c|}{S-T} &
  \multicolumn{1}{c|}{A-T} &
  \multicolumn{1}{c|}{S-A-T} &
  \multicolumn{1}{c|}{SDN} &
  \multicolumn{1}{c|}{NFV} &
  NS \\ \hline
\cite{rinaldi2020non} &
  Review on the characteristics and architectures of non-terrestrial networks, and their role in 3G, 4G and 5G ecosystems, highlighting the main contributions on NTNs and the research efforts conducted by the 3GPP. &
  \multicolumn{1}{c|}{$\checkmark$} &
  \multicolumn{1}{c|}{$\checkmark$} &
  \multicolumn{1}{c|}{$\checkmark$} &
  \multicolumn{1}{c|}{$\partial$} &
  \multicolumn{1}{c|}{$\partial$} &
  $\times$ \\ \hline
\cite{Liu2018} &
  Survey on space-air-ground integrated networks  (SAGIN) focusing on related works in system integration design, resource allocation, mobility management and routing, as well as optimization and performance analysis.&
  \multicolumn{1}{c|}{$\checkmark$} &
  \multicolumn{1}{c|}{$\checkmark$} &
  \multicolumn{1}{c|}{$\checkmark$} &
  \multicolumn{1}{c|}{$\partial$} &
  \multicolumn{1}{c|}{$\partial$} &
  $\times$ \\ \hline
\cite{A.2} &
Survey on the evolution of integrated terrestrial and non-terrestrial networks from 5G to 6G, from the perspective of IoT and MEC networks, mmWave and THz spectrum bands, as well as ML applications.  &
  \multicolumn{1}{c|}{$\checkmark$} &
  \multicolumn{1}{c|}{$\checkmark$} &
  \multicolumn{1}{c|}{$\checkmark$} &
  \multicolumn{1}{c|}{$\partial$} &
  \multicolumn{1}{c|}{$\partial$} &
  $\times$ \\ \hline
\cite{B.1.1} &
Survey on SDN paradigm presenting its key principles, and detailing the building blocks of an SDN architecture.&
  \multicolumn{1}{c|}{$\times$} &
  \multicolumn{1}{c|}{$\times$} &
  \multicolumn{1}{c|}{$\times$} &
  \multicolumn{1}{c|}{$\checkmark$} &
  \multicolumn{1}{c|}{$\partial$} &
  $\times$ \\ \hline
\cite{B.1.2} &
Review on network slicing in the 5G era, explaining its main concepts, use cases and enablers and describing RAN and core network slicing. &
  \multicolumn{1}{c|}{$\times$} &
  \multicolumn{1}{c|}{$\times$} &
  \multicolumn{1}{c|}{$\times$} &
  \multicolumn{1}{c|}{$\partial$} &
  \multicolumn{1}{c|}{$\partial$} &
  $\checkmark$ \\ \hline
  \cite{B.1.5} &
Review on NFV architecture, design considerations and implementations, and standardization efforts, while discussing related notions including cloud computing and SDN. &
  \multicolumn{1}{c|}{$\times$} &
  \multicolumn{1}{c|}{$\times$} &
  \multicolumn{1}{c|}{$\times$} &
  \multicolumn{1}{c|}{$\partial$} &
  \multicolumn{1}{c|}{$\checkmark$} &
  $\times$ \\ \hline
\cite{B.1.8} &
    Survey on works in 5G network slicing using SDN and NFV, emphasizing on network slicing architectures, management and orchestration, and practical implementations developed in industry and academia.  &
  \multicolumn{1}{c|}{$\times$} &
  \multicolumn{1}{c|}{$\times$} &
  \multicolumn{1}{c|}{$\times$} &
  \multicolumn{1}{c|}{$\checkmark$} &
  \multicolumn{1}{c|}{$\checkmark$} &
  $\checkmark$ \\ \hline
\cite{B.2.3.1} &
  Review on efforts in SDN-based and NFV-based UAV networks, providing taxonomies of the works based on application scenarios enabled by SDN/NFV in UAV networks. &
  \multicolumn{1}{c|}{$\times$} &
  \multicolumn{1}{c|}{$\checkmark$} &
  \multicolumn{1}{c|}{$\times$} &
  \multicolumn{1}{c|}{$\checkmark$} &
  \multicolumn{1}{c|}{$\checkmark$} &
  $\times$ \\ \hline
\cite{B.2.3.2} &
Survey on studies in software-defined satellite networks, highlighting three satellite network architectures (single, two, and three-layer architecture) based on the integration of LEO, MEO, and GEO satellites.
   &
  \multicolumn{1}{c|}{$\checkmark$} &
  \multicolumn{1}{c|}{$\times$} &
  \multicolumn{1}{c|}{$\times$} &
  \multicolumn{1}{c|}{$\checkmark$} &
  \multicolumn{1}{c|}{$\times$} &
  $\times$ \\ \hline
\cite{B.2.3.3} &
  Survey on works in network slicing with UAVs focusing on the roles of UAVs in different categories of 5G uses cases (eMBB, mMTC, URLLC). 
  &
  \multicolumn{1}{c|}{$\times$} &
  \multicolumn{1}{c|}{$\checkmark$} &
  \multicolumn{1}{c|}{$\times$} &
  \multicolumn{1}{c|}{$\times$} &
  \multicolumn{1}{c|}{$\times$} &
  $\checkmark$ \\ \hline
This work &

Survey on network virtualization technologies in 6G integrated TN-NTNs considering the three NTNs segments and the three main virtualization technologies, namely SDN, NFV and network slicing, while highlighting the role of AI algorithms. &
  \multicolumn{1}{c|}{$\checkmark$} &
  \multicolumn{1}{c|}{$\checkmark$} &
  \multicolumn{1}{c|}{$\checkmark$} &
  \multicolumn{1}{c|}{$\checkmark$} &
  \multicolumn{1}{c|}{$\checkmark$} &
  $\checkmark$ \\ \hline
\end{tabularx}
\caption{Summary and comparison of related surveys ("$\checkmark$": topic covered, "$\partial$": topic partially covered, "$\times$": topic not covered).}
\label{tab:Survey_Comparison}
\end{table*}

\begin{figure}
  \centering
  \includegraphics[width=0.85\linewidth]{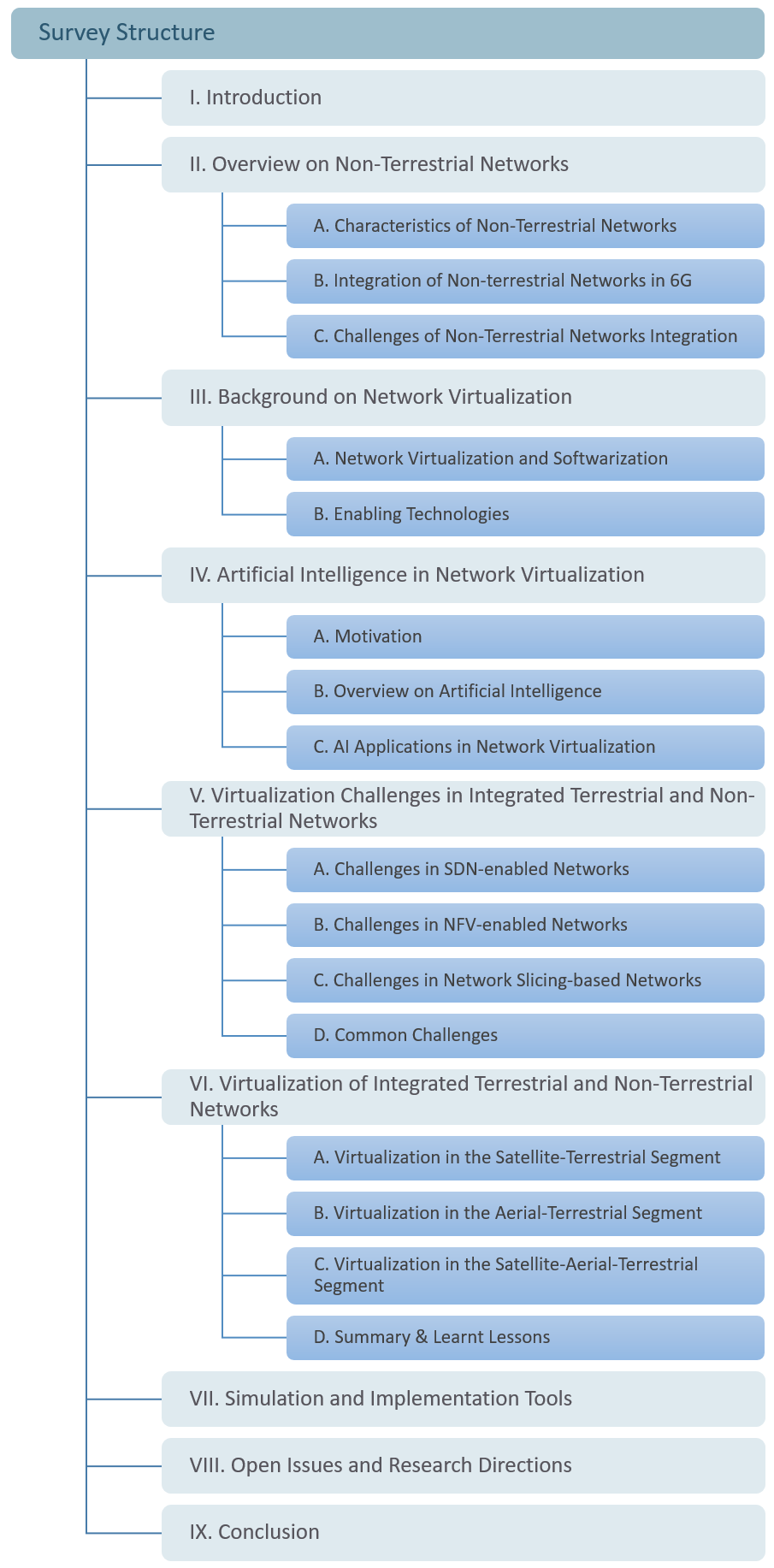}
\caption{Survey Structure.}
\label{fig:Survey_Structure}
\end{figure}

The remainder of this paper is organized as follows. Section \ref{Overview_NTN} gives an overview on non-terrestrial networks indicating their unique characteristics, as well as the key drivers, the application scenarios, and the challenges of their integration in 6G. In section \ref{Bckground_Vir}, we explain the fundamentals of network virtualization and its main enabling technologies i.e. SDN, NFV and network slicing. Section \ref{AI_Vir} highlights the role of AI models in network virtualization discussing the motivation and the primary research areas where AI algorithms are often used in SDN, NFV, and network slicing. Section \ref{Challenges_Vir_TN_NTN} deals with the most prevalent challenges facing the implementation of virtualization technologies in integrated TN-NTNs. Section \ref{Taxonomy_Vir_TN_NTN} is dedicated to reviewing the relevant contributions investigating the application of virtualization technologies in integrated networks, where a comprehensive taxonomy based on a four-level classification is provided. Section \ref{Sim_tools} summarises the simulation and implementation tools used to test, validate, and evaluate the performance of networks employing virtualization technologies. In section \ref{Open_issues}, we identify several open issues and discuss potential research directions for advancing the adaptation of virtualization technologies in next-generation networks. Finally, section \ref{Conclusion} concludes the paper, and the survey structure is depicted in Fig.\ref{fig:Survey_Structure}.

\section{Overview on Non-Terrestrial Networks}\label{Overview_NTN}
In this section, we provide an overview on non-terrestrial networks highlighting the unique characteristics of non-terrestrial platforms including satellites, HAPS and UAVs. We also present the key drivers, application scenarios, and challenges of non-terrestrial networks integration in 6G.

\subsection{Characteristics of Non-Terrestrial Networks}
Non-terrestrial networks (NTNs) are composed of two types of platforms, namely, aerial platforms including unmanned aerial vehicles (UAVs) and high altitude platform stations (HAPS), and spaceborne platforms including non-geostationary earth orbit (NGEO) (low earth orbit (LEO), medium earth orbit (MEO)) and geostationary earth orbit (GEO) satellites. Table \ref{tab:CompNTNTplatforms} provides a comparison between different non-terrestrial (NT) platforms in terms of altitude, visibility window, propagation delay, coverage, and energy supply. NTs are accessed through earth gateway stations which connect the NT platforms to the core network and Very Small Aperture Terminals which can be specific satellite terminals or 3GPP User Equipments (UEs). In TN-NTN architectures, two types of links can be identified: service links and feeder links. On one hand, the service link is established when terrestrial or non-terrestrial platforms provide services to NT nodes or end users. On the other hand, the feeder link connects NT nodes to terrestrial gateways \cite{rinaldi2020non}.

\begin{table*}[h]
\centering
\begin{tabular}{|c|c|c|c|c|c|}
\hline
\multicolumn{1}{|l|}{}                                                & GEO Satellite                                                     & MEO Satellite                                                     & LEO Satellite                                                     & HAPS                                                              & UAV                                                       \\ \hline
Altitude Range                                                        & 35786 km                    & 7000 $-$ 25000 km           & 300 $-$ 1500 km              & around 20 km               & $\leq$ 10 km  \\ \hline
Mobility  &    stationary                                                              &       medium fast                                                          &        fast                                                           &       quasi-stationary     &          very fast                                                \\ \hline
\begin{tabular}[c]{@{}c@{}}Propagation delay\\ (one way)\end{tabular} & about 270 ms                     & about 100 ms            & \textless 40 ms                   &  medium                                                                 &       medium                                                    \\ \hline
Coverage                                                              & up to 3500 km                                                   & up to 1000 km                                                    & up to 1000 km                                                    & around 60 km              &  small                                                         \\ \hline
Energy supply                                                         & \begin{tabular}[c]{@{}c@{}}Solar panel\\ and battery\end{tabular} & \begin{tabular}[c]{@{}c@{}}Solar panel\\ and battery\end{tabular} & \begin{tabular}[c]{@{}c@{}}Solar panel\\ and battery\end{tabular} & \begin{tabular}[c]{@{}c@{}}Solar panel\\ and battery\end{tabular} & Lithium battery                                           \\ \hline
\end{tabular}
\caption{Comparison of the characteristics of NTN platforms \cite{kurt2021vision,rinaldi2020non,Liu2018,wang2021blockchain,zhang2020survey}.}
\label{tab:CompNTNTplatforms}
\end{table*}

\begin{figure*}[h]
        \centering
        \begin{subfigure}[b]{0.475\textwidth}
            \centering
            \includegraphics[width=0.65\textwidth]{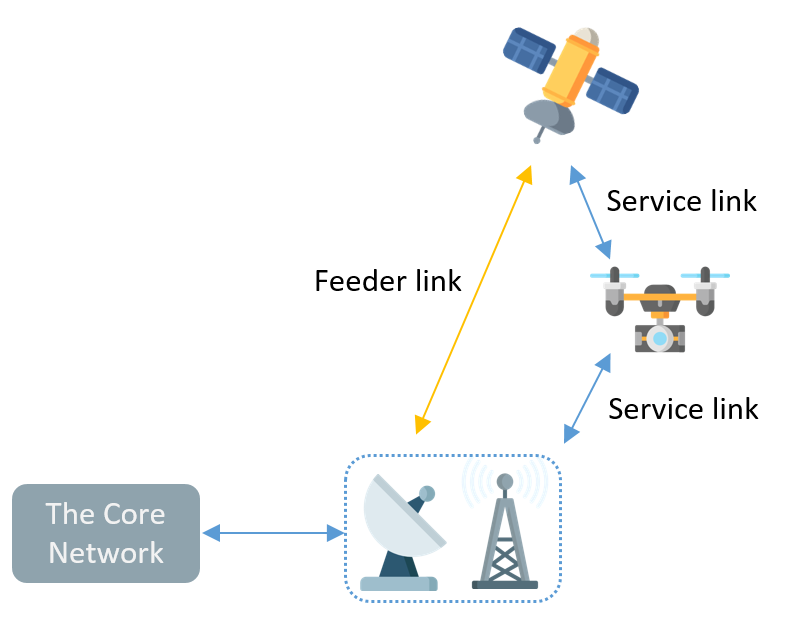}
            \caption{User NT node.}    
            \label{fig:User NT node}
        \end{subfigure}
        \hfill
        \begin{subfigure}[b]{0.475\textwidth}  
            \centering 
            \includegraphics[width=0.8\textwidth]{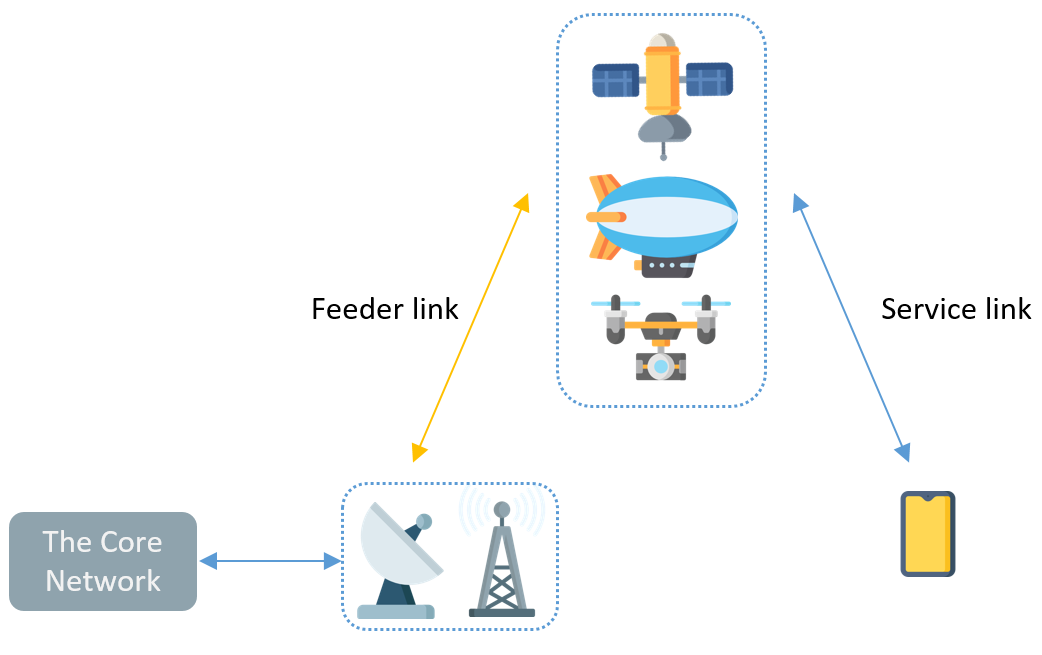}
            \caption{Relay NT node for end users.}   
            \label{fig:Relay NT node for UE}
        \end{subfigure}
        \vskip\baselineskip
        \begin{subfigure}[b]{0.46\textwidth}   
            \centering 
            \includegraphics[width=\textwidth]{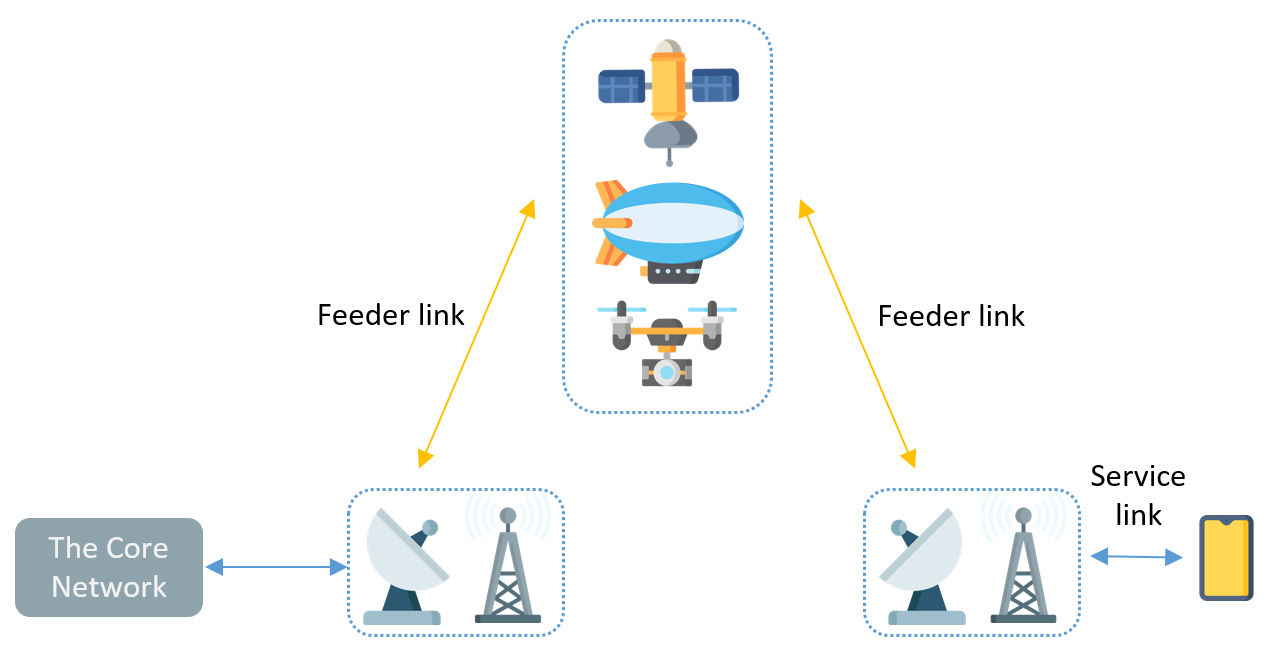}
            \caption{Relay NT node for backhauling.}    
            \label{fig:Relay NT node for B}
        \end{subfigure}
        \hfill
        \begin{subfigure}[b]{0.49\textwidth}   
            \centering 
            \includegraphics[width=\textwidth]{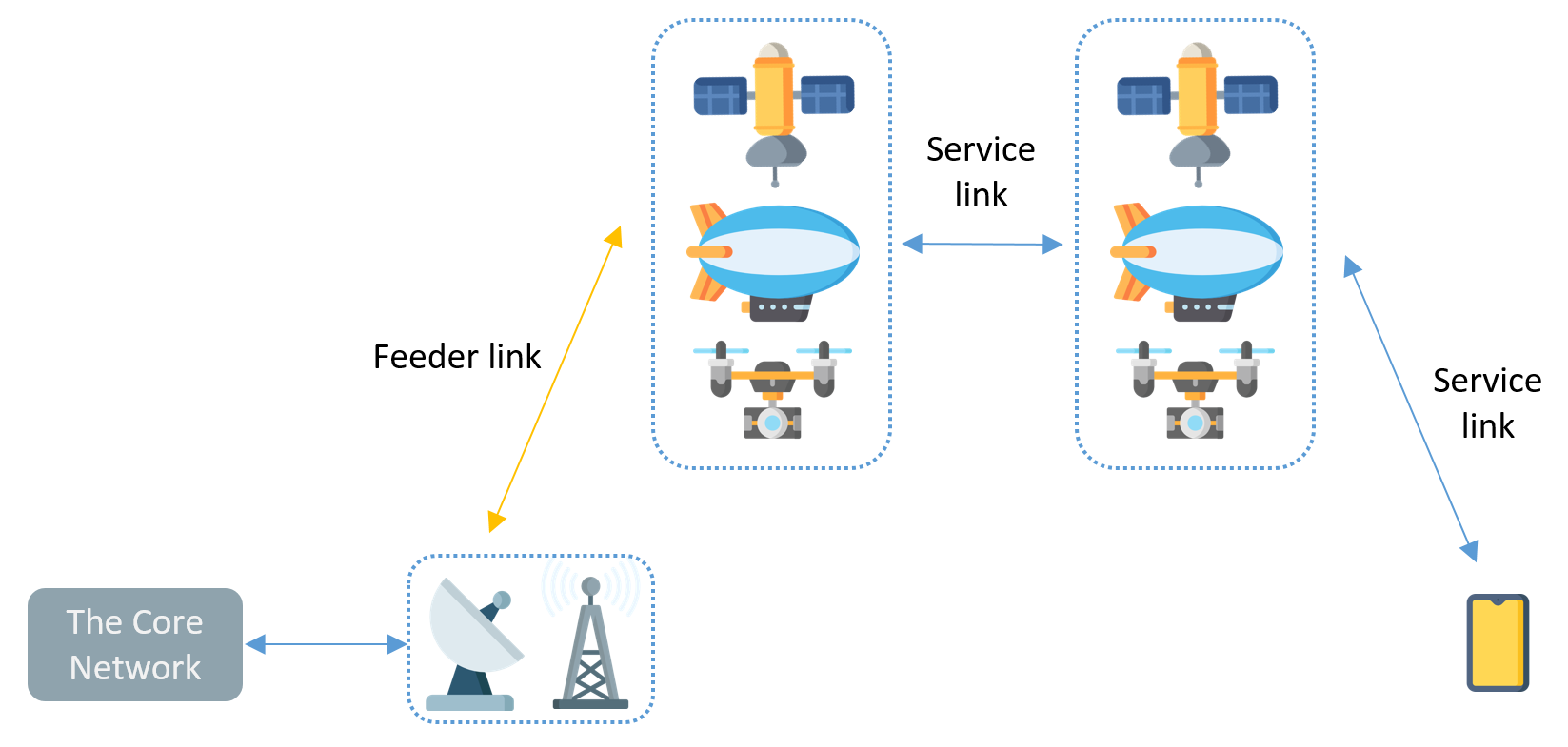}
            \caption{BS NT node.}    
            \label{fig:BS NT node}
        \end{subfigure}
        \caption{Different roles of NTN platforms in integrated TN-NTN.}
    \label{fig:NTNrole}
\end{figure*}

Non-terrestrial nodes can play a variety of roles when integrated into the functioning of terrestrial networks to serve a particular application, as illustrated in Fig. \ref{fig:NTNrole}. In general, the NT node can be a user, a relay or a base station (BS) \cite{A.2,rinaldi2020non}. First, in the case where it acts as a user, the NT platform is served through terrestrial BSs (TBSs). For example, a UAV can be served directly by a TBS or by a satellite relaying data from a terrestrial gateway as shown in Fig. \ref{fig:User NT node}. Second, the NT platform, with transparent payload, can act as a relay for two goals. on one hand, it can offer backhaul services by connecting a TBS to the core network through feeder links as depicted in Fig. \ref{fig:Relay NT node for B}. On the other hand, the NT node can enable connectivity by relaying data from TBSs to end users as illustrated in Fig.\ref{fig:Relay NT node for UE}. Third, the NT node can play the role of a BS serving terrestrial UEs or NT platforms as indicated in Fig. \ref{fig:BS NT node}. Hence, the NT should support regenerative payload with sufficient computing and processing capabilities.    

Resulting from the high altitude and mobility of NT nodes, non-terrestrial networks are distinct from conventional terrestrial networks by a number of key features mainly in terms of signal propagation, coverage and handovers, Doppler effect and platform deployment  \cite{A.2,zhang2020survey,rinaldi2020non,wang2021blockchain}. NT nodes, especially GEO and MEO satellites, are located at large distances from terrestrial end users. As a result, NTN communications suffer from longer propagation delay and higher pathloss in comparison to their terrestrial counterparts. Such NTN feature presents a bottleneck for application scenarios where low or even ultra-low latency is a critical requirement. Moreover, as shown in Table \ref{tab:CompNTNTplatforms}, NT nodes have different coverage areas which leads to different frequencies of handovers. For instance, because of their mobility, NGEO satellites have variable coverage, characterized by their visibility window, resulting in periodic and frequent handovers, while GEO satellites have large and stable coverage. Meanwhile, handovers occur in TNs during the movement of users between cells, due to the small and fixed coverage of TBSs. Furthermore, although Doppler effects exist in both types of networks (TN and NTN), the Doppler shifts induced by the high mobility of NT platforms in NTNs, primarily LEO satellites, are greater than those caused by user mobility in TNs. Finally, while the deployment of TNs is an expensive and long-term investment, rendering it an unfavorable choice in certain cases, such as remote areas connectivity, deploying NTNs can be an appealing alternative in such scenarios where aerial platforms can be deployed quickly and temporarily at cheap expense. Additionally, although satellites have long-term deployment and are more expensive than UAVs or HAPS, they offer vast coverage areas compared to aerial and terrestrial nodes.

\subsection{Integration of Non-terrestrial Networks in 6G}
The integration of terrestrial networks (TN) and non-terrestrial networks (NTN) gave birth to a new paradigm of networks characterized by a three-layered architecture composed of ground, air, and space segments. Such networks are referred to as integrated TN-NTN, space-air-ground integrated networks (SAGIN), or ground-air-space (GAS) integrated networks. Each SAGIN segment has benefits and limitations which are summarized in Table \ref{tab:SAGINseg+-}.

\begin{table}[ht]
\centering
\begin{tabular}{|c|c|c|}
\hline
Segment & Benefits                                                                                      & Limitations                                                                                             \\ \hline
Space   & \begin{tabular}[c]{@{}c@{}}- Large coverage\\ - Broadcast/multicast\\ capabilities\end{tabular}                                                                                & \begin{tabular}[c]{@{}c@{}}- High mobility\\ - Long propagation delay\\ - Limited capacity\end{tabular} \\ \hline
Air     & \begin{tabular}[c]{@{}c@{}}- Large coverage\\ - Flexible deployment\\ - Low cost\end{tabular} & \begin{tabular}[c]{@{}c@{}}- High mobility\\ - Low reliability\\ - Limited capacity\end{tabular}                             \\ \hline
Ground  & \begin{tabular}[c]{@{}c@{}}- High data rates \\ - Abundant resources\end{tabular}                                                                             & \begin{tabular}[c]{@{}c@{}}- Limited coverage\\ - Vulnerability to natrual\\ disasters\end{tabular}     \\ \hline
\end{tabular}
\caption{Benefits and limitations of SAGIN segments \cite{Liu2018,wang2021blockchain,SAT.C.SDN.8_B.2.2}}
\label{tab:SAGINseg+-}
\end{table}

Numerous applications with different Quality of Service (QoS) requirements will be supported by 6G networks. Because of their distinct characteristics from TNs, NTNs can complement 6G terrestrial networks to meet the needs of various use cases. In essence, \textit{service ubiquity}, \textit{continuity}, and \textit{scalability} are the main key drivers for TN-NTN integration \cite{A.2,tong_zhu_2021,zhang2020survey,rinaldi2020non}:

\begin{itemize}
    \item \textit{Service ubiquity}: airborne and space platforms can cost-efficiently deliver ubiquitous services by covering remote and rural locations, expanding 6G networks coverage. 
    \item \textit{Service continuity}: NTN nodes offer continuous services for IoT devices or onboard mobile vehicles to enhance 6G service reliability. 
    \item \textit{Service scalability}: NTNs facilitate 6G service scalability with broadcast and multicast capabilities, ensuring streaming content delivery to wide regions and data offloading to network edges.
\end{itemize}

The efficient integration of the three segments is expected to enable a wide range of use cases, particularly in the 6G era. In \cite{A.2}, six categories of 6G integrated TN-NTN use cases are envisioned:  
\begin{enumerate}
    \item \textit{Ubiquitous Internet} can be achieved by integrating NTN access points, such as LEO satellites and airborne platforms, into the terrestrial Internet allowing Internet services availability everywhere on the planet.    
    \item \textit{Pervasive intelligence} is enabled by AI for networking and networking for AI. Not only can space/air nodes provide a global dataset to improve the performance of AI-based solutions, but they can also serve as computing and storage units, facilitating AI-based network management through edge AI. 
    \item \textit{Integrated localization sensing and communications (JSAC) services} are key enabling technologies of 6G networks. By offering reliable line-of-sight (LoS) links and information on the device's location and orientation in 3D fashion, the NTN platforms can improve the accuracy of sensing and localization measurements and allow context-aware communications.
    \item \textit{Beyond a visual LoS (BVLoS) connected UAVs} can be supported by integrated terrestrial and satellite networks to expand the control and reachability of UAVs beyond a visual LoS. This would result into improvements in the reliability, throughput, and coverage of aerial networks.
    \item \textit{Aerial Interactive telepresence} allows virtual human presence via UAVs in scenarios where physical human presence can be dangerous or costly. It can be improved via augmented reality (AR) technology to offer haptic interations in a 3D environment and through TN-NTN integration for seamless connectivity.     
    \item \textit{Convergence of networking and computing} can be attained through NT nodes which can provide computing services and perform coordination between network edge units in order to achieve computing-aware networking.  
\end{enumerate}

\subsection{Challenges of Non-Terrestrial Networks Integration} 
The integration of NTNs into 6G networks faces several challenges particularly in terms of network management, network interoperability, QoS requirements assurance, mobile nodes dynamic networking, network security, as well as business models aspects.

\begin{enumerate}
    \item Network management: the large number of connected devices present in integrated TN-NTNs have a diversity in terms of configuration and control interfaces, as well as hardware and software specifications. As a result, network management is highly complex and flexible network reconfiguration is difficult \cite{SAT.C.SDN.8_B.2.2,Liu2018}. 
    \item Network interoperability: in current communication systems, operators provide vertically integrated stacks \cite{geraci2022integrating}, resulting in limited interoperability especially in the context of integrated TN-NTNs. 
    \item QoS requirements assurance: integrated TN-NTNs, in the 6G era, are expected to provide a huge variety of services with different QoS requirements, particularly in terms of latency, reliability and throughput. Efficient and dynamic resource allocation should be carried out to ensure QoS provisioning for each service \cite{SAT.C.SDN.8_B.2.2}. 
    \item 3D mobility management: the mobility of NTN platforms results into variation of resource availability and a high frequency of handovers, which requires dynamic networking and resource allocation \cite{SAT.C.SDN.8_B.2.2,geraci2022integrating}.  
    \item Network security: in 6G networks, security and privacy are extremely important, however, as integrated networks feature dynamic topologies, open links, and mobile nodes, enabling high levels of security is a challenging task \cite{Liu2018}. Besides conventional security techniques, secure communications based on quantum technologies can improve network security and data privacy \cite{A.3}.
    \item Business models aspects: since multiple business actors can be included in integrated TN-NTNs service delivery to end-users, new business models should be developed to identify the roles of each party and the relationships between different entities \cite{B.2.3}.
\end{enumerate}

The aforementioned issues of TN-NTN integration can be solved using network virtualization and softwarization paradigms by adopting their enabling technologies, such as network slicing, software-defined networking (SDN) and network function virtualization (NFV).

\section{Background on Network Virtualization}\label{Bckground_Vir}
This section covers the fundamentals of network virtualization where we present its basic concepts and its main enabling technologies including software-defined networking, network function virtualization, network slicing.

\subsection{Network Virtualization and Softwarization}
Network virtualization and softwarization are two innovative paradigms introduced in 5G networks to enable network reconfigurability, programmability, and flexibility, by separating the network functionalities and the underlying hardware \cite{B.1.4,B.1.8}.

\begin{itemize}
    \item \textit{Network softwarization:} Softwarization defines the concept where network functionalities run on software rather than hardware severing the software-hardware coupling. As a result, updating existing functions or adding new functionalities is realized by updating the software which increases the network flexibility and reduces the capital expenditures (CAPEX) and operating expenses (OPEX) \cite{B.1.4,B.1.8}.
    \item \textit{Network virtualization:} Virtualization in networking is the concept of creating virtual instances, defined by abstracted software-based representations, of the network entities and network hardware and software resources. This allows the software to run on commodity hardware rather than specific equipment \cite{B.1.4,B.1.2,B.1.8}. Network virtualization is based on three main principles, namely abstraction, co-existence, and isolation \cite{B.3.1}. First, the abstraction creates virtual instances of network components, including, nodes and links, and network resources masking the physical infrastructure's specifics. Second, the co-existence allows multiple virtual networks to share the same physical infrastructure. Third, the isolation ensures the independent functioning of the various virtual networks that share the same physical infrastructure \cite{B.3.1,B.1.3}. Network virtualization offers simplified network management and scalability, flexible service provisioning and efficient resource utilization. It, also, provides service-centric networking and guarantees QoS requirements. Virtualization can be realized on different levels including node, link, and resource levels as well as the network level. 
\end{itemize}

The two technologies, i.e. network virtualization and network softwarization, complement each other where while the latter decouples networks functionalities from the underlying hardware creating a softwarization layer above the physical infrastructure layer, the former constructs virtual instances of the network abstracting its components and resources, forming a higher virtualization layer that facilitates network management for developers in the application layer.

\subsection{Enabling Technologies} 
Implementing network softwarization and virtualization in next-generation networks requires multiple enabling technologies including software-defined networking, network function virtualization, network slicing, as well as cloud and edge computing \cite{B.1.4,B.1.2,B.1.3,B.3.1,B.1.8}. We focus on the first three main technologies in this survey.

\subsubsection{Software-Defined Networking (SDN)}
While conventional networks have inflexible decentralized architecture due to the coupling of the data and control planes, SDN is a networking paradigm that separates the two planes and implements the network control logic in a logically centralized fashion. To promote network flexibility, programmability and reconfigurability, SDN is based on four key concepts \cite{SDN2014survey,SDNsurvey2019application,kafetzis2022SDN,B.1.1}:
\begin{itemize}
    \item The separation of the control and data planes.
    \item The logical centralization of the control logic in external SDN controller.
    \item The flow-based packet forwarding decisions.
    \item The network programmability through software applications which run on top of the controller.
\end{itemize}

\begin{figure}[ht]
\centering\includegraphics[width=0.45\textwidth]{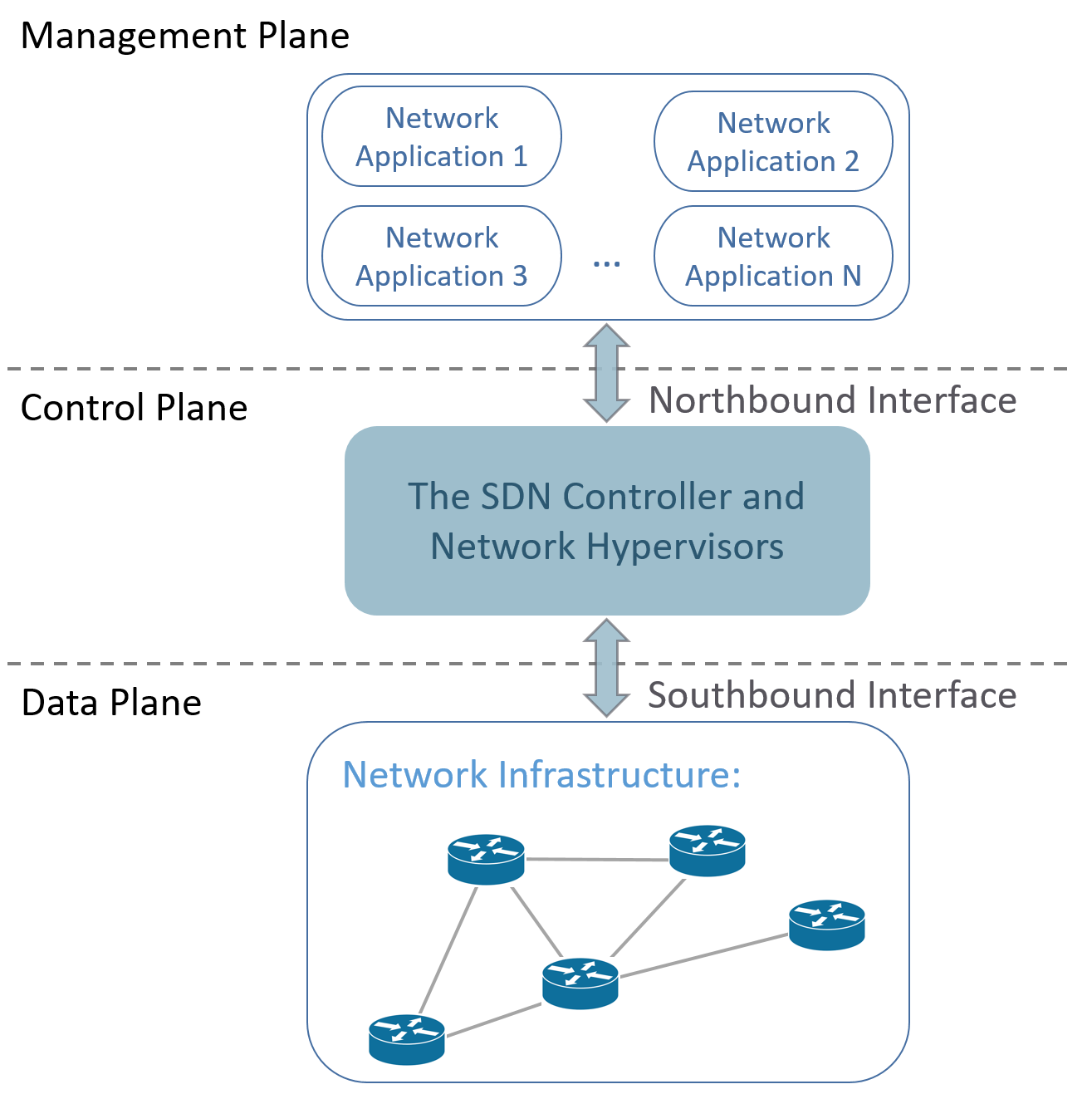}    \caption{Illustration of SDN architecture.}
    \label{fig:SDNconcept}
\end{figure}

We note that logically centralized network control does not imply
its physical centralization. Additionally, SDN can be identified as a network architecture with three planes, as illustrated in Fig.\ref{fig:SDNconcept}: 
 \begin{itemize}
     \item The data plane includes the network infrastructure and the southbound interfaces \cite{B.1.1}. With the aforementioned SDN principles, the networking devices in the physical infrastructure become simple packet forwarding devices without any intelligence. In order to control and communicate with these data plane elements, the SDN controller uses the southbound interfaces defined as standard and open application programming interfaces (APIs) and highlighting the data/control planes decoupling. Multiple southbound APIs can be found in the literature particularly OpenFlow \cite{mckeown2008openflow} is the most used protocol in SDN architectures.   
     
     \item The control plane is composed of the network hypervisors, the SDN controller and the northbound interfaces \cite{B.1.1,SDN2014survey}. Firstly, the network hypervisors enable the virtualization of the SDN architecture allowing the multi-tenacy and the slicing of the OpenFlow-based infrastructure. Secondly, the SDN controller, also known as the Network Operating System (NOS) is the key component in the SDN paradigm. It is a software platform running on commodity hardware offering abstractions and necessary resources for developers simplifying the programming of data plane devices. By logically centralizing the network intelligence, the NOS offers a global view of the network and solves the issues of traditional networks in terms of flexibility, reconfiguration and programmability. Thirdly, the northbound interfaces are APIs that enable the abstraction of the instructions, employed by the southbound APIs for the programming of forwarding elements. They are provided by the SDN controller for the application developers in the management plane.
     
     \item The management plane contains the network applications that defines the control logic which will be enforced by the control plane and executed by the data plane \cite{B.1.1}. The network applications in the SDN architecture can be divided in five categories namely traffic engineering, applications related to mobile and wireless networks, network monitoring and measurement applications, security-oriented applications and data centers networking. 
 \end{itemize}

Therefore, in the SDN architecture, a network policy is defined by the management plane, enforced by the control plane and executed by the data plane. For example, in order to send packets from source S to destination D, the network application in the management plane should select the routing path, and command the NOS in the control plane to set the corresponding forwarding rules which will be used by the data plane devices to route the packets from S to D \cite{B.1.1}.

\subsubsection{Network Function Virtualization (NFV)}

For the deployment of network functions (NFs) such as firewalls, Intrusion Detection Systems (IDSs) and Network Address Translators (NATs), conventional networks utilize middleboxes which are hardware equipments designed for specific purposes. This results into inflexible networks where the implementation of a new network function is expensive and time-consuming. NFV is based on the idea of separating the NFs from the underlying hardware on which they are running \cite{B.1.5,B.1.9}. Using virtualization techniques, virtual network functions (VNFs) are created and implemented on Virtual Machines (VMs). As a result, the CAPEX and OPEX are significantly reduced and new services can be deployed with higher flexibility and shorter time to market \cite{B.1.2}. 

\begin{figure}[ht]
\centering\includegraphics[width=0.47\textwidth]{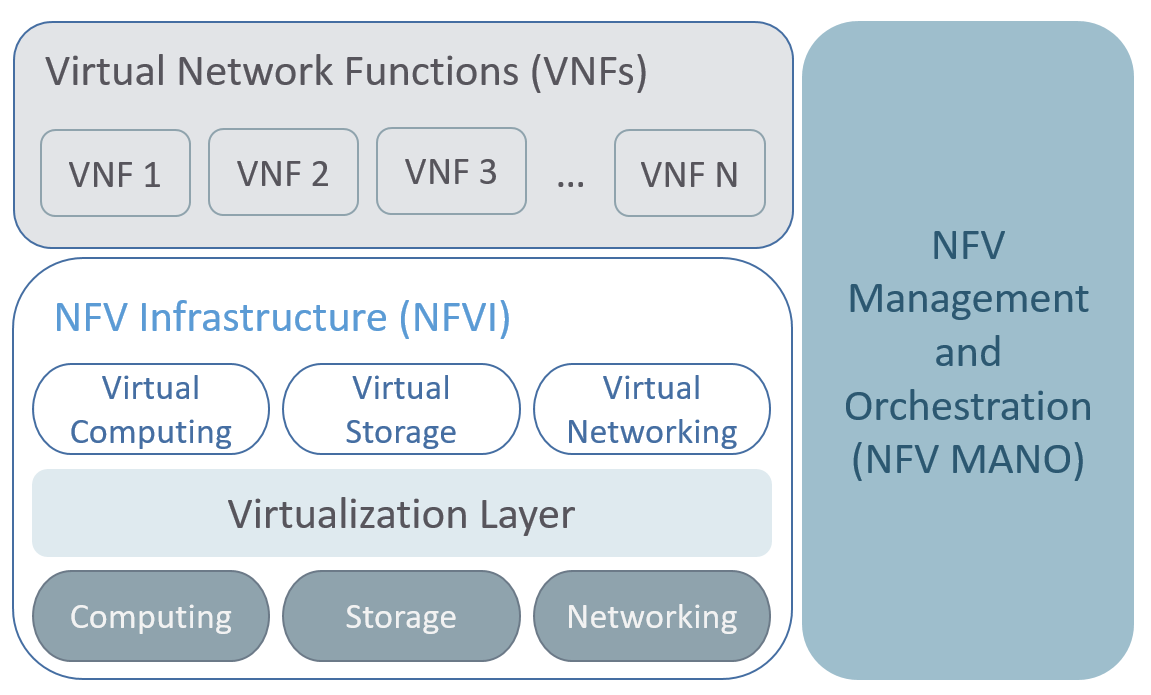}    
\caption{Illustration of NFV architecture \cite{ETSI_NFV}.}
    \label{fig:NFVarchitecture}
\end{figure}

In \cite{ETSI_NFV}, the European Telecommunications Standards Institute (ETSI) describes the NFV architecture containing three key components as shown in Fig.\ref{fig:NFVarchitecture}: the virtual network functions (VNFs) which are the software-based implementation of the NFs, the network function virtualization infrastructure (NFVI) composed of the physical and virtual resources needed for the NFV implementation and the NFV management and orchestration (NFV MANO) which ensures the VNFs provision and manages the life cycle of the resources and the VNFs \cite{B.1.2,B.1.5}.

\subsubsection{Network Slicing}

In 2015, The Next Generation Mobile Networks (NGMN) Alliance introduced network slicing in 5G networks as part of their 5G white paper \cite{el2015ngmn}. Network slicing enables multiple virtual networks to operate on shared physical infrastructure providing multi-tenant software-oriented networks \cite{B.1.2}. It is defined by the 3GPP as a technology that allows operators to build customized networks to offer optimized solutions for various market scenarios with different performance requirements \cite{B.1.2,3gpp2016study}. 

Network slicing is based on several key principles including automation, isolation, customization, elasticity, programmability, end-to-end (E2E) property, and hierarchical abstraction \cite{B.1.2,B.1.8,khan2020network}, defined as follows:
\begin{itemize}
    \item Automation permits third parties to request the creation of a slice with the needed Service Level Agreements (SLA) defining the desired requirements without manual intervention or fixed contractual agreements offering on-demand network slicing configuration. 
    \item Isolation guarantees that each tenant obtains the desired  performance and security requirements by properly specifying the level of resource separation.
    \item Customization ensures efficient utilization of the resources allocated for each tenant in order to satisfy their service requirements.  
    \item Elasticity assures that, with varying network parameters, the resource allocation of each network slice can meet the specified service requirements under varying network conditions. 
    \item Programmability permits third parties to manage the resources allocated to their slice using open APIs, which enables the automation, customization and elasticity properties of the network slicing.
    \item End-to-end is a network slicing property that facilitates service delivery from service providers to end-users by unifying different network layers and heterogeneous technologies.
    \item Hierarchical abstraction offers different levels of abstraction by repeating the resource abstraction in a hierarchical manner allowing multiple network slice services to be built on top of each other.
\end{itemize}

\begin{figure*}[ht]
\centering\includegraphics[width=0.8\textwidth]{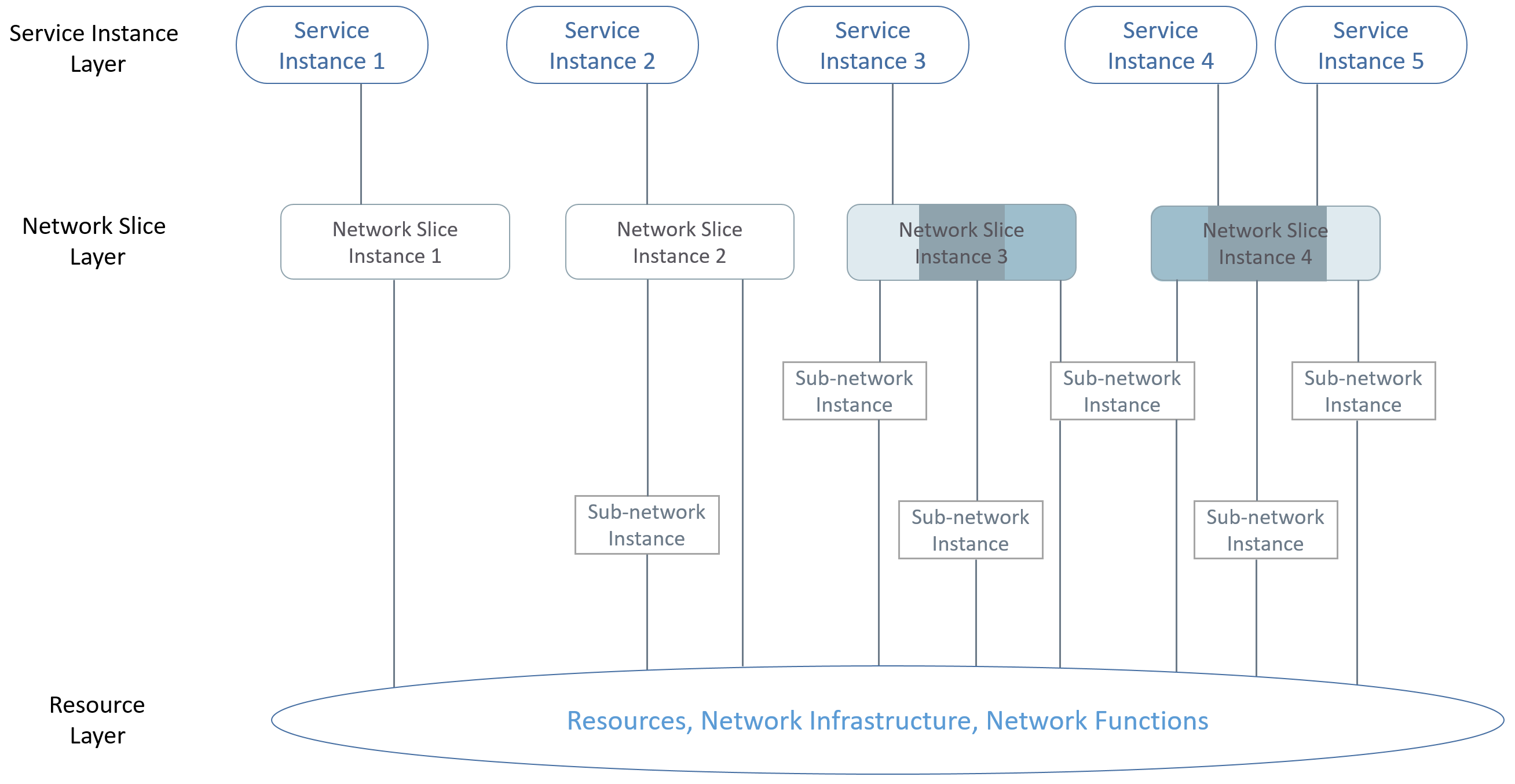}    \caption{Illustration of network slicing architecture by NGMN \cite{alliance2016description}.}
    \label{fig:NSillustration}
\end{figure*}

The principles of network slicing are implemented through its three-layered architecture which is described by the NGMN alliance in \cite{alliance2016description}. The three layers are the service instance layer, the network slice instance layer, and the resource layer as illustrated in Fig.\ref{fig:NSillustration}. First, the service instance layer comprises the services offered by either the network operator or by third parties such as application providers and verticals, where each service is defined by a service instance. Second, the network slice instance layer includes network slice instances where each instance refer to a set of network functions and resources that form a complete logical network customized to satisfy specific performance requirements demanded by the services instances. A network slice instance is created by the network operator using the network slice blueprint and it can be shared by several service instances. Additionally, it can include a number greater or equal to zero of sub-network instances which can be shared by other network slices. A sub-network instance is a collection of network functions and resources that do not necessarily constitute a complete logical network. Finally, the resource layer contains the network functions and the physical and logical resources offered by the network infrastructure.   

The 5G/6G network slicing can be based on different architecture configurations. While some advocate for a two-domains structure including the Core network (CN) and the Radio Access network (RAN), others adopt a three-domains architecture where the Transport network (TN) is linking the RAN to the CN. In this work, we consider the second network architecture where the network slicing can be carried out in three domains \cite{SAT.AI.NS.1_B.4.3.6, B.4.2, B.1.8}:
\begin{itemize}
    \item CN Slicing involves the virtualization, isolation and customization of the main core network functions such as the user plane function (UPF), the session management function (SMF), the policy control function (PCF) and the access and mobility management function (AMF) \cite{B.1.2}. Using the NFV technology, these functions can implemented as virtual network functions (VNFs), hence the main objectives in CN slicing includes the optimization of VNF embedding, SFC provisioning and virtual resource allocation to deliver different services for multiple slices. 
    \item TN Slicing revolves around the virtualization, isolation and customization of the transport domain resources which is composed of the physical infrastructure (routers, switchers, gateways, links, etc) responsible for data transmission \cite{SAT.AI.NS.1_B.4.3.6, B.2.3.3}. The SDN paradigm can be employed to facilitate the TN slicing performing resource allocation and path splitting and reconfiguration to satisfy QoS requirements of various TN slices.   
    \item RAN Slicing refers to the virtualization, isolation and customization of the radio access components such as the base stations, the antennas, and other radio equipments that provides wireless connectivity to end-users \cite{SAT.AI.NS.1_B.4.3.6, B.2.3.3}. Since computation and storage are moving towards the edge network in 5G/6G, the RAN can not only include communication (networking) resource, but also computing and caching resources. Thus, the RAN slicing involves the management and orchestration of different resources, as well as the device/user association meeting QoS requirements and adapting to the network changes.
\end{itemize}

In order to deliver various 6G applications, end-to-end network slicing should be carried out. It involves the creation and management of complete slices dedicated for a specific service from the core network passing by the transport network to the radio access network \cite{B.4.2,AI_Comb_1}. E2E slice admission control, E2E slice resource management and orchestration, and E2E slice lifecycle management are the main building blocks of E2E network slicing. 

In order to achieve the co-existence of various network slices providing multiple services with different performance requirements, network management and orchestration is another major component in network slicing \cite{B.1.2,B.1.7,khan2020network}. It can be divided into two layers: the service management layer and the network slice control layer \cite{B.1.2}. While the latter deals with resource management and network slice management and orchestration, the former handles service operations including abstraction, admission control and creation. 

\begin{figure}[ht]
\centering\includegraphics[width=0.35\textwidth]{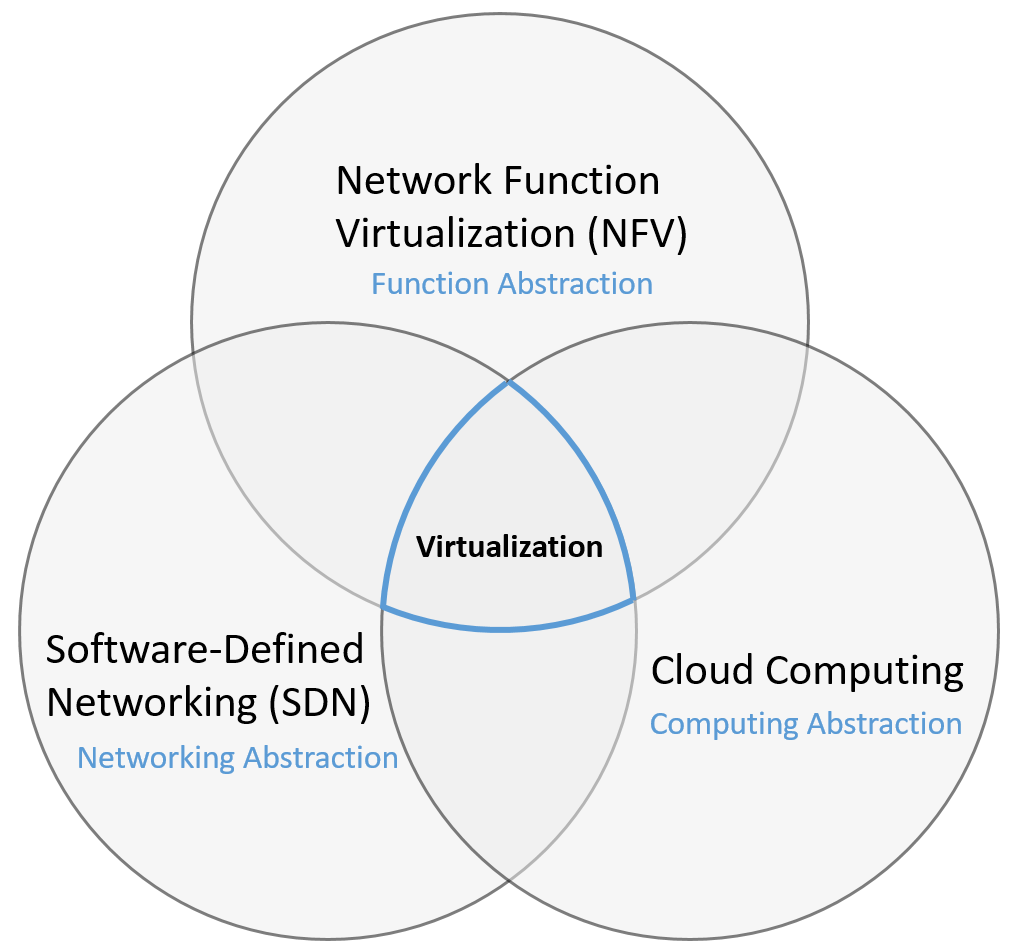}    
\caption{SDN, NFV and Cloud computing \cite{B.1.5}.}
    \label{fig:SNC}
\end{figure}

The key enabling technologies of network slicing include hypervisors, virtual machines and containers, software-defined networking and network function virtualization as well as cloud and edge computing \cite{B.1.2,B.1.7,B.1.8,khan2020network}. In particular, the combination of SDN, NFV and cloud computing results into a holistic network virtualization, as depicted in Fig.\ref{fig:SNC}. While SDN presents an abstraction of the networking resources, the NFV provides an abstraction of the network functions and the cloud computing abstracts the computational resources \cite{B.1.5}. The technologies can benefit from each other to achieve network virtualization which would offer higher flexibility, lower cost, more optimized resource utilization and better global network performance.

Nevertheless, the implementation of these technologies in next generation networks is still in its infancy as it faces multiple challenges. For instance, issues related to the network management and orchestration, the optimization of resource allocation and the network security necessitate the use of powerful tools such as AI and ML algorithms.

\section{Artificial Intelligence in Network Virtualization}\label{AI_Vir} 

As a key enabler of 6G, AI is expected to play a major role in the advancement of next-generation networks. In this section, we present an introduction on the applications of AI in the realm of network virtualization. In particular, we discuss the rationale behind the use of AI models in 6G networks where conventional approaches are not able to offer the required levels of efficiency and optimality. We also give an overview on AI techniques including supervised, unsupervised and reinforcement learning. Additionally, we briefly highlight the primary research areas where AI algorithms are often used in the context of virtualization technologies, namely SDN, NFV, and network slicing.       

\subsection{Motivation} 
With the large number of users, the diversified applications and the integrated terrestrial and non-terrestrial networks, 6G networks become substantially larger, more dynamic, and heterogeneous, which increases the complexity of realizing efficient network virtualization as well as network management, resource allocation and traffic prediction, among others \cite{B.3.1,SAT.AI.NS.1_B.4.3.6}. Consequently, conventional methods can no longer provide the necessary efficiency and optimality required for the proper operation of the network \cite{B.4.3.4_B.1.6,B.4.3.2}. In fact, traditional approaches are typically model-based which imposes several limitations. First, they require an a priori knowledge of the network traffic which is not suitable for highly dynamic networks \cite{B.4.3.4_B.1.6,B.4.3.7}. Second, they are intractable and computationally demanding for large-scale networks. Third, they may provide sub-optimal solutions depending on the statistical models accuracy \cite{B.4.3.2}. Meanwhile, providing model-free algorithms with low computational complexity after offline training \cite{B.4.3.4_B.1.6}, AI-based methods not only solve the issues of conventional approaches, but also introduces network management automation and improves network performance \cite{MLmeetNMOinEdge,B.4.2.2,B.4.3.7,B.4.4.3}. Hence, they are adequate for future 6G networks. On one hand, reinforcement learning (RL) techniques are efficient for decision making tasks in dynamic environments which facilitates network and resource management and orchestration \cite{B.4.2.1}. On the other hand, supervised and unsupervised learning algorithms have shown supremacy in terms of prediction and classification problems, which promotes for proactive decision making and resource allocation \cite{SAT.AI.NS.1_B.4.3.6,B.4.9}. Moreover, Federated learning (FL) solves the issues of data privacy and reduces communication costs by allowing distributed learning where the models train on local data, then only the learnt features are sent for aggregation \cite{SAT.AI.NS.1_B.4.3.6,B.4.2.1}.

\subsection{Overview on Artificial Intelligence}
Artificial Intelligence (AI) is a discipline of computer science that seeks to develop intelligent machines and systems capable of thinking and acting like humans. These smart machines would have the ability to carry out tasks such as learning, decision-making, and perception, which usually involve human intellect. AI is a broad field that includes not only learning-based approaches like machine learning (ML) and deep learning (DL) but also non-learning-based techniques such as rule-based systems \cite{DL_book,AI_book,AI_book2}. Using well-defined rules provided by the programmers, non-learning methods excel at solving explicit problems, while they perform poorly when dealing with sophisticated, less-structured tasks like speech and image recognition. Meanwhile, ML and DL can learn how to accomplish a certain task autonomously, allowing them to thrive in the face of complex problems. That is why ML and DL have drawn the attention of researchers from numerous domains ranging from finance and biology to wireless communications and networking. 

ML is a subfield of AI where the machine is trained to learn patterns in provided data, without explicit programming, in order to solve a specific problem \cite{DL_book,Bishop_book}. Four main categories of ML algorithms can be distinguished:
\begin{itemize}
    \item Supervised learning: The algorithm is trained on a labeled dataset, known as the training set, where data points are annotated with the target values. The supervised ML algorithm learns a mapping function, between the input data points and their target outputs, which can predict the output labels for previously unseen inputs. Supervised learning algorithms are typically used for classification and regression problems. Examples of such algorithms include linear regression, logistic regression, support vector machines (SVMs), decision trees, and neural networks \cite{ML_book,Bishop_book}.
    \item Unsupervised learning: The algorithm is trained on an unlabeled dataset, where inputs are provided without target outputs. The unsupervised ML algorithm learns to identify the hidden patterns and structures in the data. Unsupervised approaches are employed for different purposes such as clustering, dimensionality reduction, and data visualization. Such methods comprise K-means, hierarchical clustering, principal component analysis (PCA), and locally linear embedding (LLE) \cite{ML_book,Bishop_book}.
    \item Semi-supervised learning: This method combines both supervised and unsupervised learning by training the algorithm on a dataset that contains both labeled and unlabeled data. Semi-supervised techniques are beneficial when labeled data is limited or costly to collect \cite{ML_book}.
    \item Reinforcement learning (RL): The RL algorithm learns through the interaction with the environment. Based on the knowledge it gathers from observing its environment, an agent learns to select the actions which will maximize a certain reward. The objective is to learn a policy maximizing long-term rewards. Q-Learning, SARSA, and Actor-Critic are examples of RL algorithms that are commonly utilized for decision-making in dynamic environments \cite{ML_book,DL_book,RL_book}.
\end{itemize}

DL is a subfield of ML that that involves training artificial neural networks to solve sophisticated problems. These neural networks are composed of multiple layers of interconnected neurons that process data hierarchically, by extracting higher-level features in each layer to generate the final output. DL models have demonstrated outstanding performance in complicated tasks such as image and speech recognition, natural language processing, and game play. However, they usually require large training datasets, and high computational resources. Examples of commonly used deep neural networks (DNNs) architectures include convolutional neural networks (CNNs) utilized in computer vision, recurrent neural networks (RNNs) used for sequential data processing tasks, and generative adversarial networks (GANs) employed for new data generation \cite{bengio2017deep}.

In the field of wireless communications, ML and DL algorithms have been adopted to solve a variety of problems such as resource allocation and management, network optimization, channel prediction, traffic forecasting, and network security \cite{fourati2021artificial,jiang2016machine,chen2019artificial,sheth2020taxonomy,bhattacharyyamachine}.

\subsection{AI Applications in Network Virtualization} 
AI is envisioned to invade every aspect of 6G networks including network virtualization \cite{tong_zhu_2021,B.3.1}. In particular, ML algorithms show great potential in solving SDN, NFV and network slicing issues, where traditional methods are no longer efficient in future dynamic, heterogeneous and large-scale networks. Several surveys are reported in the literature reviewing the applications of AI techniques in virtualization \cite{B.4.2,B.4.3.4_B.1.6,B.4.7,B.4.4,B.4.5,B.4.9,B.4.3.5_B.4.10,B.4.3.3_B.4.11,B.4.4.3}. Here, we briefly highlight the main research directions where ML algorithms are commonly employed in the context of SDN, NFV and network slicing, as summarized in Table \ref{tab:AI_Virtualization}.

\begin{table*}[]
\centering
\begin{tabularx}{\textwidth}{ p{0.125\textwidth}  p{0.17\textwidth}  p{0.41\textwidth}  X }
\cline{2-4}
 &
  Applications &
  AI approaches &
  References \\ \hline
  
  &
  Controller Placement &
  Unsupervised and supervised learning techniques (K-means, neural networks, decision trees...) &
   \cite{B.4.5,B.4.4,AI_Comb_11}
   \\ \cline{2-4} 
   \multirow{-2}{*}{\begin{tabular}[c]{@{}l@{}}Software-Defined \\ Networking (SDN)\end{tabular}}&
  Routing Optimization &
  RL algorithms, Supervised learning (LSTM, linear regression, Naive Bayes...) &
   \cite{B.4.4,B.4.7,B.4.5,AI_SDN_2,AI_SDN_10}
   \\ \hline

   \begin{tabular}[c]{@{}l@{}}Network Function \\ Virtualization (NFV)\end{tabular}&
  VNF and SFC Deployment &
  RL and Deep RL approaches &
\cite{AI_NFV_10,AI_NFV_11,AI_NFV_12,AI_NFV_13,AI_NFV_21,AI_NFV_22,AI_Comb_6,AI_Comb_8,B.4.2.2,AI_NFV_7}
   \\ \hline
   
  &
  Slice Admission Control &
  RL and Deep RL approaches &
  \cite{B.4.3.4_B.1.6,B.4.2,B.4.3.3_B.4.11,B.4.3.5_B.4.10, B.4.3.1_B.4.8,AI_NS_11}
   \\ \cline{2-4}
   
  \multirow{-2}{*}{Network Slicing}&
  E2E Network Slicing &
  RL algorithms, DNNs, Federated learning approaches &
   \cite{B.4.2,AI_NS_2,AI_NS_6,AI_NS_29,AI_Comb_1}
   \\ \hline

   &
  Traffic Prediction and Classification &
  Classification methods (decision tree, random forest, SVM...), Regression methods (linear regression and LSTM...), Deep neural networks (RNNs and CNNs), RL algorithms &
  \cite{B.4.5,B.4.4,B.4.2,B.3.1,B.4.3.4_B.1.6,AI_SDN_4,AI_SDN_7,AI_NS_8}
   \\ \cline{2-4} 
 
  &
  Resource Management and Orchestration &
  Prediction techniques (graph neural networks, LSTM, k-nearest neighbors...), RL and Deep RL approaches & 
    \cite{B.3.1,SAT.AI.NS.1_B.4.3.6,B.4.2.2,B.4.2.1, B.4.9, B.4.3.1_B.4.8,B.4.3.3_B.4.11,B.4.3.5_B.4.10,AI_SDN_5,AI_SDN_6,AI_NS_3,AI_NS_4,AI_NS_5,AI_NS_7,AI_NS_9,AI_NS_12,AI_NS_17,AI_NS_19,AI_NS_21,AI_NFV_1,AI_NFV_3,AI_NFV_6,AI_NFV_5,AI_NFV_18,AI_NFV_19,AI_NFV_20,AI_NFV_23,AI_Comb_12}
   \\ \cline{2-4} 
   
  \multirow{-5}{*}{\begin{tabular}[c]{@{}l@{}}Common \\ Applications\end{tabular}}&
  Network Security &
  Classification algorithms (SVM, deep neural networks, random forests..), RL algorithms, Hidden Markov Models & 
  \cite{B.4.5,B.4.4,SAT.AI.NS.1_B.4.3.6,B.4.3.5_B.4.10,B.4.2.2, AI_SDN_1,AI_SDN_9,AI_Comb_5,AI_NS_1,AI_NS_2,AI_NS_27,AI_NS_13,AI_NFV_16}
   \\ \hline
\end{tabularx}
\caption{AI Applications in Network Virtualization.}
\label{tab:AI_Virtualization}
\end{table*}

\subsubsection{AI Applications in SDN}
In the context of SDN-based networks, unsupervised and supervised learning techniques such as K-means, neural networks, and decision trees can be used to solve the controller placement problem by predicting the optimal locations of the SDN controller using traffic distribution \cite{B.4.4,AI_Comb_11}. Moreover, reinforcement learning algorithms can be employed in routing optimization, where the controller, which is responsible for traffic flows control and routing, can be considered as an agent in a decision-making RL algorithm, interacting with the environment described by the network and learning to select the routing paths that optimize specified metrics such as delay, packet loss rate, and energy efficiency \cite{B.4.9,AI_SDN_2}. Additionally, supervised learning models including LSTM, linear regression, and Naive Bayes among others are combined with heuristic algorithms to offer dynamic routing with optimized network performance metrics such as reduced delay and improved QoE. Traffic prediction and classification, resource management, and network security issues can be addressed and optimized by applying ML algorithms \cite{B.4.5,B.4.4,B.4.9}.

\subsubsection{AI Applications in NFV}

In NFV-enabled networks, AI models are utilized for NFV management and orchestration, VNF and SFC deployment, as well as network security \cite{B.4.2.1,B.4.2.2,AI_NFV_10,AI_NFV_16}. In particular, The SFC and VNF embedding problem, involving the mapping, configuration and placement of VNFs at suitable hosting locations for service provision, can be formulated as a decision-making task where RL and Deep RL agents can be used to obtain optimal VNF placement and configuration strategies \cite{AI_NFV_12} enabling automated and dynamic SFC and VNF deployment which improves resource utilization efficiency and service delivery.

\subsubsection{AI Applications in network slicing}

The applications of AI in network slicing include slice admission control, slice traffic prediction, slice resource management and orchestration, End-to-End network slicing, and network security \cite{B.4.3.4_B.1.6,SAT.AI.NS.1_B.4.3.6,B.4.3.5_B.4.10,B.4.3.1_B.4.8,B.4.2,B.4.3.3_B.4.11}. Slice admission control is a decision-making task where the algorithm decides whether to accept or deny a new slice request in multi-tenancy networks, taking into account resource availability and QoS requirements \cite{B.4.2,B.4.3.3_B.4.11}. To improve network efficiency and provide slice admission automation, RL and Deep RL approaches are used to learn optimal admission strategies to optimize a specified objective such as profit maximization, resource utilization enhancement, and utility maximization. Meanwhile, ML algorithms can enable automatic, intelligent and proactive end-to-end network slicing, where reinforcement, deep, and federated learning can be adopted for end-to-end slice admission control, end-to-end slice resource management and orchestration, and end-to-end slice lifecycle management.    

\subsubsection{Common AI Applications}

Applying AI approaches in traffic prediction and classification, resource management and orchestration, and network security is common to the three virtualization technologies. On the one hand, traffic prediction and classification is mainly considered in SDN for proactive and efficient resource management and optimized routing, and in network slicing to enhance slice resource utilization and lifecycle management, minimize SLA violations and ensures fairness in terms of resource allocation to each slice. While classification methods including decision tree, random forest, and SVM, as well as deep neural networks, particularly RNNs and CNNs, are employed to identify and classify different types of network traffic flows, ML algorithms used for regression problems such as linear regression and Long short-term memory (LSTM) are adopted to predict future network traffic \cite{B.4.5,B.4.4,B.4.2,B.3.1,B.4.3.4_B.1.6}.
On the other hand, AI-based resource management and orchestration is adopted in SDN, NFV and network slicing offering efficient resource utilization, dynamic resource allocation and optimized network performance. Various ML algorithms can be utilized in this context. For instance, graph neural networks, LSTM, and k-nearest neighbors are used for NFV resource prediction, while model-free RL and Deep RL approaches are adopted to dynamically optimize VNF resource allocation and automate VNF management functionalities \cite{B.4.2.1, AI_NFV_5}. Also, since the SDN and slice resource allocation problems can be considered as optimization problems, they can be solved by model-free RL or Deep RL algorithms \cite{B.4.3.4_B.1.6,B.4.2} offering efficient, adaptive and intelligent resource management. Meanwhile, the utilization of AI models such as deep neural networks and RL agents can improve network security by autonomously and proactively detecting and mitigating cyber-attacks and malicious activities in networks based on virtualization technologies \cite{SAT.AI.NS.1_B.4.3.6,B.4.3.5_B.4.10,B.4.2.2}. ML-based classification algorithms including support vector machines and random forests can identify and detect malicious activities such as Distributed Denial of Service (DDoS) attacks by analyzing the network traffic. In SDN-enabled networks, the controller can automatically identify the appropriate strategies for network protection in a real-time fashion, using reinforcement learning \cite{B.4.4,B.4.5,AI_SDN_1}. In addition, Intrusion Detection System (IDS) can employ ML algorithms such as Hidden Markov Models for attacks prediction to proactively protect the network.

\section{Virtualization Challenges in Integrated Terrestrial and Non-Terrestrial Networks}\label{Challenges_Vir_TN_NTN}

Numerous challenges face the development of virtualization technologies in integrated terrestrial and non-terrestrial networks. In this section, we discuss the most prevalent issues that scholars have studied and reported in the literature. In SDN-enabled integrated networks, for example, the primary concerns explored are SDN controller placement, routing optimization, handover management, and resource allocation; in NFV-enabled networks, on the other hand, the emphasis is on VNF placement, SFC embedding, and virtual resource management. Meanwhile, the main focus in the design of network slicing-based networks pertains to the user association, aerial UAV slicing, and RAN resource management. Additionally, traffic scheduling and offloading, and network security and resilience are further common challenges among these technologies, that are examined in the virtualization of TN-NTNs. The majority of these issues can be formulated as NP-hard graph-based optimization problems. Researchers attempted to solve them employing not only conventional optimization techniques but also AI algorithms to cope with the unique characteristics of these networks.

\subsection{Challenges in SDN-enabled Networks}
Although the SDN paradigm facilitates TN-NTN seamless integration and improves network flexibility, employing SDN in integrated networks still faces several challenges. Compared to terrestrial networks, the SDN controller placement, resource management, and routing problems become more complex due to the dynamic and large-scale topology. These problems can also be jointly considered with the satellite gateway placement and the UAV positioning problems, to enhance network performance, which further increases their complexity. The satellites handover management is another major issue in SDN-enabled integrated networks, because of their high mobility.

\subsubsection{Controller placement problem}

    Acting as the brain of SDN-based networks, the controller not only provides a logically centralized intelligence and a global view of the network but also enables network flexibility and programmability which facilitates its management. Hence, to ensure efficient and optimized network operation, the SDN controller should be strategically positioned within the network topology. The task of obtaining this strategic positioning is known as the SDN controller placement problem (CPP) \cite{das2019survey}. The objective is to determine the optimal locations for deploying the controllers within the network infrastructure to efficiently manage the network traffic and achieve different objectives such as meeting the diverse requirements of various applications.  
    
    Depending on the size, topology, and performance requirements of the network, the SDN control structure can be organized into three controller configurations \cite{das2019survey,B.2.3.2}:
    \begin{itemize}
        \item Single controller configuration: the entire network management relies on a single centralized SDN controller enabling simplified implementation, reduced complexity, and minimal resource utilization. However, it is not suitable for large-scale networks because of the resilience and scalability limitations as well as the single-point failure. 
        \item Multi-controllers configuration: it alleviates the limitations of the single controller structure by deploying multiple SDN controllers within the network to manage different regions or sub-networks. These controllers work cooperatively to manage their respective sub-networks, leading to improved network performance and scalability. By distributing the control plane functions, the multi-controllers structure provides enhanced fault tolerance and adaptability to the increasing traffic demands in larger networks at the cost of increased complexity, resource consumption, and coordination overhead. 
        \item Hierarchical multi-controllers configuration: it extends the multi-controller approach by organizing the controllers into a hierarchical structure, allowing better management of larger and more complex networks. In this setup, higher-level controllers oversee the entire network by coordinating the communication between lower-level controllers that manage specific domains, improving network scalability and flexibility.
    \end{itemize}

    The controller placement problems can be classified according to the adopted controller configuration. Also, given the network characteristics, two main types of SDN controller placement problem can be distinguished \cite{NTN.C.SDN.2.4}:
    \begin{itemize}
        \item Static CPP: the optimization of controllers' locations is performed only once during the initial design of the network, and their positions do not change throughout the network's lifetime. It assumes that the network conditions are time-invariant leading to adaptability and scalability challenges, especially in highly dynamic non-terrestrial networks. 
        \item Dynamic CPP: the controller placement continuously adapts to the changing network topology, traffic patterns, and service requirements. By adjusting the controllers' locations on a continuous or periodic basis, based on the network status, the dynamic approach provides adaptive, scalable, and optimized controller placement.
    \end{itemize}
   
    The integrated TN-NTNs are generally modeled as a graph describing its topology where the graph nodes are the network components which can be the end users, the controllers, the switches, or the satellite gateways, etc, and the graph edges present the communications links between them. Using the graph-based representation of the network, the SDN controller placement problem is formulated as an optimization problem, and its solution involves the number of controllers, their locations, and the controller-switches assignments \cite{das2019survey}. To achieve different QoS requirements, various optimization objectives are considered in the CPP such as latency minimization, network reliability maximization, and deployment and management cost minimization. Traditional optimization techniques based on heuristic and meta-heuristic approaches, as well as AI-based methods including clustering and reinforcement learning models, are adopted to solve the CPP optimization problem in integrated TN-NTNs.
    
    Different evaluation metrics are employed to assess the performance of the controller placement algorithms for example network latency, network reliability, load balancing, average flow setup time, as well as computational complexity and running time \cite{B.2.3.2}.

\subsubsection{Routing optimization}

Routing is a fundamental concept in networking entailing the optimal selection of paths to transfer data packets from a source to a destination. In SDN-based routing, the controller is an integral component, being the central entity that controls and selects paths to route traffic flows. Hence, one classification for routing algorithms in SDN-enabled integrated TN-NTNs can be based on the control structure i.e. single controller, multi-controllers, and hierarchical multi-controllers configurations. Another classification for routing mechanisms can be according to the type of routing which can be single- or multi-path routing. While the single-path routing algorithms provide only one path between a source and a destination, the multi-path routing schemes determine multiple paths to transmit the data between two network nodes.

Based on a graph representation for the integrated TN-NTNs topology, the routing problem can be formulated as an optimization problem with the goal of obtaining optimal routing paths while meeting specific network requirements. In SDN-enabled integrated TN-NTNs, various optimization objectives are considered to design routing mechanisms such as congestion and cost minimization, network link utilization maximization, load balancing, etc \cite{B.2.3.2}. To solve such optimization problem, graph algorithms based on heuristics and shortest-path methods are employed as conventional optimization techniques, while ML classifiers and RL algorithms are adopted as AI-based approaches. The routing schemes are assessed using evaluation metrics such as latency, packet drop rate, throughput, bandwidth utilization, and load balancing.

\subsubsection{Satellites handover management}

Handover management is the process of transferring the connection of a communication device from one network node to another to ensure seamless and uninterrupted connectivity. In terrestrial networks, a handover occurs when users travel between cells, while in integrated TN-NTNs, handovers occur not only because of the user's mobility but also because of the mobility of NT platforms causing in particular the limited visibility window of NGEO satellites. In general, Handover strategies can be categorized as either hard or soft schemes \cite{chowdhury2006handover}. Hard handover methods involve releasing the current link before establishing the subsequent connection, whereas soft handover techniques retain the current link until establishing the next link. In case of integrated satellite-terrestrial networks, satellites handover management strategies can be categorized based on the handover link \cite{NTN.C.SDN.4.1}:
\begin{itemize}
    \item Satellite handover: it refers to the transfer of the connection from one satellite to another.
    \item Spotbeam handover: it takes place between the multiple beams of the same satellite.
    \item Inter-satellite links (ISL) handover: it occurs when links between satellites in neighboring orbits are temporarily lost resulting in the handover of the current connections relying on these ISLs.
\end{itemize}

The design of handover mechanisms involves solving a graph-based optimization problem to determine an optimal strategy describing the decision-making process when a handover occurs. To guarantee seamless connectivity and maintain user experience, different optimization objectives are considered such as handover frequency and drop-flow minimization, as well as RSSI and UE utility maximization. Game theory and heuristic-based methods are adopted to derive optimized handover mechanisms in integrated S-T networks. The average number and latency of handovers, the transmission quality, the throughput, and the user QoE are used as evaluation metrics for these handover management strategies.

 \subsubsection{Resource management}
 Resource management in SDN can be classified into two categories. The first category is control plane resource management which mainly focuses on the allocation of the computing resources of network hypervisors in multi-tenancy SDN-based networks \cite{B.4.9}. The second category deals with the independent or joint management of data plane resources including networking, computing, and caching (storage). In the literature, researchers focus on data plane resource management in SDN-enabled integrated terrestrial and non-terrestrial networks. It can be formulated as an optimization problem aiming to determine optimal resource allocation strategies taking into account different objectives such as network utility and revenue maximization and latency and energy consumption minimization. In case of UAV-terrestrial networks, the SDN resource management can be jointly considered with the UAV placement to boost network efficiency \cite{AT.C.SDN.3.1}. The problem can be solved through conventional optimization techniques (convex optimization and heuristics) and AI-based approaches (deep RL methods), and assessed using multiple metrics including throughput, latency, energy consumption, and network utility.

\subsection{Challenges in NFV-enabled Networks}
The virtualization of network functions through NFV offers unprecedented network flexibility and reduced deployment costs. Nonetheless, the use of NFV technology in integrated TN-NTNs poses major challenges particularly with respect to VNF placement, SFC embedding and virtual resource management. These problems are typically formulated as graph-based optimization problems which become NP-hard problems with multiple constraints in the integrated networks scenario because of the dynamic environment, the large-scale topology and the limited on-board resources of NTN platforms.

\subsubsection{VNF placement}
The virtualization of network functions through the NFV paradigm allows the flexible deployment of VNFs within the network. This results into the VNF placement (VNF-P) problem which refers to the strategic positioning of VNFs in the network's physical and/or virtual infrastructure to satisfy the network requirements \cite{B.1.9}. The goal is to obtain the optimal locations for the VNFs deployment to ensure efficient resource utilization, optimized traffic routing and diversified service provisioning. The VNF-P is pivotal in NFV-based networks because it significantly impacts the scalability, the reliability and the deployment costs of network services. 

The VNF placement problem is formulated as an optimization problem based on the network graph representation. Typical VNF-P problem formulations include the Linear Programming (LP), the Integer Linear Programming (ILP) and the Mixed Integer Linear Programming (MILP) formulations. Different objectives can be considered in solving the VNF-P problem such as the minimization of network cost and end-to-end delay, the maximization of overall network payoff maximization and the optimization of resource utilization while satisfying QoS constraints. The VNF-P is usually an NP-hard problem especially in dynamic large-scale networks like the integrated TN-NTNs \cite{bari2016orchestrating}. That is why optimization techniques based on heuristic search algorithms and game theory methods are commonly employed to solve such problems \cite{NTN.C.NFV.2}. Various evaluation metrics are utilized to assess the performance of the VNF-P strategies, for example the algorithmic complexity and running time, the network cost and service deployment delay, and the number of function nodes and energy consumption.

\subsubsection{SFC embedding}

The concept of Service Function Chaining (SFC) is employed by network service providers and infrastructure providers to deliver customized services satisfying specific QoS requirements \cite{hantouti2020service}. To do so, multiple VNFs, also known as Service Functions (SFs) are invoked following a predefined order, imposed by the SFC, thereby directing the network traffic through the ordered SFs to deliver a specific service. The SFC embedding involves the mapping and allocation of physical and/or virtual network resources in order to execute a sequence of SFs, called a service function chain, within the infrastructure. The objective is to determine the optimal SF placement and establish the appropriate connections to build the chain, while meeting the SFC constraints and ensuring optimized network performance.

The SFC embedding is graph-based optimization problem aiming to find the optimal resource allocation and forwarding paths to execute the desired service chain considering the resource availability and QoS requirements to achieve efficient service delivery, and optimized resource utilization. The goal of the SFC optimization varies according to the network application, for example, service delivery latency and resource consumption minimization with load balancing, as well as number of completed tasks maximization, cost minimization and revenue maximization can be targeted in the design of an SFC embedding strategy. The SFC embedding problem is typically solved using game theory and heuristics based approaches. Cost and revenue average ratio, overall service delivery latency, service acceptance rate, reliability and resource utilization efficiency can be used as evaluation metrics. 

\subsubsection{Virtual resource management}

The virtual resource management in NFV-based networks is a crucial factor in guaranteeing optimal network performance and efficient resource utilization while satisfying QoS requirements \cite{NTN.C.NFV.7}. It involves determining an optimal allocation strategy of virtualized resources, namely the communication, computing and caching resources to different network nodes. The resource allocation problem can be formulated as optimization problem with different objectives such as resource consumption minimization, revenue maximization and resource utilization efficiency maximization. The problem can be solved using conventional optimization techniques such as heuristics-based search algorithms and evaluated using multiple metrics including resource consumption, delay, task completion ratio and execution time.

\subsection{Challenges in Network Slicing-based Networks}
Network slicing relies on virtualization technologies including SDN and NFV, to offer customized and isolated network slices. While SDN provides logically centralized control enabling dynamic traffic and resource management and enhancing network programmability, NFV virtualizes network functions creating tailored service chains for different slices and improving network flexibility \cite{AT.C.NS.2}. To meet the increased demands for various 6G applications with diversified QoS requirements, researchers are attempting to adopt network slicing in integrated TN-NTNs. Nonetheless, this is a challenging task since multiple issues emerge due to the unique features of integrated networks. For instance, RAN resource management and device/user association become more complex problems mainly because of the large-scale network and highly dynamic environment. Additional challenges arising with the application of network slicing in networks involving non-terrestrial platforms include traffic scheduling and offloading, and aerial UAV slicing.

\subsubsection{RAN resource management}
RAN resource management is a major challenge in network slicing-based networks, especially for integrated TN-NTNs since the resource allocation problem becomes more complex due to SAGIN characteristics. The RAN resource management involves two stages \cite{B.4.3.1_B.4.8}:
\begin{itemize}
    \item RAN resource reservation, also known as inter-slice resource management, where resources including networking, computing, and caching are allocated to each network slice based on their specific service demands and requirements.
    \item RAN resource orchestration, also referred to as intra-slice resource management, where the reserved resources are managed and allocated to end users in each slice.
\end{itemize}

The resource management problem in both stages can be considered as an optimization problem with the goal of obtaining optimal resource allocation policies to achieve a specific objective, or multiple objectives such as slice cost minimization, network utility maximization, and energy and resource consumption minimization, while satisfying QoS requirements. Various methods can be employed to solve the resource allocation problem in network slicing-based integrated networks, including conventional optimization techniques (convex optimization, queuing theory and Lyapunov optimization) and AI-based approaches (RL agents, DNNs and LSTM networks). Slice request acceptance and recovery ratios, user service completion time, energy consumption, slice and overall cost, and QoS level of satisfaction are used as performance metrics for RAN resource management algorithms.

\subsubsection{Device/user association}

In the context of network slicing, the device/user association refers to the process of assigning a device or a user to a specific network slice based on user service requirements and network conditions \cite{B.4.3.4_B.1.6, AT.C.NS.2}. Compared to terrestrial networks, the issue is more complex in integrated TN-NTNs not only because of the large-scale and heterogeneous network topology but also due to the high mobility caused by the users and non-terrestrial BSs (satellite, HAPS and UAVs). To ensure desired user QoS satisfaction and efficient overall resource utilization, device/user association is formulated as an optimization problem with different objectives such as overall cost and resource consumption minimization. Conventional and ML-based optimization methods can be employed to classify user requests and assign the appropriate slice to each user while meeting QoS constraints. Device/user association strategies can be evaluated using various metrics, for example, user acceptance ratio, slice costs and resource utilization efficiency.

\subsubsection{Aerial UAV slicing}

Aerial UAV slicing is a special case in network slicing-based integrated TN-NTNs. In fact, to deliver connectivity services in different use case scenarios, UAVs are usually remotely controlled. As a result, network slicing in UAV-assisted networks, also referred to as UAV slicing involves the creation of at least two slices \cite{AT.C.NS.4}:
\begin{itemize}
    \item The UAV control slice is used to control the movements of the UAVs and it usually has similar characteristics as the URLLC slice.
    \item The UAV payload slice is utilized to provide diversified communication services including mobile broadband connectivity, machine-type communications, among others.
\end{itemize}

Furthermore, other RAN network slicing issues, including device/user association and RAN resource management, can be jointly studied with UAV deployment to optimize resource utilization and improve network performance in UAV-assisted networks \cite{AT.C.NS.2,AT.C.NS.7}. This complicates the optimization problems and necessitates the development of more advanced optimisation algorithms.

\subsection{Common Challenges}
In addition to the aforementioned issues which are specific to each virtualization technology, common challenges associated with the adaptation of SDN, NFV, and network slicing techniques in integrated networks include traffic scheduling and offloading, as well as network security and resiliency.

\subsubsection{Traffic scheduling and offloading} 
Traffic scheduling and offloading represent major challenges in the integration of terrestrial and non-terrestrial networks. While scheduling involves orchestrating and prioritizing data transmissions to meet specified QoS requirements and optimize network performance, offloading refers to the redirection of tasks and traffic between the non-terrestrial and terrestrial components. Computation offloading and data traffic offloading are two categories of the offloading problem, where computing tasks and data traffic are transferred from one node to another, respectively \cite{AT.C.SDN.5.4}. The scheduling and offloading can be individually or jointly studied in SDN-enabled and/or network slicing-based integrated networks as reported in the literature. They are formulated as separate or joint optimization problems with different objectives including power consumption and latency minimization, as well as network utility and reliability maximization. Both traditional and AI-enabled solutions such as game theory-, convex optimization-, and RL-based techniques are employed in the design of scheduling and offloading strategies and the metrics of throughput, delay, load balancing and energy consumption are utilized in their evaluation.

\subsubsection{Network security and resiliency}
Although adopting virtualization technologies including SDN, NFV and network slicing can enhance network performance and efficiency, security and resiliency remain key challenges in the integration of terrestrial and non-terrestrial networks. The issues mainly involve the detection and mitigation of different types of attacks and anomalies, the preservation of data privacy, and the fast network recovery from failures and other disruptions, to safeguard data traffic and network components and ensure service continuity \cite{SAT.AI.NS.5}. Researchers attempt to enhance network security and resilience either by considering these aspects in the architectural design of the networks, or by designing techniques based on conventional and/or AI approaches aimed at making the network more robust and secure.

\section{Virtualization of Integrated Terrestrial and Non-Terrestrial Networks}\label{Taxonomy_Vir_TN_NTN}

\begin{figure*}
    \begin{center}
        \resizebox{0.95\textwidth}{!}{%
            \begin{forest}
                for tree={
                    grow'=east,
                    forked edges,
                    draw,
                    rounded corners,
                    node options={
                        align=center },
                    text width=2.7cm,
                    anchor=west,
                }
                [Virtualization of Integrated Terrestrial and Non-Terrestrial Networks, fill=gray!25, parent
                [Satellite-Terrestrial (S-T) Segment, for tree={fill=yellow!25, child}
                [SDN-enabled Networks
                [Architectures and experimental implementations: \cite{NTN.C.SDN.1.1,NTN.C.SDN.1.2,NTN.C.SDN.1.3,NTN.C.SDN.1.4,NTN.C.SDN.1.5,NTN.C.SDN.1.6,NTN.C.SDN.1.8,NTN.C.SDN.1.9,NTN.C.SDN.5.2,NTN.C.SDN.5.16, NTN.C.Comb.4,NTN.C.SDN.5.7,NTN.C.SDN.5.8,NTN.C.SDN.5.12,NTN.C.Comb.10_B.2.3,NTN.C.Comb.11} ,text width=12.65cm] 
                [Traditional approaches
                    [Controller placement: \cite{NTN.C.SDN.2.3, NTN.C.SDN.2.4, NTN.C.SDN.2.6, NTN.C.SDN.2.9, NTN.C.SDN.2.10, NTN.C.SDN.2.2, NTN.C.SDN.2.2(1), NTN.C.SDN.2.5, NTN.C.SDN.2.7, NTN.C.SDN.2.8, NTN.C.SDN.2.11, NTN.C.SDN.2.12, NTN.C.SDN.2.13}] 
                    [Routing optimization: \cite{NTN.C.SDN.3.1, NTN.C.SDN.3.4, NTN.C.SDN.3.5, NTN.C.SDN.3.6, NTN.C.SDN.3.7, NTN.C.SDN.3.8, NTN.C.SDN.5.5, NTN.C.SDN.5.15,NTN.C.SDN.3.2, NTN.C.SDN.3.3, NTN.C.SDN.3.9, NTN.C.SDN.5.6, NTN.C.SDN.5.9}] 
                    [Handover management: \cite{NTN.C.SDN.4.1, NTN.C.SDN.4.2, NTN.C.SDN.4.3, NTN.C.SDN.4.4, NTN.C.SDN.4.4(1)}]
                    [Network security and resiliency: \cite{NTN.C.SDN.5.11, NTN.C.Comb.13_NTN.C.SDN.4.1, NTN.C.SDN.5.13}]
                    [Traffic scheduling and offloading: \cite{NTN.C.SDN.5.4}]]
               [AI-based approaches 
                    [Controller placement: \cite{NTN.AI.SDN.2, NTN.AI.SDN.5}]
                    [Routing optimization: \cite{NTN.AI.SDN.4,NTN.AI.SDN.6}]
                    [Resource management: \cite{NTN.AI.SDN.3}]
                    [Network security and resiliency: \cite{NTN.AI.SDN.1,NTN.AI.SDN.7}]] 
                ]
                [NFV-enabled Networks
                [Traditional approaches
                    [VNF placement: \cite{NTN.C.NFV.8, NTN.C.NFV.4, NTN.C.NFV.10, NTN.C.NFV.14, NTN.C.NFV.16, NTN.C.NFV.1, NTN.C.NFV.2, NTN.C.NFV.17,NTN.C.Comb.1_B.2.1}]
                    [SFC embedding: \cite{NTN.C.NFV.13, NTN.C.NFV.5, NTN.C.NFV.15, NTN.C.NFV.3, NTN.C.NFV.3(1)}]
                    [Virtual resource management: \cite{NTN.C.NFV.6, NTN.C.NFV.18, NTN.C.NFV.7, NTN.C.NFV.19}]]
                ]
                [Network Slicing-based Networks
                [Traditional approaches
                    [Traffic scheduling and offloading: \cite{NTN.C.NS.2, NTN.C.NS.4}]
                    [RAN resource management: \cite{NTN.C.NS.5}]
                    [CN slicing: \cite{NTN.C.NS.3}]]
                [AI-based approaches
                    [RAN resource management: \cite{NTN.AI.NS.1, NTN.AI.NS.2, NTN.AI.NS.4}]
                    [Device/user association: \cite{NTN.AI.NS.3}]]
                ]               
                ]
                [Aerial-Terrestrial (A-T) Segment, for tree={fill=blue!10,child}
                [SDN-enabled Networks
                [Architectures and experimental implementations: \cite{AT.C.SDN.1.1, AT.C.SDN.1.2, AT.C.SDN.1.3, AT.C.SDN.1.4,AT.C.SDN.1.5, AT.C.SDN.1.6, AT.C.SDN.1.7, AT.C.SDN.1.8,AT.C.SDN.1.9, AT.C.SDN.1.10,AT.C.SDN.1.11,AT.C.SDN.1.12, AT.C.SDN.1.13, AT.C.SDN.1.14},text width=12.65cm]
                [Traditional approaches
                    [Controller placement: \cite{AT.C.SDN.6.2}]
                    [Routing optimization: \cite{AT.C.SDN.2.1,AT.C.SDN.2.3,AT.C.SDN.2.4,AT.C.SDN.2.5,AT.C.SDN.2.6} ]
                    [Resource management: \cite{AT.C.SDN.3.1,AT.C.SDN.3.3}]
                    [Traffic scheduling and offloading: \cite{AT.C.SDN.5.1, AT.C.SDN.5.3, AT.C.SDN.5.4,AT.C.SDN.3.3}]
                    [Network security and resiliency: \cite{AT.C.SDN.4.1}]
                    ]
                [AI-based approaches
                    [Routing optimization: \cite{AT.AI.SDN.4}]
                    [Resource management: \cite{AT.AI.SDN.3}]
                    [Network security and resiliency: \cite{AT.AI.SDN.2}]] 
                ]
                [NFV-enabled Networks
                [Architectures and experimental implementations: \cite{AT.C.NFV.1, AT.C.NFV.2, AT.C.NFV.3, AT.C.NFV.9},text width=12.65cm] 
                [Traditional approaches
                    [VNF placement: \cite{AT.C.NFV.6}]
                    [SFC embedding: \cite{AT.C.NFV.4, AT.C.NFV.5}]
                    [Network security and resiliency: \cite{AT.C.NFV.7}]]
                ]
                [Network Slicing-based Networks
                [Traditional approaches
                    [Joint RAN resource management and UAV deployment: \cite{AT.C.NS.2,AT.C.NS.7}]
                    [RAN resource management: \cite{AT.C.NS.5}]]
                [AI-based approaches
                    [UAV slicing: \cite{AT.AI.NS.4}]
                    [RAN resource management: \cite{AT.AI.NS.1, AT.AI.NS.2, AT.AI.NS.3}]
                    [Traffic scheduling and offloading: \cite{AT.AI.NS.1}]]
                ]
                ]
                [Satellite-Aerial-Terrestrial (S-A-T) Segment, for tree={fill=orange!25, child}
                [SDN-enabled Networks
                [Architectures and experimental implementations: \cite{SAT.C.SDN.1,SAT.C.SDN.3,SAT.AI.SDN.3,SAT.C.SDN.7,SAT.C.SDN.8_B.2.2} ,text width=12.65cm]
                [Traditional approaches
                    [Routing optimization: \cite{SAT.C.SDN.2,SAT.C.SDN.6}]]
                [AI-based approaches
                    [Controller placement: \cite{SAT.AI.SDN.4}]
                    [Resource management: \cite{SAT.AI.SDN.2}]
                    [Traffic scheduling and offloading: \cite{SAT.AI.SDN.1}]]
                ]
                [NFV-enabled Networks
                [Traditional approaches
                    [VNF placement: \cite{SAT.C.NFV.2,SAT.C.NFV.3}]
                    [SFC embedding: \cite{SAT.C.NFV.1,SAT.C.NFV.4,SAT.C.NFV.5}]]
                ]
                [Network Slicing-based Networks
                [Traditional approaches
                    [Joint RAN resource management and UAV deployment: \cite{SAT.C.NS.1, SAT.C.NS.3}]
                    [Device/user association: \cite{SAT.C.NS.1}]]
                [AI-based approaches
                    [RAN resource management: \cite{SAT.AI.NS.3, SAT.AI.NS.6}]
                    [Joint RAN resource management and UAV deployment: \cite{SAT.AI.NS.6}]
                    [Network security and resiliency: \cite{SAT.AI.NS.2, SAT.AI.NS.5}]]
                ]
                ]     
                ]
            \end{forest}    
        }
    \end{center}
\caption{Taxonomy of Integrated Terrestrial and Non-Terrestrial Networks Virtualization.}
\label{fig:Tax_NTN_Vir}
\end{figure*}
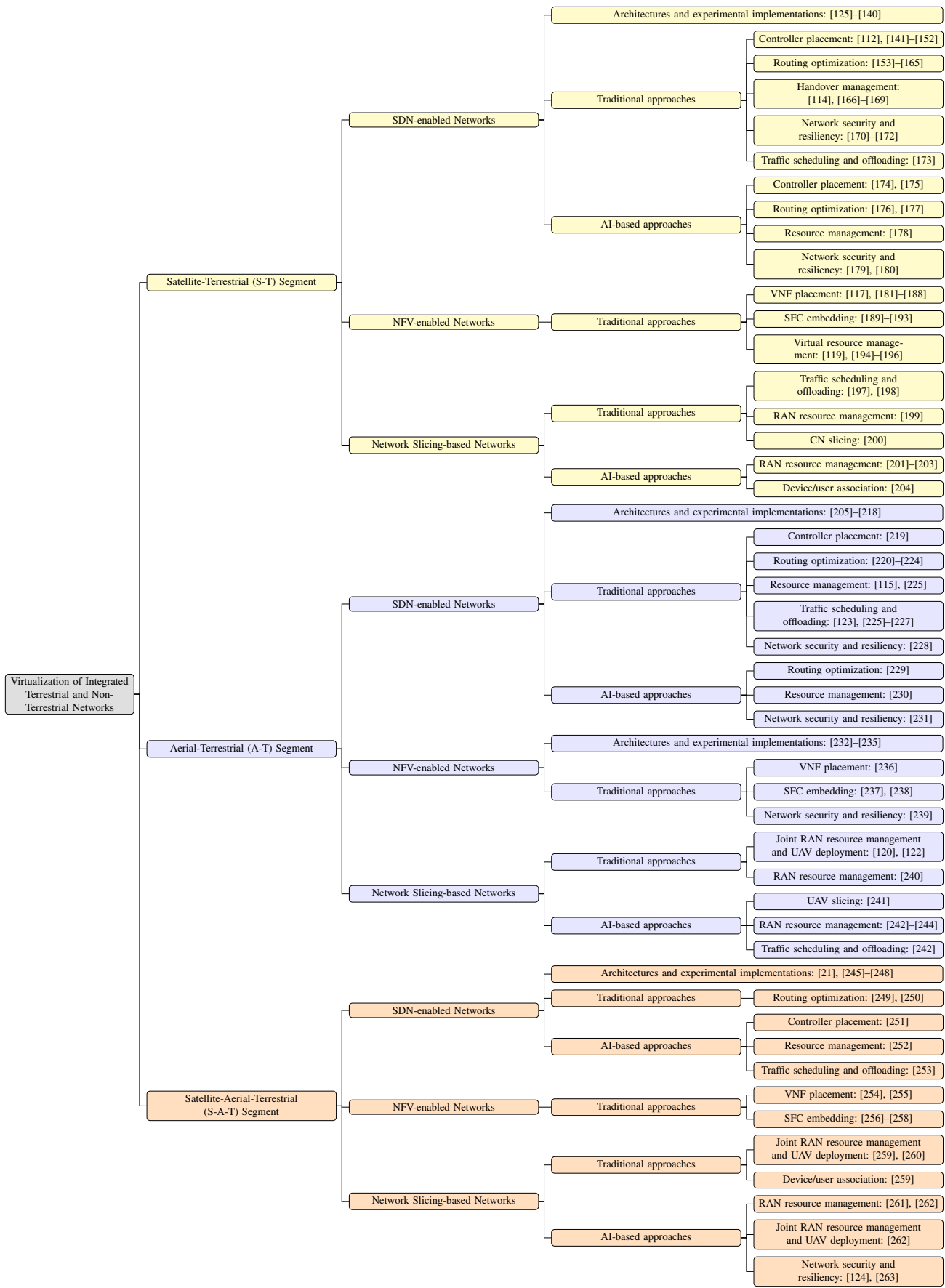

After discussing the main challenges of implementing virtualization technologies in integrated TN-NTNs in the previous section, we delve further into the relevant contributions that have been reported in the literature in this section. As shown in Fig.\ref{fig:Tax_NTN_Vir}, we provide a comprehensive taxonomy illustrating the classification of the documented works on four levels. First, we consider the level of TN-NTNs integration resulting into three categories representing the Satellite-Terrestrial, Aerial-Terrestrial, and Satellite-Aerial-Terrestrial segments. Then, we concentrate on the primary virtualization technology on which the reported work focuses, yielding three types of networks: SDN-, NFV-, and network slicing-based networks. The next classification level involves the type of the study presented by the authors where they either examine architectural considerations and experimental implementations or tackle the aforementioned issues employing conventional or AI-enabled methods. As a result, the contributions are divided into three classes: architectural and experimental implementations, traditional approaches, and AI-based approaches. Lastly, the reported works are further classified according to the category of the addressed problem, associated with the adaptation of virtualization technologies in integrated networks.

\tikzset{parent/.style={align=center,text width=4cm,rounded corners=3pt},
    child/.style={align=center,text width=6cm,rounded corners=3pt}
    }

\subsection{Virtualization in the Satellite-Terrestrial Segment}

\subsubsection{SDN-enabled Networks} \hfill\\
Leveraging the SDN paradigm, the integration of satellite and terrestrial networks unlocks a plethora of opportunities to enhance network performance, optimize network operation, and offer seamless, resilient, and scalable connectivity. Researchers have recently been dedicating their efforts to developing SDN-enabled integrated Satellite-Terrestrial networks. While several works explored the key architectural considerations and experimental implementations for incorporating SDN with terrestrial and satellite infrastructures, other studies focused on addressing the challenges of adopting SDN concepts in such networks, utilizing traditional or AI-enabled methodologies \cite{B.2.3.2, NTN.C.SDN.1.3, NTN.C.SDN.5.14, NTN.AI.SDN.6,NTN.C.SDN.5.3,NTN.C.Comb.12}. The controller placement problem, routing, and handover and resource management are the main issues in SDN-enabled networks in the Satellite-Terrestrial segment.

\textbf{\textit{ a) Architectures and experimental implementations:}} The SDN paradigm was first introduced into satellite network architectures in \cite{NTN.C.SDN.1.5} to improve network efficiency and flexibility. Several works focused on the characterization of SDN-based architectures in the S-T segment for specific use case scenarios. For example, the authors of \cite{NTN.C.SDN.1.1} propose an SDN-enabled architecture for Operationally Responsive Space (ORS) satellite networks used in post-disaster communication. They provide a description of the network in the form of a graph-based meta-model which can be employed to solve networking problems taking into consideration the network heterogeneity, topology dynamicity, and satellite mobility. Moreover, to offer multimedia broadcast communications, an architecture that combines SDN with the concept of Information-Centric Networking (ICN), is proposed in \cite{NTN.C.SDN.1.2}. Taking into account the dynamic environment and the limited resources of LEO satellites, the SDN controllers are placed on GEO and MEO satellites, and based on this architecture, heuristic caching schemes are proposed for efficient content retrieval. Another application-oriented SDN-based architecture is developed in \cite{NTN.C.SDN.1.9} and \cite{NTN.C.Comb.11} for broadband communications. The authors of \cite{NTN.C.SDN.1.9} present a flexible and reconfigurable broadband satellite network (FRBSN) architecture based on network virtualization and SDN principles. Also, using a time-evolving resource graph (TERG), they propose an optimized resource management strategy. In \cite{NTN.C.Comb.11}, a cloud-based architecture for SDN/NFV-enabled integrated satellite-terrestrial networks is introduced with a detailed analysis of its functionalities. In addition, using a terrestrial SDN controller, the Software Defined Space-Terrestrial Integrated Network (SD-STIN) architecture, introduced in \cite{NTN.C.SDN.1.8}, promotes ubiquitous global connectivity. The researchers discuss the issues of the proposed integrated S-T architecture involving mobility management, resource allocation, routing, and security, and suggest potential solutions. \par 

Meanwhile, other efforts were concentrated on the implementation aspects of SDN-based integrated S-T networks. Utilizing simulation tools and emulation platforms, they simulate the architecture's environment and implement SDN-based laboratory testbeds to evaluate network performance. For instance, the OpenFlow protocol was used in \cite{NTN.C.SDN.1.6} to implement and validate a prototype of SERvICE, the proposed SD framework for integrated space-terrestrial communication. In the context of delay-tolerant networks, the authors not only introduce and implement the SERvICE framework but also provide two QoS-based heuristic algorithms for routing and bandwidth allocation. Moreover, to study the feasibility of the HetNet architecture in \cite{NTN.C.SDN.1.3}, the EmuStack emulation platform was utilized to assess the proof of concept prototype. Enabled by SDN and NFV, HetNet is a flexible network architecture based on ICN and Locator/ID split concepts, offering routing scalability, heterogeneous network convergence, mobility support, and efficient content delivery. The network simulator NS3 and the OpenFlow protocol are extended in \cite{NTN.C.SDN.1.4} to implement the SDN-based integrated S-T architecture which includes hierarchical controllers for heterogeneous resource management, and to evaluate the designed multipath routing algorithm. Additionally, the feasibility of OpenFlow protocol in Satellite-Terrestrial networks was studied in \cite{NTN.C.SDN.5.8} using a terrestrial SDN controller. The authors employed the Linktropy mini2 emulator to emulate the satellite-terrestrial channel and the Trema framework to design the OpenFlow controller. The Mininet environment is a widely used simulation tool to develop SDN-enabled networks. The researchers in \cite{NTN.C.SDN.5.12} utilized the Mininet environment, the POX SDN controller, and the OpenFlow protocol to study the network performance of the proposed SDN-based S-T architecture in massive multimedia content delivery applications. In the context of Satellite-Terrestrial mobile backhaul networks, researchers implement an SDN-based laboratory testbed of a satellite-terrestrial integration solution. They used the satellite channel emulator OpenSAND, the Ryu SDN controller, and the OpenFlow protocol to enable SDN-based traffic engineering applications\cite{NTN.C.SDN.5.7}. Furthermore, in \cite{NTN.C.Comb.10_B.2.3}, an SDN-based satellite-terrestrial networks architecture is proposed with an implementation roadmap using extended OpenFlow protocol, and network management strategies are discussed for performance optimization. \par

In network virtualization, Virtual Network Embedding (VNE) refers to the process of mapping a set of virtual network requests onto a physical network infrastructure while guaranteeing efficient resource utilization and satisfying performance requirements. The VNE problem is studied in the context of SDN-based integrated S-T networks in \cite{NTN.C.SDN.5.2, NTN.C.SDN.5.16, NTN.C.Comb.4}. In \cite{NTN.C.SDN.5.2} and \cite{NTN.C.SDN.5.16}, the authors design and evaluate VNE algorithms through implemented SDN-based laboratory testbeds. On one hand, the Mininet environment and the Ryu SDN controller are employed in \cite{NTN.C.SDN.5.2} for the evaluation of VNE algorithms performance, in highly dynamic LEO Satellite-Terrestrial backhaul networks. On the other hand, a dynamic topology-aware VNE algorithm was proposed in \cite{NTN.C.SDN.5.16} and the STK toolkit, the Ryu SDN controller, and the OpenFlow protocol are used for the SDN-based testbed implementation. Table \ref{tab:SDN_ST_architectures_exp} summarizes the efforts in terms of proposed architectures and experimental implementations in SDN-enabled integrated Satellite-Terrestrial networks.

\begin{table*}
\centering
\begin{tabularx}{\textwidth}{ p{0.0375\textwidth} p{0.135\textwidth} p{0.145\textwidth} p{0.2\textwidth} X }
\hline 
Ref.                 & Controller placement & Use case scenario & Implementation Tools & Comments \\ \hline

\cite{NTN.C.SDN.1.1} &
Ground station &
Post-disaster communication &
  N/A &
Describe the satellite network as a graph-based meta-model to solve  networking problems in future works
   \\ \hline
   
\cite{NTN.C.SDN.1.2} &  
 GEO and MEO satellites  & 
 Multimedia broadcast communication &  
N/A &    
Propose an ICN/SDN-based architecture with caching schemes for efficient content retrieval
\\ \hline

\cite{NTN.C.SDN.1.3} & 
Ground station &          
N/A &    
EmuStack emulation platform &
Propose a flexible network architecture based on ICN and Locator/ID split concepts
\\ \hline

\cite{NTN.C.SDN.1.4} &  
GEO and MEO satellites and ground station &          
N/A & 
Extended NS3 simulator, OpenFlow protocol &
Propose an architecture based on hierarchical controllers for heterogeneous resource management  
\\ \hline

\cite{NTN.C.SDN.1.5} &  
GEO satellites &          
N/A &    
N/A &
Describe the first software-defined satellite network architecture
\\ \hline

\cite{NTN.C.SDN.1.6} &  
GEO satellites and ground station &          
Delay Tolerant Networks &    
OpenFlow protocol &
Provide two QoS-based algorithms for routing and bandwidth allocation
\\ \hline

\cite{NTN.C.SDN.1.8} &  
Ground station &          
Global connectivity &    
N/A &
Discuss the issues of the proposed integrated satellite-terrestrial architecture
\\ \hline

\cite{NTN.C.SDN.1.9} &  
Ground station &          
Broadband communications  &    
N/A &
Propose a flexible and reconfigurable broadband satellite network architecture with optimized resource management 
\\ \hline

\cite{NTN.C.SDN.5.2} &  
Non-GEO satellites or ground station &          
Satellite-Terrestrial backhaul networks &    
Mininet environment, Ryu controller  &
Implement SDN-based laboratory testbed to evaluate VNE algorithms performance  
\\ \hline

\cite{NTN.C.SDN.5.7} &  
GEO and MEO satellites &          
Satellite-Terrestrial mobile backhaul networks &    
OpenSAND emulator, Ryu controller, OpenFlow protocol &
Implement SDN-based laboratory testbed to enable SDN-based traffic engineering applications   
\\ \hline

\cite{NTN.C.SDN.5.8} &  
Ground station &          
N/A &    
OpenFlow, Linktropy mini2 emulator, Trema framework &
Study the feasibility of OpenFlow protocol in Satellite-Terrestrial networks  
\\ \hline

\cite{NTN.C.SDN.5.12} &  
GEO satellite &          
Massive multimedia content delivery &    
Mininet environment, POX controller, OpenFlow protocol &
Study the network performance in massive multimedia content delivery applications  
\\ \hline

\cite{NTN.C.SDN.5.16} &  
Ground station &          
N/A &    
STK toolkit, Ryu controller, OpenFlow protocol &
Implement SDN-based laboratory testbed to validate the feasibility of VNE algorithms  
\\ \hline

\cite{NTN.C.Comb.10_B.2.3} &
Ground station, GEO and LEO satellites &
N/A &
Extended OpenFlow protocol &
Propose an SDN-based satellite-terrestrial networks architecture with network management strategies and implementation roadmap
\\ \hline

\cite{NTN.C.Comb.11} &
Ground station &
Broadband communications &
OpenSAND emulator, OpenFlow protocol &
Introduce a cloud-based architecture for SDN/NFV-enabled integrated satellite-terrestrial networks 
\\ \hline

\end{tabularx}
\caption{SDN-enabled integrated Satellite-Terrestrial networks: Architectures and experimental implementations.}
\label{tab:SDN_ST_architectures_exp}
\end{table*}

\textbf{\textit{b) Traditional approaches:}} In SDN-enabled satellite-terrestrial networks, the majority of research works mainly focus on addressing the problems of controller placement, routing optimization, and handover management using traditional approaches. Meanwhile, traffic offloading and network security and resilience are considered only in a handful of studies.
    \begin{itemize} 
         \item Controller placement problem: Because of the network dynamics and the limited resources, the SDN controller placement problem becomes more complex in integrated satellite-terrestrial networks and the CPP solutions developed for terrestrial networks cannot be adopted. Therefore, efforts have been dedicated to addressing this issue using mainly multi-controllers, and hierarchical multi-controllers configurations. 

        On the one hand, the works in \cite{NTN.C.SDN.2.2, NTN.C.SDN.2.2(1), NTN.C.SDN.2.5, NTN.C.SDN.2.7, NTN.C.SDN.2.8, NTN.C.SDN.2.11, NTN.C.SDN.2.12, NTN.C.SDN.2.13} focus on designing controller placement techniques in networks with multi-controllers structure while considering different types of CPPs. In \cite{NTN.C.SDN.2.2}, the static CPP is formulated as a mixed integer linear program (MILP), for the joint optimization of the average control path reliability and the controller to gateway latency. The problem is solved using a heuristic greedy algorithm, yielding near-optimal solutions. In addition, the dynamic CPP is studied in \cite{NTN.C.SDN.2.5} and \cite{NTN.C.SDN.2.11} with the objective of average flow setup time minimization and traffic load minimization, respectively. The authors of \cite{NTN.C.SDN.2.5} present a model for network traffic taking into account spatial and temporal variations of user demands, and the dynamic CPP is formulated as an Integer Linear Programming (ILP) problem. The solution is based on the Python Gurobi framework and it outperforms the static technique in LEO constellation-based network. Meanwhile, in \cite{NTN.C.SDN.2.11}, two algorithms are designed to solve the dynamic CPP using non-convex integer programming problem formulation. By adopting a regularization framework and a randomized rounding, an approximate regularization-based online algorithm (AROA) is introduced offering global optimal solutions using only current load information. Then, the heuristic regularization-based online algorithm (HROA) is proposed for large-scale networks. Both algorithms are compared to existing works and results show that AROA and HROA have better performance in terms of network scalability and control overhead.

        Moreover, in the case of integrated satellite-terrestrial networks, the SDN controller placement problem can be jointly considered with the satellite gateway placement problem \cite{NTN.C.SDN.2.7,NTN.C.SDN.2.8}. This would result in a multi-objective optimization problem which can be solved to ensure seamless integration, optimized resource utilization, and enhanced network performance. The joint controller and gateway placement problem (JCGPP) is considered in \cite{NTN.C.SDN.2.7} and \cite{NTN.C.SDN.2.8}, with the objective of network reliability maximization while meeting the latency requirement. The optimization problem in \cite{NTN.C.SDN.2.7} is solved using the proposed simulated annealing and clustering hybrid algorithm (SACA) which provides approximate optimal results with lower computational complexity compared to the enumeration algorithms. Meanwhile, the joint controller and gateway placement problem in \cite{NTN.C.SDN.2.8} is formulated as an ILP and solved using two meta-heuristic algorithms, namely a double simulated annealing algorithm (SASA) and a genetic algorithm-based approach. The simulations show that these algorithms outperform SACA in terms of accuracy and computational complexity. 
        
        Researchers in \cite{NTN.C.SDN.2.12} focus on the static controller placement with dynamic controller-switch assignment. Considering the dynamic network topology and the limited bandwidth in LEO satellite networks, they formulate the CPP as a mixed integer programming problem with the goal of latency minimization. Using the proposed shortest-path based dynamic assignment (SPDA) technique, the controllers are statically placed and the switches are dynamically assigned. The results demonstrate that the SPDA performs better than static and dynamic CPPs especially in highly dynamic networks. Moreover, the authors of \cite{NTN.C.SDN.2.13} consider the update controller problem where the controllers are relocated given the network topology while minimizing the expansion cost. The proposed update controller algorithm is based on a simulated annealing heuristic approach and it offers update cost reduction while maintaining network performance.
        
        On the other hand, the efforts reported in \cite{NTN.C.SDN.2.3, NTN.C.SDN.2.4, NTN.C.SDN.2.6, NTN.C.SDN.2.9, NTN.C.SDN.2.10} address the controller placement problem in the case of hierarchical multi-controllers configuration. The static CPP is studied in \cite{NTN.C.SDN.2.3} with the objective of joint cost minimization and stability enhancement. Considering a three-layers control structure with super and slave controllers, a slave controller selection strategy is proposed and validated in terms of switch-to-controller and controller-to-controller delays. Additionally, the dynamic CPP with hierarchical control architecture is examined in \cite{NTN.C.SDN.2.6, NTN.C.SDN.2.9, NTN.C.SDN.2.10}. First, \cite{NTN.C.SDN.2.6} formulates the CPP as a mixed non-integer linear programming problem taking into account topology dynamics, network latency, and reliability as well as load balancing and controller handover. The proposed heuristic algorithm, named the dynamic controller placement and adjustment (DCPA) algorithm minimizes the cost of controller deployment and management. Using NS3 simulations, the authors prove that the DCPA presents better load balancing and response time performance than the SACA and ASOP algorithms \cite{NTN.C.SDN.2.4}. Second, \cite{NTN.C.SDN.2.9} considers the factors of satellite mobility and workload variations to design an adaptive controller placement and assignment (ACPA) algorithm minimizing the management cost. The algorithm is built on the control relation graph (CRG) technique, which quantifies the control overhead in the satellite network. Through emulation, the CRG-based scheme outperforms existing works \cite{NTN.C.SDN.2.5, NTN.C.SDN.2.7}. Third, the dynamic CPP is modeled as a capacitated facility location problem in \cite{NTN.C.SDN.2.10}, with the goal of networking response latency minimization while satisfying dynamic network demands. The on-demand dynamic approximation algorithm (ODAA) is proposed to obtain an approximate solution to the placement problem. Lastly, the authors of \cite{NTN.C.SDN.2.4} investigate both the static and dynamic controller placement problems, aiming to minimize the cost of controller deployment and assignment. They design a heuristic-based method, named Accelerate Particle Swarm Optimization (APSO) algorithm to solve the CPPs and they evaluate it based on delay, jitter, controller load, reliability and cost for both static and dynamic scenarios.

        \item Routing algorithms: Although multiple efficient routing algorithms have been designed for terrestrial networks, they cannot be directly adopted in integrated S-T networks. Because of the limited resources, the large-scale networks, and the highly dynamic environment, routing becomes more difficult in the satellite-terrestrial segment. Thus, developing routing mechanisms that can adapt to these characteristics is a crucial task in S-T integration. SDN-based routing algorithms can be categorized based on the control structures given the importance of the controller in the routing process. Single- and multi-controllers configurations are mainly adopted for the design of routing algorithms in SDN-enabled S-T networks. 
 
        Routing optimization in networks with SDN single controller is studied in \cite{NTN.C.SDN.3.1, NTN.C.SDN.3.4, NTN.C.SDN.3.5, NTN.C.SDN.3.6, NTN.C.SDN.3.7, NTN.C.SDN.3.8, NTN.C.SDN.5.5, NTN.C.SDN.5.15}. A congestion-aware load balancing routing algorithm is proposed in \cite{NTN.C.SDN.3.1} to optimally distribute traffic load and minimize link congestion. The proposed scheme outperforms the well-known Dijkstra algorithm and the Explicit Load Balancing technique in terms of latency, packet drop rate, and throughput. Another work that focuses on load balancing optimization is reported in \cite{NTN.C.SDN.3.4}. It mitigates the problems of load imbalance and congestion in software-defined satellite networks through a multipath TCP-based load balancing-aware routing method. Using the Mininet environment, the Ryu SDN controller, and the OpenFlow protocol, the simulation results demonstrate that the proposed solution reduces the delay and improves the throughput of the network. 
        
        The multipath TCP (MPTCP) routing technique is also used in \cite{NTN.C.SDN.3.8, NTN.C.SDN.5.5, NTN.C.SDN.3.5} while satisfying different optimization objectives. On the one hand, the network utility is maximized in \cite{NTN.C.SDN.3.8}. Based on SDN cooperated multipath TCP (scMPTCP), two algorithms are developed to select and adjust sub-flow routes while avoiding other sub-flows bottleneck and adapting to the network load dynamics. The mechanisms are implemented in Mininet framework with Ryu SDN controller and the results demonstrate higher throughput performance compared to non-overlapping and shortest path methods. On the other hand, the optimization objective in \cite{NTN.C.SDN.5.5} is network link utilization maximization. By dividing the data into multiple flows and transmitting them on multiple paths, the designed algorithm improves the throughput and the transmission efficiency, which is proven through OpenFlow-based simulations. In addition, the goal of the segment control-based MPTCP path selection algorithm in \cite{NTN.C.SDN.3.5} is the joint cost minimization and traffic flow maximization. The proposed solution combines the ideas of the segment control technology, which minimizes the control overhead, with the SDN paradigm in LEO satellite networks. This results in reducing the network delay, enhancing the transmission reliability and efficiency, and improving the bandwidth utilization, as shown by the experimental results. 
        
        Moreover, the authors of \cite{NTN.C.SDN.3.6} employ the concepts of segment routing in SDN-enabled CubeSat networks while minimizing the link cost. They propose an online segment routing-based algorithm to compute routes in a near-optimal manner. Compared to the shortest path-based route computing technique, the proposed algorithm reduces the control traffic volume and offers better flow demand satisfaction. The link cost minimization is also considered in \cite{NTN.C.SDN.3.7} for large-scale integrated terrestrial and LEO satellite networks. The Depth First Search (DFS) technique and the Dijkstra algorithm are combined to design a dynamic routing algorithm that outperforms the DFS method in terms of delay and packet drop rate. 
        
        Furthermore, the software-defined multicast routing is studied in \cite{NTN.C.SDN.5.15} for large-scale multimedia LEO satellite networks. With the goal of bandwidth saving maximization, the authors employ graph theory to build a multicast routing algorithm based on a multi-layer rectilinear Steiner tree (ML-RST). Through experimental tests, they show that the designed routing strategy enhances bandwidth saving compared to other multicast tree-based routing techniques. 

        Developing routing algorithms in SDN-enabled satellite-terrestrial networks with multi-controllers structures is reported in \cite{NTN.C.SDN.3.2, NTN.C.SDN.3.3, NTN.C.SDN.3.9, NTN.C.SDN.5.6, NTN.C.SDN.5.9}. A dynamic routing algorithm based on Inter-Satellite Link (ISL) attributes is introduced in \cite{NTN.C.SDN.3.2} for LEO satellite networks. Taking into account the effect of the ISL attributes on link quality, the authors derive a mathematical model for path utility and maximize it to obtain the optimal routes. This increases the routing adaptability and reliability, and improves the network performance. The optimization of QoS requirements is the focus of \cite{NTN.C.SDN.3.3} when designing the routing algorithm fybrrLink for SDN-based integrated S-T networks, using Bresenham and Dijkstra algorithms. Topology monitoring and flow transfer schemes are also proposed. The NS3 simulator is used to test fybrrLink and results show its superiority over existing works in terms of latency, packet loss rate, average route finding time, and load balancing. 
        
        Routing algorithms with multi-objective optimization are proposed in \cite{NTN.C.SDN.3.9} and \cite{NTN.C.SDN.5.9}. The authors of \cite{NTN.C.SDN.3.9} propose an end-to-end service-oriented fragment-aware routing algorithm for LEO satellite-terrestrial networks. To obtain routing paths that optimize the load balancing, latency, and wavelength fragment, they employ a heuristic approach based on an ant colony. Compared to Dijkstra and SADR \cite{yan2016sadr}, the proposed method presents better load balancing, and reduced wavelength fragmentation and bandwidth utilization. The joint network overhead minimization and the transmission reliability maximization are considered in the design of a multipath selection algorithm in \cite{NTN.C.SDN.5.9}. The problem is formulated as a non-linear binary programming problem and solved using a particle swarm optimization based heuristic algorithm. The routing strategy is evaluated in terms of resource utilization, retransmission probability, and throughput.

        \item Handover management: The majority of the research on handover management for SDN-based integrated satellite-terrestrial networks documented in the literature focuses on satellite handover \cite{NTN.C.SDN.4.1, NTN.C.SDN.4.3, NTN.C.SDN.4.4, NTN.C.SDN.4.4(1)}. In \cite{NTN.C.SDN.4.1}, a handover strategy based on game theory is proposed in SDN-enabled LEO satellite-terrestrial networks. Using a bipartite graph to represent the connections between ground mobile terminals and LEO satellites, the authors develop a potential game-based handover algorithm maximizing the utility of mobile terminals. They also design a userspace maximization based terminal random access algorithm for load balancing. The simulation results show that the solution improves SNR quality with reduced average handover number, while maintaining network load balance. A seamless handover algorithm for software-defined satellite-terrestrial networks is proposed in \cite{NTN.C.SDN.4.3}. Leveraging the global network view of the SDN controller, the handover strategy is designed with the goal of selecting the UE-satellite link with the highest received signal strength indicator (RSSI). Compared to the hard and the hybrid handover schemes in \cite{chowdhury2006handover} and \cite{gkizeli2001hybrid}, the seamless handover demonstrates increased throughput, reduced handover latency, and higher level of user QoE. 
        
        The authors of \cite{NTN.C.SDN.4.4} and \cite{NTN.C.SDN.4.4(1)} concentrate on the problem of flow table management during handovers in SDN-based S-T networks. Considering the limited flow table space and the increased flow table size caused by frequent handovers, they design a heuristic timeout strategy-based mobility management algorithm and a dynamic classified timeout algorithm aiming to minimize the handover drop-flow and the flow table size, respectively. They conduct experimental tests using STK toolkit and evaluate the proposed algorithm in terms of drop-flow rate, flow table size, and transmission quality. The traffic gateway handover is considered in \cite{NTN.C.SDN.4.2} where the traffic is reallocated between the satellite gateways. A handover control strategy is developed based on the Smart Gateway Diversity (SGD) management logic with the objective of minimizing the number of reallocated groups of user beams while a traffic load balance among the gateways. The simulation results demonstrate an improvement in the aggregated throughput and SNR quality for Extremely High Frequency (EHF) satellite networks.

        \item Other research directions: While most of the works on SDN-enabled satellite-terrestrial networks focus on addressing the aforementioned issues, other efforts consider different research directions. For example, data traffic offloading and spectrum management are studied in \cite{NTN.C.SDN.5.4} for integrated S-T networks. To achieve cooperation and competition between terrestrial cellular networks and satellite networks, a traffic offloading and spectrum sharing scheme based on auction theory is proposed for multicast multimedia communications maximizing the utility of the satellite and the MNO while leveraging the SDN controller's capabilities. The simulation results validate the performance of the proposed traffic offloading technique. Moreover, network security in SDN-enabled satellite-terrestrial networks is examined in \cite{NTN.C.SDN.5.11}. The authors propose a two-step secure dynamic access method in a hierarchical multi-controllers architecture. They design a node reputation evaluation mechanism and a resource-aware dynamic access technique to guarantee network security and enhance network QoS. The results show improved data transmission rates. Furthermore, researchers in \cite{NTN.C.SDN.5.13} and \cite{NTN.C.Comb.13_NTN.C.SDN.4.1} investigate traffic engineering to improve resilience in SDN-based integrated satellite-terrestrial backhaul networks. They design a traffic engineering application to optimally allocate the available terrestrial and satellite capacity while maximizing the network utility. The simulations demonstrate improvement in terms of bit rate and global network utility. 
        
    \end{itemize}
        
\begin{table*}[]
\centering
\begin{tabularx}{\textwidth}{p{0.07\textwidth} p{0.0375\textwidth}
p{0.11\textwidth} p{0.08\textwidth} p{0.17\textwidth}  X p{0.16\textwidth}}
\hline
Research focus & Ref. & Controller configuration & Type of problem & Optimization objective & Proposed solution & Evaluation metrics \\ \hline


&
\cite{NTN.C.SDN.2.2} & 
Multi-controllers & 
Static CPP &  
Path reliability and controller to gateway latency optimization &    
Heuristic greedy CP algorithm &
Average control path reliability 
\\ \cline{2-7}

&
\cite{NTN.C.SDN.2.3} & 
Hierarchical multi-controllers& 
Static CPP & 
Joint cost minimization and stability enhancement& 
Slave controller selection strategy& 
 Switch/controller to controller delays  
\\ \cline{2-7}

&
\cite{NTN.C.SDN.2.4} & 
Hierarchical multi-controllers& 
 Dynamic, static CPP& 
Cost minimization & 
Accelerate Particle Swarm based Dynamic and static CP schemes & 
 Delay, jitter, controller load, reliability and cost 
  \\ \cline{2-7}

&
\cite{NTN.C.SDN.2.5} &
Multi-controllers &
Dynamic CPP& 
Average flow setup time minimization & 
CP algorithm based on Python Gurobi framework & 
Average flow setup time and number of controllers
\\ \cline{2-7}

&
\cite{NTN.C.SDN.2.6} &
Hierarchical multi-controllers &
Dynamic CPP& 
Controller deployment cost minimization & 
Heuristic based CP algorithm & 
Average response time and load balancing 
\\ \cline{2-7}

&
\cite{NTN.C.SDN.2.7} &
Multi-controllers &
JCGPP & 
Network reliability maximization & 
Simulated annealing and clustering hybrid JCGP algorithm & 
Average latency and reliability
\\ \cline{2-7}

Controller placement &
\cite{NTN.C.SDN.2.8} &
Multi-controllers &
JCGPP & 
Network reliability maximization & 
 Double simulated annealing and Genetic algorithms based JCGP technique & 
 Average reliability and running time 
\\ \cline{2-7}

&
\cite{NTN.C.SDN.2.9} &
Hierarchical multi-controllers &
Dynamic CPP& 
Management cost minimization & 
Control relation graph based dynamic CP algorithm & 
Response time and load balancing
\\ \cline{2-7}

&
\cite{NTN.C.SDN.2.10} &
Hierarchical multi-controllers&
Dynamic CPP& 
Networking response latency minimization &
On-demand dynamic CP approximation algorithm &
 Average flow setup time and response latency 
\\ \cline{2-7}

&
\cite{NTN.C.SDN.2.11} &
Multi-controllers &
Dynamic CPP & 
Traffic load minimization& 
Regularization-based online dynamic CP algorithms  &
 Control overhead, scalability and latency 
\\ \cline{2-7}

&
\cite{NTN.C.SDN.2.12} &
Multi-controllers &
Static CPP/ dynamic assignment  &
Latency minimization &
Shortest-path based CP  and dynamic assignment method  &
 Average latency and switch to controller delay
\\ \cline{2-7}

&
\cite{NTN.C.SDN.2.13} &
Multi-controllers  &
Controller updating  &
Expansion cost minimization &
 Simulated annealing based heuristic update algorithm  &
 Average flow setup time, reliability, load balancing
\\ \hline


&
\cite{NTN.C.SDN.3.1} & 
Single controller &
Single-path routing  &
Congestion minimization and load balancing &
 Congestion-aware load balancing routing algorithm  &
Latency, packet drop rate and throughput
\\ \cline{2-7}

&
\cite{NTN.C.SDN.3.2} &
Multi-controllers &
Single-path routing & 
Path utility maximization&
ISL attributes-based dynamic routing algorithm   &
Delay, packet drop rate and throughput
\\ \cline{2-7}

&
\cite{NTN.C.SDN.3.3} & 
Multi-controllers & 
Single-path routing & 
QoS requirements optimization& 
QoS-aware routing algorithm & 
Latency, packet loss rate, average route finding time and load balancing
\\ \cline{2-7}

&                
\cite{NTN.C.SDN.3.4}&
Single controller &
Multi-path routing & 
Load balancing&
Multipath TCP based load balancing algorithm &
Delay and throughput
\\ \cline{2-7}

&
\cite{NTN.C.SDN.3.5}&
Single controller &
Multi-path routing & 
Cost minimization and traffic flow maximization  &
Segment control-based MPTCP path selection algorithm  &
Delay, packet drop rate and bandwidth utilization
\\ \cline{2-7}

&
\cite{NTN.C.SDN.3.6}&
Single controller &
Single-path routing & 
Link cost minimization &
Online segment routing based algorithm &
Demand satisfaction, average link utilization and control traffic volume 
\\ \cline{2-7}

&
\cite{NTN.C.SDN.3.7}&
Single controller &
Single-path routing & 
Link cost minimization &
\begin{tabular}[c]{@{}l@{}} DFS and Dijkstra based \\ dynamic routing algorithm \end{tabular} &
Delay and packet drop rate
\\ \cline{2-7}

&
\cite{NTN.C.SDN.3.8}&
Single controller &
Multi-path routing & 
Network utility maximization &
Load and bottleneck aware sub-flow route selection algorithm &
Aggregated throughput
\\ \cline{2-7}

Routing optimization &
\cite{NTN.C.SDN.3.9}&
Multi-controllers &
Single-path routing & 
Optimization of load balancing, latency and wavelength fragment  &
 End-to-end service-oriented ant colony-based heuristic routing algorithm  &
 Latency, bandwidth utilization, load balancing 
\\ \cline{2-7}

&
\cite{NTN.C.SDN.5.5}&
Single controller &
Multi-path routing & 
Network link utilization maximization  &
 Mutipath TCP routing algorithm  &
Throughput
\\ \cline{2-7}

&
\cite{NTN.C.SDN.5.9}&
Multi-controllers &
Multi-path routing&
Joint network overhead and transmission reliability optimization &
 Particle swarm optimization based multipath selection algorithm &
Resource utilization, retransmission probability and throughput 
\\ \cline{2-7}

 &
\cite{NTN.C.SDN.5.15}&
Single controller &
Single-path routing & 
Bandwidth saving maximization&
 ML-RST-based multicast routing algorithm &
Average delay, bandwidth consumption
\\ \hline


&
\cite{NTN.C.SDN.4.1} &  
Single controller &
Satellite handover &
Mobile terminal utility maximization &
Potential game-based handover strategy&
Average handover number and SNR quality 
\\ \cline{2-7} 

&
\cite{NTN.C.SDN.4.3} &  
Single controller &
Satellite handover &
RSSI maximization &
Maximum RSSI link selection based handover strategy &
 Handover latency, user QoE and throughput
\\ \cline{2-7} 

 &
\cite{NTN.C.SDN.4.4} \cite{NTN.C.SDN.4.4(1)} &  
Single controller &
Satellite handover &
Handover drop-flow minimization &
Heuristic timeout strategy-based mobility management algorithm &
Drop-flow rate, flow table size, transmission quality \\ \cline{2-7} 

Handover management&
\cite{NTN.C.SDN.4.2} & 
Single controller &
Traffic gateway handover &
Number of reallocated groups of user beams minimization &
 SGD network configuration based handover control strategy &
 Aggregated throughput and SNR quality 
\\ \hline

Traffic offloading &
\cite{NTN.C.SDN.5.4} & 
Single controller &
Data offloading &
 Utility maximization &
 Auction theory-based traffic offloading and spectrum sharing scheme  &
 Maximum expected utility 
\\ \hline

\end{tabularx}
\caption{SDN-enabled integrated Satellite-Terrestrial networks: Traditional approaches.}
\label{tab:SDN_ST_C}
\end{table*}

\begin{table*}[]
\centering
\begin{tabularx}{\textwidth}{p{0.08\textwidth} p{0.0375\textwidth}
p{0.11\textwidth} p{0.08\textwidth} p{0.17\textwidth}  X p{0.15\textwidth}}
\hline
Research focus & Ref. & Controller configuration & Type of problem & Optimization objective & Proposed solution & Evaluation metrics \\ \hline


&
\cite{NTN.AI.SDN.2} &
Multi-controllers &
JCGPP &
Network reliability maximization  &
Simulated annealing partition-based K-Means JCGP algorithm&
 Latency, network reliability and complexity
 \\ \cline{2-7} 

Controller placement &
\cite{NTN.AI.SDN.5} &
Hierarchical multi-controllers &
Dynamic CPP &
Joint optimization of flow setup delay, load balancing and switching cost &
 Multi-agents deep Q-learning based dynamic CP algorithm &
 Delay, load balance and switch number 
\\ \hline

&                       
\cite{NTN.AI.SDN.4} &
Single controller &
Single-path routing &
Congestion minimization &
Dynamic MADDPG-based congestion control mechanism &
 Content delivery rate, throughput and delay 
\\ \cline{2-7}

Routing optimization &                
\cite{NTN.AI.SDN.6}&
Multi-controllers &
Single-path routing &
QoS optimization &
Ensemble SVR-based QoS-aware dynamic routing strategy &
 Delay, packet loss rate and throughput
\\ \hline

Resource management&
\cite{NTN.AI.SDN.3}&
Multi-controllers &
Data plane resources &
 Network utility maximization &
Deep Q-learning resource allocation algorithm&
Network utility per resource
\\ \hline

&
\cite{NTN.AI.SDN.1}&
Single controller &
N/A &
DDoS attack detection &
SVM-based attack detection algorithm &
Accuracy, false alarm rate and F1 score
\\ \cline{2-7}

Network security &
\cite{NTN.AI.SDN.7}&
Multi-controllers &
N/A &
Data privacy improvement and attack detection &
Attack detection technique based  on FL and ML models &
 Accuracy, F1-score, precision and recall 
\\ \hline

\end{tabularx}
\caption{SDN-enabled integrated Satellite-Terrestrial networks: AI-based approaches.}
\label{tab:SDN_ST_AI}
\end{table*}

\textbf{\textit{c) AI-based approaches:}} The combination of AI techniques and SDN paradigm in the satellite-terrestrial segment is aimed to solve networking issues including controller placement, resource management, routing, and security. 
In \cite{NTN.AI.SDN.2}, the joint controller and gateway placement problem is solved using an AI-based approach in a multi-controllers integrated satellite-terrestrial network architecture. A simulated annealing partition-based K-means (SAPKM) algorithm is designed to maximize the network reliability while satisfying the latency constraint. Compared to the enumeration algorithms and SACA, the method proposed in \cite{NTN.C.SDN.2.7}, SAPKM show better performance in terms of latency, network reliability and computational complexity. Meanwhile, a deep reinforcement learning technique is employed in \cite{NTN.AI.SDN.5} to design a dynamic controller placement scheme in a multi-controllers configuration based satellite network. To obtain the optimal controllers' locations, and controller-switches assignments, the proposed multi-agents deep Q-learning realizes the joint optimization of the flow setup delay, the load balance and the switching cost. The simulation results demonstrate the superiority of the RL-based approach over K-means in terms of delay, load balance, and switch number.

The authors of \cite{NTN.AI.SDN.4} focus on routing optimization in SDN-enabled satellite-terrestrial networks. In particular, they develop a dynamic congestion control mechanism based on reinforcement learning for massive data delivery applications. The proposed multi-agent deep deterministic policy gradient (MADDPG) algorithm aims to avoid congestion and improve network adaptability. It achieves reduced delay and enhanced content delivery rate and throughput compared to the delay-based path-specified congestion control protocol. Supervised learning can also be used in the optimization of routing algorithms as demonstrated in \cite{NTN.AI.SDN.6} where the ensemble Support Vector Regression (SVR) algorithm is adopted in the design of a QoS-aware dynamic routing strategy, called Intelligent QoS Routing (IQR). The designed routing solution employs ensemble SVR for traffic prediction in order to improve the global network views of the multiple controllers which enhances the IQR routing performance. Compared to the algorithms in \cite{NTN.C.SDN.1.6}, the IQR shows improvement in terms of delay, packet loss rate, and throughput while assuring better QoS. 
    
Because of their high performance in classification tasks, AI models are mainly used for attack detection to enhance network security in SDN-based integrated satellite-terrestrial networks. In \cite{NTN.AI.SDN.1}, the Support Vector Machine (SVM) algorithm is adopted to detect DDoS attacks taking into account the time-variance of satellite networks. Based on Mininet simulations, the proposed detection technique outperforms other traditional approaches in terms of detection accuracy, false alarm rate, and F1 score. Combined with a federated learning framework, other ML classifiers such as decision trees and random forests are employed in \cite{NTN.AI.SDN.7} for traffic classification and attack detection while improving data privacy in SDN-enabled S-T IoT networks. The OpenMined-based FL security solution is implemented and experimental trials show an improvement in accuracy, F1-score, precision, and recall. 

AI-enabled resource management approaches are showing promising results because of their efficient decision-making, especially in dynamic environments. Deep Q-learning models, in particular, demonstrate high performance in resource management and orchestration in SDN-based integrated satellite-terrestrial networks. In \cite{NTN.AI.SDN.3}, a deep Q-learning algorithm is proposed to jointly and dynamically allocate the networking, computing, and caching resources with the objective of network utility maximization. The proposed resource allocation scheme shows increased network utility per resource.

Tables \ref{tab:SDN_ST_C} and \ref{tab:SDN_ST_AI} summarize the research efforts on traditional and AI-enabled solutions for SDN-enabled integrated Satellite-Terrestrial networks indicating the research focus, the objective, the controller configuration, the type of problem, the proposed solution, as well as the evaluation metrics for each work.

\subsubsection{NFV-enabled Networks} \hfill\\
By creating virtual network functions (VNFs) that can run on commodity hardware, NFV technology enhances network agility and reduces the CAPEX and OPEX \cite{kaur2022review}. Combining NFV with integrated Satellite-Terrestrial networks offers efficient resource utilization, flexible network functions deployment, and improved service provisioning. NFV-enabled integrated satellite and terrestrial networks are based on different network architectures, such as SDN/NFV-enabled satellite-terrestrial networks where both SDN and NFV paradigms are adopted \cite{NTN.C.Comb.2, NTN.C.NFV.1, NTN.C.NFV.8}, and satellite-terrestrial edge/cloud computing networks where NFV is combined with mobile edge and cloud computing technologies \cite{NTN.C.NFV.2, NTN.C.NFV.17}. Considering such integrated network architecture, researchers focus on tackling the main issues of integrating NFV in S-T networks, including VNF placement \cite{NTN.C.NFV.8, NTN.C.NFV.4, NTN.C.NFV.10, NTN.C.NFV.14, NTN.C.NFV.16, NTN.C.NFV.1, NTN.C.NFV.2, NTN.C.NFV.17}, SFC deployment \cite{NTN.C.NFV.13, NTN.C.NFV.5, NTN.C.NFV.15, NTN.C.NFV.3, NTN.C.NFV.3(1)} and virtual resource management \cite{NTN.C.NFV.6, NTN.C.NFV.18, NTN.C.NFV.7, NTN.C.NFV.19}.

     \textbf{\textit{Traditional approaches: }} To solve the NFV issues in integrated satellite-terrestrial networks, researchers employ conventional optimization techniques. 
\begin{itemize} 
    \item VNF placement problem: Compared to terrestrial networks, the VNF-P problem becomes more complex and difficult to solve in integrated terrestrial-satellite networks due to their dynamic, time variant and large-scale topology \cite{NTN.C.NFV.1, NTN.C.NFV.2}. That is why VNF-P solutions developed for terrestrial networks cannot be employed in the Satellite-Terrestrial segment. Hence, efforts have been dedicated to derive optimal solutions for the VNF-P problem in such highly mobile networks. A dynamic heuristic-based VNF placement strategy is proposed in \cite{NTN.C.NFV.10} with the objective of end-to-end delay minimization in NFV-enabled terrestrial and LEO CubeSats networks. The looking-ahead horizon time-based optimization problem is formulated as an ILP problem and solved using three heuristic-based algorithms including Simulated Annealing, Tabu Search, and Genetic Local Search algorithms. The Greedy method is utilized as a benchmark to evaluate the proposed solutions in terms of service deployment delay, computing resources consumption, and processing time. The results show the superiority of the Simulated Annealing-based VNF placement strategy. The service provisioning delay minimization is also considered in \cite{NTN.C.NFV.1} and \cite{NTN.C.NFV.2} for the design of dynamic VNF placement schemes. On the one hand, the authors of \cite{NTN.C.NFV.1} propose a dynamic security VNFs deployment strategy in SDN/NFV-enabled satellite-terrestrial networks. While satisfying the QoS and system constraints, the mechanism dynamically and optimally places the virtual network functions that provide security services such as intrusion detection and prevention, secure transmission channel, and authentication, across the satellite infrastructure. They formulate the VNF-P problem as ILP, solve it using the Tabu Search technique, and validate the solution feasibility in terms of service provisioning delay and running time through simulations. On the other hand, Gao et al. \cite{NTN.C.NFV.2} develop a dynamic distributed VNF placement (D-VNFP) algorithm that jointly minimizes the bandwidth cost and the service end-to-end delay for satellite edge cloud networks serving IoT users. They adopt the integer nonlinear programming (INLP) problem formulation and design the distributed VNF-P algorithm by combining the Viterbi algorithm with a path selection scheme. These two techniques are utilized to search for VNF placement strategies, for the  user requests, in satellite edge servers and in cloud data center servers, respectively. The simulation results demonstrate that the proposed solution shows reduced network bandwidth cost and service end-to-end delay compared to the Viterbi algorithm while having similar performance as the game theory-based approach \cite{he2019game}. Adopting the same integrated S-T network architecture that combines the NFV technology with edge and cloud computing, Gao et al. aim to dynamically allocate network resources for the VNF deployment while maximizing the overall network payoff in \cite{NTN.C.NFV.17}. Using the INLP problem formulation and the potential game theory, they propose a decentralized resource allocation technique (PGRA) to obtain optimal VNF-P strategies. Compared to the benchmarks, the Viterbi and Greedy algorithms, the PGRA-based placement scheme demonstrates minimized service delay, bandwidth cost, and energy consumption.
    
    While the aforementioned studies focus only on solving the VNF-P problem, others advocate the joint optimization of VNF placement and flow routing (VNF-PR) because of the direct impact of the VNF-P on the routing process. In \cite{NTN.C.NFV.8}, software defined time evolving graph is used to describe the network topology capturing its time-variance and accordingly design a VNF deployment and routing strategy. The VNF-PR is considered as a multi-slot ILP with the objective of minimizing the network cost under resource constraints. The time-slot decoupled algorithm (TDA) is proposed as a heuristic-based solution for the VNF-PR problem, and it shows improved total network cost with lower computational complexity compared to the optimal solution obtained through a numerical solver. Leveraging the user services information provided by the satellite mission planning, the authors of \cite{NTN.C.NFV.14} propose a VNF placement and traffic routing algorithm for NFV-based satellite ground station networks. They utilise the INLP to formulate the network resource utilization minimization problem and the Greedy algorithm and the IBM CPLEX platform to solve it. They implement two location-aware resource allocation (LARA) based VNF-PR algorithms considering three different network architectures, VL2 \cite{greenberg2009vl2}, BCube \cite{guo2009bcube} and Fat-Tree \cite{leiserson1985fat}. The results show that the two proposed solutions have similar performance reducing the average resource utilization compared to the Greedy approach. Yang et al. examine the joint VNF deployment and flow routing for NFV-based space information networks in \cite{NTN.C.NFV.4} and \cite{NTN.C.NFV.16}, with two optimization objectives. On the one hand, they maximize the network flow while satisfying the SFC constraints of the mission flow in \cite{NTN.C.NFV.4}. They exploit the group sparse structure of the mission flow and formulate the group sparse joint VNFs deployment and flow routing (GS-VNF-R) problem as a convex optimization problem. Because of the large-scale property of the integrated S-T networks, solving the GS-VNF-R problem using conventional convex optimization techniques results into high computational complexity. Hence, the authors propose a group sparse based block-successive upper-bound minimization method of multipliers (BSUM-M-GS) algorithm, that can obtain optimal solutions with lower complexity. They validate the proposed VNF-PR strategy in terms of the network maximum flow, the averaged number of active function nodes, and the algorithmic complexity. On the other hand, Yang et al. develop a QoS-aware joint VNFs deployment and flow routing strategy that maximizes the number of completed missions under SFC constraints in \cite{NTN.C.NFV.16}. The proposed VNF-PR scheme outperforms existing methods, namely the heuristic sequential allocation \cite{zhang2017network} and the fixed VNF deployment schemes in terms of mission complete ratio and number of function nodes. Furthermore, the VNF-P problem is jointly considered with the virtual link mapping problem in \cite{NTN.C.Comb.1_B.2.1} to design a dynamic heuristic-based VNE (D-ViNE) algorithm. The problem is formulated as an ILP problem with the objective of jointly maximizing the service revenue and minimizing the costs of power consumption and VNF deployment. Compared with existing works, D-ViNE shows improved average service revenue, and reduced power consumption and VNF deployment costs.

    \item SFC embedding: In NFV-enabled integrated satellite-terrestrial networks, the dynamic and large-scale topology, the long propagation delays, and the limited on-board satellite resources increase the complexity of SFC embedding, rendering the terrestrial-based solution inadequate. Therefore, researchers are dedicating their efforts to build SFC embedding suitable for the satellite-terrestrial segment. For instance, the authors of \cite{NTN.C.NFV.15} design a multiple SFC embedding scheme for multi-service delivery for ultra-dense LEO satellite-terrestrial networks. The goal is to minimize the service delivery latency while considering the SFCs competition and resource sharing. They formulate the problem as a non-cooperative game and propose three algorithms based on potential game. The proposed best response, adaptive play, and stochastic learning (SL) algorithms show reduced overall service delivery latency. The end-to-end delay minimization is also studied in \cite{NTN.C.NFV.5} in the context of multi-domain SFC for SDN/NFV-enabled satellite-terrestrial networks. To deliver certain services, the required SFs can be distributed across multiple administrative domains which necessitates the establishment of multi-domain SFCs. Thus, the authors propose a multi-domain SFC mapping algorithm based on a heuristic approach and combined with a cooperative inter-domain path calculation technique. The simulation results show that compared to the non-cooperative method, the proposed scheme offer reduced end-to-end delay with similar bandwidth utilization.

    Furthermore, in \cite{NTN.C.NFV.3} and \cite{NTN.C.NFV.3(1)}, an SFC mapping approach based on the concepts of SF multiplexing and SFC merging is introduced for satellite-terrestrial hybrid cloud networks. With the goal of the resource consumption minimization, the SF multiplexing and SFC merging reduce the number of active VMs on cloud servers, by mapping new SFs to VMs that have previously executed them and combining new SFCs with identical SFCs already executed in the clouds. The proposed solution is implemented as a proof-of-concept prototype in the HetNet architecture \cite{NTN.C.SDN.1.3} and the results demonstrate that the SFC embedding scheme minimizes the cost and revenue average ratio compared to Greedy methods. Lastly, a load balancing-aware SFC deployment strategy is proposed in \cite{NTN.C.NFV.13}. The objective is to minimize the VNF migration cost while balancing the load of service chains. The optimization problem is modeled as a hidden Markov model and solved using the MLB-Viterbi algorithm. The proposed optimization algorithm outperforms the  multi-resource load balancing scheme \cite{wang2017multi} and the multi-steps algorithm \cite{riggio2015virtual} in terms of satellite nodes load rate and migration cost. 

    \item Virtual resource management: The optimized allocation of virtual resources in NFV-based networks is a crucial factor in ensuring optimal network performance satisfying QoS requirements and efficient resource utilization \cite{NTN.C.NFV.7}. In integrated satellite-terrestrial networks, the task becomes more challenging because of the dynamic and heterogeneous environment \cite{NTN.C.NFV.6}. A joint MEC caching placement and power allocation scheme for NFV/MEC-based satellite-terrestrial networks in proposed \cite{NTN.C.NFV.7} to jointly manage the networking and caching resources. The designed technique is based on the Mayfly algorithm \cite{zervoudakis2020mayfly} and jointly maximizes the revenue and minimizes the power consumption. The results show that it outperforms the Greedy and the particle swarm optimization methods in terms of power consumption and total system utility function. Meanwhile, the authors of \cite{NTN.C.NFV.6} develop a resource management strategy based on the idea of user intent while optimizing the resource distribution in SDN/NFV-enabled satellite networks. The intent-driven resource management CoX mechanism follows a decomposition process of the intent, defined as a series of network actions, to obtain optimal resource allocation policies. The method is validated through simulation results proving reduced delay and resource costs. Moreover, Jia et al. integrate the concept of VNF orchestration, involving the dynamic management and coordination among multiple VNFs, in the design of resource management algorithms in \cite{NTN.C.NFV.18} and \cite{NTN.C.NFV.19}. They formulate the problem as a communication resource consumption minimization problem using ILP in \cite{NTN.C.NFV.19} and solve it using the Dantzig-Wolfe decomposition, the branch-and-bound algorithm, and the column generation method. Then, they concentrate on solving the problem of satellite-to-satellite resource consumption minimization in \cite{NTN.C.NFV.18} adopting the same approach. Jia et al. employ the communication resource consumption, task completion ratio, execution time, and energy consumption to demonstrate the effectiveness of the proposed algorithms.

\end{itemize}

    Table \ref{tab:NFV_ST} summaries the relevant research on VNF placement, SFC embedding and virtual resource management in NFV-enabled integrated satellite-terrestrial networks indicating the adopted network architecture, the targeted optimization objective, the proposed solution, and the evaluation metrics.

\begin{table*}[]
\centering
\begin{tabularx}{\textwidth}{p{0.1\textwidth} p{0.0375\textwidth}
p{0.15\textwidth} p{0.16\textwidth} X p{0.2\textwidth}}
\hline
Research focus & Ref. & Network architecture & Optimization objective & Proposed solution & Evaluation metrics \\ \hline


&
\cite{NTN.C.NFV.8} & 
SDN/NFV-enabled satellite networks & 
Network cost minimization &
TDA-based VNF deployment and routing strategy  &
 Total network cost and computational complexity
 \\ \cline{2-6}

&
\cite{NTN.C.NFV.4} & 
NFV-based space information networks &
Network flow maximization &
Group sparse joint VNFs deployment and flow routing strategy   &
 Network maximum flow, averaged number of active nodes
 \\ \cline{2-6}

&
\cite{NTN.C.NFV.10} &
NFV-enabled LEO CubeSats networks &
End-to-end delay minimization &
Dynamic heuristic based VNF placement strategy  &
Service deployment delay, computing resources consumption, and processing time 
 \\ \cline{2-6}

&
\cite{NTN.C.NFV.14} & 
NFV-based satellite ground station networks  &
 Network resource utilization minimization   &
 LARA-based VNF placement and traffic routing algorithm  &
Average resource utilization
\\ \cline{2-6}

&
\cite{NTN.C.NFV.16} & 
NFV-based space information networks  &
 Number of completed missions maximization  &
 QoS-aware joint VNFs deployment and flow routing strategy  &
 Mission complete ratio and number of function nodes 
\\ \cline{2-6}

&
\cite{NTN.C.NFV.1} &
SDN/NFV-enabled satellite networks  &
 Sum of service provisioning delays minimization  &
 Dynamic security VNFs deployment strategy  &
 Service provisioning delay and running time 
 \\ \cline{2-6}

&
\cite{NTN.C.Comb.1_B.2.1} &
SDN/NFV-enabled satellite networks  &
 Revenue and cost optimization  &
 Dynamic heuristic VNE (D-ViNE) algorithm  &
 VNF deployment cost, service revenue, power consumption 
 \\ \cline{2-6}

&
\cite{NTN.C.NFV.2} &
Satellite edge cloud networks   &
 Joint bandwidth cost and service end-to-end delay minimization  &
Dynamic distributed VNF placement algorithm  &
 Average network bandwidth cost, service end-to-end delay, and running time 
 \\ \cline{2-6}

\multirow{-9}{*}{VNF placement} &
\cite{NTN.C.NFV.17} &
Satellite edge cloud networks   &
 Overall network payoff maximization  &
 Dynamic potential game-based VNF placement algorithm &
 Service delay, bandwidth cost, and energy consumption 
 \\ \hline


&
\cite{NTN.C.NFV.15} &
Ultra-dense LEO satellite networks  &
 Service delivery latency minimization  &
Multiple SFC embedding scheme   &
 Overall service delivery latency and convergence performance 
\\ \cline{2-6}

&
\cite{NTN.C.NFV.5} &
SDN/NFV-enabled satellite networks  &
 End-to-end delay minimization  &
 Multi-domain SFC mapping algorithm  &
 Bandwidth utilization and delay 
\\ \cline{2-6}

&
\cite{NTN.C.NFV.3} \cite{NTN.C.NFV.3(1)}  &
 Satellite-terrestrial hybrid cloud networks  &
 Resource consumption minimization  &
 SFC mapping approach based on SF multiplexing and SFC merging  &
 Cost and revenue average ratio 
\\ \cline{2-6}

\multirow{-9}{*}{SFC embedding} &
\cite{NTN.C.NFV.13} &
SDN/NFV-enabled satellite networks  &
 VNF migration cost and load balance optimization  &
 Load balancing-aware SFC deployment strategy  &
 Satellite nodes load rate and migration cost 
\\ \hline


&
\cite{NTN.C.NFV.7} &
MEC-enabled satellite networks  &
Joint power consumption and revenue optimization  &
 Joint caching placement and power allocation scheme  &
 Power consumption and total system utility function
\\ \cline{2-6}

&
\cite{NTN.C.NFV.6} &
SDN/NFV-enabled satellite networks  &
 Resource distribution optimization  &
 Intent-driven resource management  mechanism   &
Delay and resource costs 
\\ \cline{2-6}

Virtual resource management &
\cite{NTN.C.NFV.18} &
SDN/NFV-enabled LEO satellite networks  &
Resource consumption minimization  &
VNF orchestration-based resource management algorithm  &
Resource consumption, execution time, task completion ratio 
\\ \cline{2-6}

 &
\cite{NTN.C.NFV.19} &
SDN/NFV-enabled LEO satellite networks  &
 Resource consumption minimization  &
 VNF orchestration-based service provision scheme  &
 Task completion ratio, resource and energy consumption 
\\ \hline

\end{tabularx}
\caption{NFV-enabled integrated Satellite-Terrestrial networks: Traditional approaches.}
\label{tab:NFV_ST}
\end{table*}

\subsubsection{Network Slicing-based Networks} \hfill\\
By partitioning the network infrastructure, network slicing enables the creation of multiple virtual and isolated networks providing diversified applications. Incorporating such technology in the integration of satellite and terrestrial networks allows operators to achieve service ubiquity, continuity, and scalability while ensuring efficient resource utilization and enhanced network performance \cite{NTN.C.NS.1, NTN.C.NS.6, NTN.AI.Comb.1}. Nonetheless, adopting network slicing in the S-T segment is still in its infancy as research efforts in this area are limited with only a handful of studies. They mainly investigate the issues of traffic scheduling \cite{NTN.C.NS.2, NTN.C.NS.4} and resource management \cite{NTN.C.NS.5, NTN.AI.NS.1, NTN.AI.NS.2, NTN.AI.NS.4} in network slicing-based S-T networks, while employing traditional or AI-enabled methodologies.

    \textbf{\textit{a) Traditional approaches:}} In slicing-based integrated S-T networks, the traffic scheduling and offloading are simultaneously studied in \cite{NTN.C.NS.2} and \cite{NTN.C.NS.4}. On the one hand, the authors of \cite{NTN.C.NS.2} design a hybrid satellite-LTE downlink data scheduler that derives the service priorities in the same URLLC slice, while optimizing the network reliability and latency. On the other hand, computation offloading and scheduling are examined in edge computing-based satellite (SatEC) networks for  applications in \cite{NTN.C.NS.4}. A multi-objective optimization of latency, end-to-end transmission power attenuation, and computational power is formulated and solved using two heuristic algorithms, namely the multi-objective Tabu search (MOTS) and the golden-section technique. The satellite edge MOTS (SE-MOTS) method determines the offloading scheme for different slices, while the golden-section-based algorithm computes the sliced SatEC network scheduling technique for different users based on the offloading rule provided by the SE-MOTS. The strategy is validated through simulations in terms of latency, transmission power, and computational power in case of LEO and VLEO satellites. 
    
    Core network slicing is considered in \cite{NTN.C.NS.3}, where the authors propose an on-demand resource allocation method for VNF and SFC provisioning. They formulate the slicing problem as a MILP problem with the objective of resource consumption minimization and solve it using the AIMMS optimization framework. The proposed OnDReAMS shows improved performance for the end-to-end delay, and the average number of QoS violations and accepted slice requests. Furthermore, the network slice planning, based on inter-slice RAN resource reservation, is examined in \cite{NTN.C.NS.5} for integrated satellite-terrestrial networks. Compared to terrestrial networks, network planning is more complex in the S-T segment due to the frequent handovers caused by satellite mobility. The slice planning problem is modeled as VNE and satellite handover management problems. Taking into account the optimization of latency, transmission, and computational power, four handover-based VNE schemes are designed including the “closest, max flow”, the “closest, low latency”, the “longest, max flow” and the “longest, low latency” methods. The proposed mechanisms are implemented using shortest-path algorithms in an SDN-enabled network and evaluated in terms of the number of handovers, cost, latency, and throughput.

    \textbf{\textit{b) AI-based approaches: }} In the context of network slicing-based networks, AI models are mainly employed to solve issues related to resource management. For instance, the intra-slice resource management is considered as a case study in \cite{NTN.AI.NS.1} to demonstrate the efficiency of AI models in addressing network slicing issues for highly dynamic integrated S-T networks. Various AI models, including CNN and DRL, are employed in the design of AI-based RAN slicing algorithms. With the goal of slice cost minimization, they optimally allocate available radio resources to the end users while meeting the QoS and slice isolation constraints. Additionally, the RAN resource orchestration of the 5G eMBB slice is studied in \cite{NTN.AI.NS.4} with the objective of providing eMBB services to train passengers via an integrated satellite-terrestrial network. Taking into account the different QoS levels required to satisfy the users' demands, the packet delivery latency is minimized to obtain the optimal resource allocation strategy for each slice. Two algorithms are designed based on queuing theory and neural networks (NN) to solve the optimization problem. The simulation results show the superiority of the NN-based resource allocation technique in terms of latency and QoS satisfaction compared to the queuing theory-based approach. Moreover, the authors of \cite{NTN.AI.NS.2} formulate the problem of joint RAN resource reservation and orchestration as a problem of joint slicing and scheduling of spectrum resources (JRSS) in integrated satellite-terrestrial vehicular networks providing delay-sensitive (DSS) and delay-tolerant services (DTS). They use stochastic optimization to model the problem minimizing the long-term system cost which includes the costs of slice reconfiguration, DTS delay, DSS requirements violations. They also develop a two-layered RL-based JRSS technique by decomposing the problem into two sub-problems; resource slicing and resource scheduling sub-problems. While the former is solved by applying an RL-based proximal policy optimization algorithm to pre-allocate the resources to multiple slices, the latter is solved using matching-based optimization approaches for the scheduling of resources to users in each slice. Compared to existing algorithms such as the best SNR and the equal bandwidth associations methods, the proposed solution shows reduced system cost and bandwidth consumption while meeting QoS constraints. Meanwhile, a network slicing framework with a dynamic ML-based user association strategy is introduced in \cite{NTN.AI.NS.3} for SDN-NFV-enabled S-T networks. The proposed scheme utilizes an ML-based ant colony optimization algorithm to classify user requests and assign the appropriate slice to each user while minimizing the delay and link cost. Compared to the shortest delay and best fit slicing schemes, the ML-based method offers efficient resource management with an increased user acceptance ratio. 
    
    Table \ref{tab:NS_ST} summarises the efforts dedicated to the adaptation of network slicing in integrated satellite-terrestrial networks through conventional and AI-enabled solutions, it highlights the research focus, objective and proposed solution for each work.

\begin{table*}[]
\centering
\begin{tabularx}{\textwidth}{p{0.11\textwidth} p{0.0375\textwidth} p{0.25\textwidth} X p{0.22\textwidth}}
\hline

 Research focus &
  Ref. &
  Objective &
  Proposed solution &
  Evaluation metrics \\ \hline
  
    \multicolumn{5}{c}{Traditional approaches} \\ \hline

   &
  \cite{NTN.C.NS.2} &
  Network reliability and latency optimization &
  Hybrid satellite-LTE downlink scheduler &
  Delay \\ \cline{2-5}

  \multirow{-2}{*}{\begin{tabular}[c]{@{}l@{}} Traffic scheduling \\ and offloading \end{tabular}} &
  \cite{NTN.C.NS.4} &
  Multi-objective optimization of latency, transmission, and computational power  &
   SE-MOTS-based offloading and sliced SatEC network scheduling strategies  &
   Latency, transmission and computational power 
   \\ \hline

  CN Slicing &
  \cite{NTN.C.NS.3} &
  Resource consumption minimization &
  On-demand resource allocation method for VNF and SFC provisioning  &
   Delay, average number of QoS violations and accepted slice requests  
   \\ \hline

 RAN resource management & 
  \cite{NTN.C.NS.5} &
  Optimization of latency, transmission, and computational power  &
   Handover management and VNE schemes for RAN resource reservation &
  Number of handovers, cost, latency, and throughput 
  \\ \hline
 
  \multicolumn{5}{c}{AI-based approaches} \\ \hline
  
    &
  \cite{NTN.AI.NS.1} & 
  Slice cost minimization &
  Various AI-based RAN slicing algorithms &
  Loss and cost values \\ \cline{2-5}
  
    &
  \cite{NTN.AI.NS.2} & 
  Long-term system cost minimization & 
  Two-layered RL-based JRSS technique &
  System cost and bandwidth consumption  \\ \cline{2-5}

    \multirow{-3}{*}{\begin{tabular}[c]{@{}l@{}} RAN resource \\ management \end{tabular}} &
    \cite{NTN.AI.NS.4} & 
  Packet delivery latency minimization  &
   QoS-aware NN-based resource allocation technique  &
   Latency and QoS satisfaction \\ \hline

 Device/user association  &
  \cite{NTN.AI.NS.3} &
  Delay and cost minimization &
  Dynamic ML-based user association strategy  &
  User acceptance ratio \\ \hline

\end{tabularx}
\caption{Network Slicing-based integrated Satellite-Terrestrial networks: Traditional and AI-based approaches.}
\label{tab:NS_ST}
\end{table*}

\subsection{Virtualization in the Aerial-Terrestrial Segment}

\subsubsection{SDN-enabled Networks} \hfill\\
The data/control planes decoupling and the centralized control of the SDN paradigm can enhance network flexibility and programmability with improved resource management efficiency. This prompted researchers to investigate the introduction of the SDN paradigm in integrated Aerial-Terrestrial networks, with an emphasis on UAV-assisted networks \cite{AT.C.SDN.6.1}. While a number of studies examined the key architectural considerations and experimental implementations for applying SDN in UAV-Terrestrial networks \cite{AT.C.SDN.1.1, AT.C.SDN.1.2, AT.C.SDN.1.3, AT.C.SDN.1.4,AT.C.SDN.1.5, AT.C.SDN.1.6, AT.C.SDN.1.7, AT.C.SDN.1.8,AT.C.SDN.1.9, AT.C.SDN.1.10,AT.C.SDN.1.11,AT.C.SDN.1.12, AT.C.SDN.1.13, AT.C.SDN.1.14}, others tackled the arising issues including routing optimization, resource management, traffic offloading, and network security, employing conventional techniques \cite{AT.C.SDN.2.1,AT.C.SDN.2.3,AT.C.SDN.2.4,AT.C.SDN.2.5,AT.C.SDN.2.6, AT.C.SDN.3.1,AT.C.SDN.3.3,AT.C.SDN.5.1, AT.C.SDN.5.3, AT.C.SDN.5.4,AT.C.SDN.4.1,AT.C.SDN.6.2} or AI-enabled approaches \cite{AT.AI.SDN.2,AT.AI.SDN.3,AT.AI.SDN.4}.

   \textbf{\textit{a) Architectures and experimental implementations:}}
Airborne platform has been employed to serve as backhaul connections or gateways for ground data transmission in different application scenarios. Researchers proposed various SDN-enabled architectures for joint aerial-terrestrial networks for mobile and WiFi connectivity applications
\cite{AT.C.SDN.1.1, AT.C.SDN.1.2, AT.C.SDN.1.3, AT.C.SDN.1.8,AT.C.SDN.1.9,AT.C.SDN.1.11}. The aerial devices are considered to replace the base stations in the terrestrial mobile network to provide connectivity to the UEs. In \cite{AT.C.SDN.1.2}, the authors consider the scenario that UAVs could serve as handover links for ground UEs moving between marco cells, or as access points to form small cells. An SDN-UAV architecture has been proposed for deploying shifting policies and network management. The macro base station on the ground acts as a controller and UAVs operate as SDN switches. Both centralized and distributed controller approaches have been developed. The distributed controller approach outperforms the centralized approach with lower handover latency, and ease for troubleshooting. Another proposed architecture SkyCore pushes further the core functionality to the UAVs \cite{AT.C.SDN.1.3}. SkyCore relocates the BS's evolved packet core (EPC) entity to run on the UAVs where the EPC functionalities are defined as lightweight SDN applications to eliminate distributed interfaces and reduce function complexity. In addition, the UAV-related control information such as flight change events and remaining battery are implemented into the SDN control. Other than serving as a backhaul, UAVs offer other applications. To further enhance the network performance in future 6G, a software-defined multi-controller-based UAV architecture has been proposed in \cite{AT.C.SDN.1.8} for mobile connectivity. Two layers of SDN controllers are placed on the ground network, and the large-scale UAV networks serve as backhaul base stations. By applying multi-layer controllers, the system will become more flexible and provide more control to different tenants. The primary controller serves as central management and secondary controllers manage specific regions. The authors of \cite{AT.C.SDN.1.11} design an SDN-based framework for UAV networks taking into account the UAVs' location and energy constraints in the SDN controller. In their proposed architecture, the management plane considers the battery level of each UAVs for data route selection in the UAV mesh network. The traffic load is distributed evenly with lower battery consumption to maintain a longer lifetime UAV network.

Moreover, some architectures proposed mixed ground and UAVs controller structure \cite{AT.C.SDN.1.1, AT.C.SDN.1.9}. In \cite{AT.C.SDN.1.9}, UAVs not only serve as data plane forwarding devices but also as SDN controllers which can be placed either on ground or aerial platforms. In this architecture, the UAV controller is responsible for controlling the location and battery storage of UAVs, while the SDN controller is responsible for network management. Meanwhile, SDN controllers can be software applications running on ground servers or co-located with switches in the aircraft in \cite{AT.C.SDN.1.1}. The authors consider a packet-based distribution layer deployed between the traditional core and radio access networks in mobile networks. The distribution layer consists of many aerial vehicles to form a complex time-dynamic multi-hop wireless mesh network. The time-dynamic distribution layer imposes additional complexity in network control. Challenges such as coordinating links between aircraft peers and handling disruption on the dynamic UAV routes are addressed by the proposed temporospatial SDN (TS-SDN) architecture. In TS-SDN, the future state of networks could be predicted by the knowledge of the dynamic nature and physical relations between UAVs and ground stations. 

Besides, post-disaster applications are considered in \cite{AT.C.SDN.1.4, AT.C.SDN.1.5}. An SDN architecture SD-UAVNet is proposed to deliver a life video surveillance service for disaster recovery combining aerial and terrestrial networks \cite{AT.C.SDN.1.4}. Different from previously mentioned architectures, UAVs could be source, relay, and controller nodes in SD-UAVNet. In this architecture, one UAV serves as the controller node to perform all control functions such as maintaining the network topology establishing the routing path, and directing UAVs to the optimal locations. Meanwhile, multiple UAVs serve as relay nodes for forwarding data packets and they are controlled by the UAV controller. Based on a ground controller, an SDN system is proposed to predict aerial gateway link outages by analyzing the aircrafts’ location and radio link performance. While the UAVs are serving as gateways for connecting disjointed networks on the ground and in the air in post-disaster and military scenarios, the aerial system is responsible for providing information such as position, antenna direction, and link quality to the SDN controller \cite{AT.C.SDN.1.5}. In this architecture, the SDN controller updates the high-level flow table to the UAVs by OpenFlow protocol, and the radio network controller controls the radio terminals by Simple Network Management Protocol (SNMP).

For vehicular networks scenario, the authors of \cite{AT.C.SDN.1.7} propose an SDN-enabled three-tier architecture where the wireless communication between ground vehicles, UAVs, and base stations enables a real-time road traffic navigation strategy. The UAVs, acting as SDN switches, and ground vehicles provide instantaneous road traffic information to the SDN controllers for suggesting the best shortest time path planning for the ground vehicles. Meanwhile, using hierarchical multi-controllers, researchers in \cite{AT.C.SDN.1.10} design an SDN-based UAV-assisted infrastructure-less architecture for vehicular ad-hoc networks where the UAVs are used to assist emergency vehicles in road incidents. They introduce a monitoring platform to analyze the UAV information with a load-balancing algorithm. In this architecture, the primary controller is located at the emergency vehicles such as fire engines. The local controller would be a ground vehicle that monitors cars nearby. The UAV controller monitors the UAV information and forwards data. In addition, an agricultural application is targeted in the design of a cloud-based softwarization architecture for UAVs and wireless sensor networks (WSN) in \cite{AT.C.SDN.1.6}. The higher layers, including UAV controller layer, WSN controller layer, orchestration layer, and application layer, are implemented into the cloud. The virtualization of the hardware infrastructure is decoupled from the control layer, improving the system's reaction to failures, and increasing reliability. 

Furthermore, architectures for UAV swarms have been proposed in \cite{AT.C.SDN.1.12, AT.C.SDN.1.13, AT.C.SDN.1.14}. An SDN architecture for battlefield UAV swarms is proposed in \cite{AT.C.SDN.1.12}. Each UAV in the swarm can act as a master or a slave in the swarm by switching on and off the onboard functions. Both the master and slave UAVs consist of application, control, and forwarding planes. The SDN controller in the control plane estimates the topology and calculates a multi-path solution meeting QoS requirements. The architecture introduced in \cite{AT.C.SDN.1.13} is also designed for military UAV swarms. The proposed UAV swarm architecture includes one controller UAV node, a set of relay UAV nodes, and a set of independent nodes. The topology and routing management are centralized at the flying controller nodes. The relay nodes establish data connections for independent nodes and the controller node. The independent nodes can be ground vehicles, soldiers troops, or UAVs performing military missions. The controller is responsible for setting up routing table rules for all nodes and managing the topology network. Another SDN-based swarm architecture is studied in \cite{AT.C.SDN.1.14} providing security features. Securing the Ad-hoc On-Demand Distance Vector (AODV) routing protocol is the primary action to prevent routing attacks. The SDN controller becomes a source of credentials and a building block for a public key infrastructure for protection.

Table \ref{tab:SDN_AT_architectures_exp} summarizes the efforts in terms of proposed architectures and experimental implementations in SDN-enabled integrated Aerial-Terrestrial networks.

\begin{table*}
\centering
\begin{tabularx}{\textwidth}{ p{0.0375\textwidth} p{0.135\textwidth} p{0.145\textwidth} p{0.2\textwidth} X }
\hline 
Ref.                 & Controller placement & Use case scenario & Implementation Tools & Comments \\ \hline

\cite{AT.C.SDN.1.1} & 
Ground and aerial platforms  &
Backhaul mesh mobile networks &
OpenFlow-inspired CDPI protocol&
Design an SDN-enabled architecture where network state is predicted using the knowledge of physical position and trajectory of UAVs
 \\ \hline

\cite{AT.C.SDN.1.2} &
Ground station &
Mobile networks & 
Mininet, OpenFlow &
Employ SDN to solve the problem of UEs handover in UAV-assisted mobile network

 \\ \hline

\cite{AT.C.SDN.1.3} &
UAVs &
LTE mobile networks &
P4 w/OpenFlow, Lagopus, Open vSwitch &
Propose the SkyCore architecture where the EPC functionalities are lightweight SDN applications running on UAVs
\\ \hline

\cite{AT.C.SDN.1.4}&
UAVs &
Post-disaster applications &
OMNeT++ &
Design an SDN-based UAV network for disaster recovery applications using multiple UAV relays and a UAV global controller
\\ \hline

\cite{AT.C.SDN.1.5}&
Ground station &
Post-disaster and military applications &
Mininet, NS-3, OpenFlow, Open vSwitch, OpenDayLight &
Develop an SDN system that predicts aerial gateway link outages by analyzing the aircrafts’ location and radio link performance
\\ \hline

\cite{AT.C.SDN.1.6}&
Cloud & 
Agricultural applications &
NodeJS &
Propose a Cloud-based softwarization architecture for UAV-based WSNs 
\\ \hline

\cite{AT.C.SDN.1.7}&
 Ground station &
 Vehicular applications &
 OMNeT++, SUMO, MobiSim &
 Design an SDN-enabled UAV-based architecture for vehicle path planning where the UAVs detect and monitor the road traffic acting as SDN switches
\\ \hline

\cite{AT.C.SDN.1.8}&
Ground station &
Mobile connectivity &
 OpenDaylight, MATLAB SimEvents &
Propose a holistic UAV system based on SDN hierarchical multi-controllers structure, with hybrid routing and adaptive load balancing algorithms 
\\ \hline

\cite{AT.C.SDN.1.9} &
Ground and aerial platforms &
WiFi connectivity &
Mininet-WiFi, POX controller &
Design an SDN architecture for UAV backbone network with monitoring platform and load balancing algorithm 
\\ \hline

\cite{AT.C.SDN.1.10} &
Ground station and UAVs &
Vehicular ad-hoc networks  &
MATLAB for numerical evaluation &
Propose an SDN-based UAV-assisted architecture using hierarchical multi-controllers for emergency vehicles assistance in road incidents
\\ \hline

\cite{AT.C.SDN.1.11}&
Ground station &
WiFi connectivity &
OpenFlow &
Design SDN-based framework for UAV networks with traffic load balancing path selection algorithm
\\ \hline

\cite{AT.C.SDN.1.12}&
 UAVs &
 Military applications &
 N/A &
 Propose a UAV swarm architecture based on SDN and message queue telemetry transport (MQTT) structure with a QoS-based multi-path routing scheme
\\ \hline

\cite{AT.C.SDN.1.13}&
UAVs &
Military applications &
 OMNet++, M3WSN &
 Propose an SDN-based centralized UAV topology management algorithms for multi-hop UAV ad-hoc networks
\\ \hline

\cite{AT.C.SDN.1.14}&
Ground stations and UAVs &
UAV swarm security applications&
OpenFlow, Ryu controller, OFSoftSwitch13, AODV &
Proposed and evaluated two sub-architectures to improve the security of UAV swarms using an ad-hoc routing protocol an in-band SDN-based architecture
\\ \hline

\end{tabularx}
\caption{SDN-enabled integrated Aerial-Terrestrial networks: Architectures and experimental implementations.}
\label{tab:SDN_AT_architectures_exp}
\end{table*}

\textbf{\textit{b) Traditional approaches:}} Routing optimization, resource management, and traffic offloading are the primary issues tackled in SDN-enabled integrated Aerial-Terrestrial networks utilising traditional techniques. Other research directions involve controller placement and network security.

\begin{itemize} 
    \item Routing optimization: In the context of SDN-based integrated A-T networks, routing algorithms are developed in \cite{AT.C.SDN.2.1,AT.C.SDN.2.3,AT.C.SDN.2.4,AT.C.SDN.2.5,AT.C.SDN.2.6} considering different controller configurations. Firstly, the single controller structure is adopted in \cite{AT.C.SDN.2.3,AT.C.SDN.2.4,AT.C.SDN.2.5} to route the network traffic of UAV-terrestrial networks to achieve different objectives. With the goal of joint throughput, delay and load balancing optimization, the authors of \cite{AT.C.SDN.2.3} design a priority-based ad-hoc routing scheme employing the Dijkstra and the Ford-Fulkerson algorithms. The simulation results shows that the proposed scheme outperforms other ad-hoc routing algorithms in terms of throughput, delay, and packet delivery ratio. Meanwhile, the end-to-end delay is minimized in \cite{AT.C.SDN.2.4, AT.C.SDN.2.5}. A resilient multi-path routing algorithm is proposed combining the Vertex Splitting method and the Dijkstra algorithm for vehicular applications. Compared to conventional shortest-path routing schemes, the proposed routing mechanism presents improved end-to-end resiliency, and outage probability at the cost of slight increase in end-to-end delay. Secondly, the authors of \cite{AT.C.SDN.2.6} consider the multi-controllers configuration in their airborne backbone network architecture. With the goal of reliability and bandwidth utilization maximization, they develop a reliable multi-path routing scheme based on segment routing (MRP-TS), combining a reliable path calculation algorithm and a bandwidth preemption method. The results demonstrate the superiority of MRP-TS compared to existing routing protocols with respect to the metrics of reliability, delay, and bandwidth utilization. Thirdly, the hierarchical multi-controllers structure is employed in \cite{AT.C.SDN.2.1} for SDN-based flying ad-hoc sensor networks. With the objective of delay minimization and reliability maximization, an ant-colony based traffic-differentiated routing algorithm is designed and validated in terms of throughput, delay, and packet dropping ratio.

    \item Resource management: Through the data/control planes separation and the logically centralized control, the SDN paradigm facilitates the management of data plane resources. In integrated Aerial-Terrestrial networks, researchers focus on taking advantages of these SDN features to optimally allocate network resources in UAV-assisted networks where the UAVs act as forwarding devices managed by the SDN controller    \cite{AT.C.SDN.3.1,AT.C.SDN.3.2,AT.C.SDN.3.3}. For instance, a hybrid computing resource allocation algorithm based on a Simulated Annealing technique and a greedy algorithm is designed in \cite{AT.C.SDN.3.3}. Leveraging its global view of the network, the SDN controller allocates the computing resources to the UAVs for the processing of its applications. The hybrid approach allows the controller to select the optimal server, which can be located on-board of the UAVs or at cloud/edge servers, with the goal of minimizing the average application latency and the UAV energy consumption while satisfying the QoS requirements. The proposed scheme outperforms the only-cloud and only-on-board processing methods with respect to the UAV energy consumption and the application latency. Moreover, the authors of \cite{AT.C.SDN.3.1} exploit the SDN controller's capabilities to jointly optimize the resource allocation, user association and 3D UAV placement for UAV-assisted cellular networks with the objective of overall users data rate utility maximization. They propose a distributed alternating maximization iterative resource allocation scheme based on successive convex optimization (SCO) and modified alternating direction method of multipliers (ADMM) techniques. The simulation results show improved throughput and network utility compared to benchmark algorithms. 

    \item Traffic scheduling and offloading: SDN-enabled Integrated UAV-terrestrial networks offers the opportunity to offload traffic and tasks from one network node to another in a flexible manner  \cite{AT.C.SDN.5.1,AT.C.SDN.5.3, AT.C.SDN.5.4}. The authors of \cite{AT.C.SDN.5.3} propose a data traffic offloading scheme aiming to offload the data of cellular subscribers from the licensed UAV link to the unlicensed WiFi link in SDN-based UAV-WiFi networks. Taking into account the UAV placement, the licensed spectrum allocation, and the cellular subscribers association and offloading to WiFi, they minimize the queuing delay of cellular subscribers and meet the delay requirements of WiFi subscribers. The designed algorithm is based on heuristics and convex optimization techniques and validated in terms of cellular subscribers average queuing delay performance. Meanwhile, the UAV charging is considered with data offloading \cite{AT.C.SDN.5.4} with the goal of network utility maximization. An SDN–enabled location-aware opportunistic data offloading and UAV charging mechanism is developed aiming to avoid congested paths and extend the UAV flight time. While the works in \cite{AT.C.SDN.5.3} and \cite{AT.C.SDN.5.4} concentrate on data traffic offloading, the researchers in \cite{AT.C.SDN.5.1} examine issue of computation offloading. They design a dynamic game theory-based computation offloading mechanism for SDN-enabled UAV-based vehicular networks. Targeting the minimization of energy consumption and execution time of computing tasks, vehicular users offload them to the flying UAVs which can either execute the computation tasks or offload them to edge servers. Compared to other strategies, the proposed solution presents reduced system cost and increased number of completed tasks per minute.

    \item Other research directions: The controller placement problem is studied in \cite{AT.C.SDN.6.2} for SDN-enabled aeronautical networks. Based on a hierarchical multi-controllers structure, two dynamic placement schemes are proposed with the objective of maximum controller load ratio minimization. The first algorithm optimally places the controllers using an enumeration technique and assigns the switches based on fastest shortest-path method, while the second CPP scheme dynamically optimizes the controllers placement and switches assignment using a Genetic algorithm. Network security is another issue that has been examined in the literature where the authors of \cite{AT.C.SDN.4.1} develop an SDN-based topology deception scheme to mitigate the target selection attack and protect key UAVs in UAV-assisted WSNs. Thanks to the centralized control, the mechanism deceives the attackers by creating a virtual topology using honeypot drones impairing their judgment. 
\end{itemize}

    \textbf{\textit{c) AI-based approaches:}} Thanks to their ability to adapt to highly dynamic environments, RL models are employed to design dynamic resource management and routing mechanisms for SDN-based UAV-terrestrial networks. On the one hand, the authors of \cite{AT.AI.SDN.3} propose a data plane resource allocation algorithm based on deep Q-learning in SDN-enabled ad-hoc UAV networks. They minimize the number of active UAVs to optimally allocate WiFi channels to end users while maintaining desired QoS and optimizing UE coverage and energy efficiency. They also validate the proposed solution through testbed experiments taking into account the QoS satisfaction, UE coverage, and power consumption as performance metrics. On the other hand, a dynamic single-path routing strategy, named the Air-to-ground Intelligent Information Pushing Optimization (AIIPO) algorithm, is developed in \cite{AT.AI.SDN.4}. The AIIPO is based on a deep Q-learning model which solves the optimization problem of throughput maximization while adapting to network changes in IoT data collection UAV networks. The simulation results show that the AIIPO outperforms benchmark methods with respect to throughput and computation complexity. Moreover, the K-means clustering model is combined with the autoregressive integrated moving average (ARIMA) algorithm in  \cite{AT.AI.SDN.2} to improve the security of data dissemination in SDN-enabled UAV-based IoT networks. K-means and ARIMA are employed with a blockchain technique to secure data transmission from IoT devices to UAVs to SDN controllers by detecting eavesdropping and malicious data and mitigating cyber-attacks on the controllers.

Table \ref{tab:SDN_AT} gives a summary on relevant works reported in the literature investigating SDN-enabled integrated Aerial-Terrestrial networks employing traditional and AI-based approaches.

\begin{table*}[]
\centering
\begin{tabularx}{\textwidth}{p{0.07\textwidth} p{0.0375\textwidth}
p{0.11\textwidth} p{0.08\textwidth} p{0.17\textwidth}  X p{0.16\textwidth}}
\hline
Research focus &
  Ref. &
  Controller configuration &
  Type of problem &
  Objective &
  Proposed solution &
  Evaluation metrics \\ \hline

  \multicolumn{7}{c}{Traditional approaches} \\ \hline

   &
   \cite{AT.C.SDN.2.1} &
   Hierarchical multi-controllers &
   Single-path routing &
   Delay minimization and reliability maximization &
    Ant-colony based traffic-differentiated routing algorithm &
   Throughput, delay, and packet dropping ratio
   \\ \cline{2-7}

  &
  \cite{AT.C.SDN.2.3} &
  Single controller &
  Single-path routing &
  Throughput, delay and load balancing optimization &
    Priority-based ad-hoc routing scheme based on Dijkstra and Ford-Fulkerson algorithms &
    Throughput, delay, and packet delivery ratio 
  \\ \cline{2-7}

   Routing optimization &
   \cite{AT.C.SDN.2.4} \cite{AT.C.SDN.2.5}&
   Single controller &
   Multi-path routing &
   Delay minimization &
   Resilient multi-path routing algorithm based on Vertex Splitting and Dijkstra methods &
   E2E resiliency, delay and outage probability
   \\ \cline{2-7}

  &
   \cite{AT.C.SDN.2.6} &
   Multi-controllers &
   Multi-path routing &
   Reliability and bandwidth utilization maximization &
    Segment routing-based reliable multi-path routing scheme &
    Reliability, delay, and bandwidth utilization 
    \\ \hline

   Resource management &
   \cite{AT.C.SDN.3.1} &
   Single controller &
   Data plane resources &
    Overall users data rate utility maximization&
    SCO and ADMM-based Iterative resource allocation scheme &
    Throughput and network utility 
    \\ \cline{2-7}

     &
    \cite{AT.C.SDN.3.3} &
   Single controller &
   Data plane resources &
    UAV energy consumption and average application latency minimization &
    Heuristics-based hybrid computing resource allocation algorithm &
    UAV energy consumption and application latency 
    \\ \hline

       &
   \cite{AT.C.SDN.5.1} &
    Single controller &
    Computation offloading & 
    Execution time and energy consumption minimization &
    Dynamic game theory-based computation offloading scheme &
    Cost and number of completed tasks per minute
    \\ \cline{2-7}

       &
   \cite{AT.C.SDN.5.3} &
   Single controller &
   Data offloading & 
   Cellular subscribers queuing delay minimization &
   Heuristic and convex optimization based cellular subscribers traffic offloading scheme &
    Cellular subscribers average queuing delay 
    \\ \cline{2-7}

 \multirow{-4}{*}{\begin{tabular}[c]{@{}l@{}} Traffic\\ scheduling \\ and \\ offloading \end{tabular}}   &
    \cite{AT.C.SDN.5.4} &
    Hierarchical multi-controllers &
    Data offloading & 
   Network utility maximization  &
    Location-aware opportunistic data offloading and UAV charging mechanism &
    Throughput, E2E delay, and handover latency 
    \\ \hline

   Controller placement &
   \cite{AT.C.SDN.6.2} &
   Hierarchical multi-controllers &
   Dynamic CPP with switches assignment &
    Maximum controller load ratio minimization &
    Enumeration-based and Genetic algorithm-based dynamic placement with fastest and dynamic assignment strategies &
    Load balancing and controller load ratio
    \\ \hline


   Network security &
    \cite{AT.C.SDN.4.1} &
    Single controller &
    N/A & 
   Target selection attack mitigation &
    Topology deception-based attack mitigation scheme  &
   Connectivity loss 
    \\ \hline

      \multicolumn{7}{c}{AI-based approaches} \\ \hline

   Resource management &
   \cite{AT.AI.SDN.3} &
   Single controller &
   Data plane resources &
    Active UAVs number minimization &
    Deep Q-learning based resource allocation algorithm  &
    QoS satisfaction, UE coverage, and power consumption
    \\ \hline

   Routing optimization &
   \cite{AT.AI.SDN.4} &
    Single controller &
    Single-path routing &
   Throughput maximization &
    Deep Q-learning based dynamic routing strategy &
   Throughput and computation complexity 
    \\ \hline

    Network security &
    \cite{AT.AI.SDN.2} &
    Multi-controllers &
    N/A &
   Eavesdropping and malicious data detection  &
    Blockchain-enabled IoT data dissemination scheme based on K-means and ARIMA algorithms &
   Accuracy, precision, recall, and f1 score
    \\ \hline

\end{tabularx}
\caption{SDN-enabled integrated Aerial-Terrestrial networks: Traditional and AI-based approaches.}
\label{tab:SDN_AT}
\end{table*}

\subsubsection{NFV-enabled Networks} \hfill\\
Adapting the NFV technology further enhances the flexibility and agility provided by the integrated aerial and terrestrial networks with reduced deployment costs \cite{B.2.3.1}. Only a few works have been reported in the literature discussing the use of NFV in the A-T segment with a focus on UAV-based networks. They provide insights on architecture and implementation considerations \cite{AT.C.NFV.1, AT.C.NFV.2, AT.C.NFV.3, AT.C.NFV.9} and propose potential solutions to issues related to VNF placement and SFC deployment \cite{AT.C.NFV.4, AT.C.NFV.5, AT.C.NFV.6} in integrated aerial-terrestrial networks. 

    \textbf{\textit{a) Architectures and experimental implementations:}} In \cite{AT.C.NFV.1}, an NFV-enabled UAV-based system is proposed to deliver different services in an ad-hoc communication network. The feasibility of the system is tested using a prototype and the results show that using lightweight VNFs increases the flexibility and cost efficiency of network service deployment over resource-limited UAVs. Another architecture combining NFV, SDN, and MEC technologies is designed in \cite{AT.C.NFV.2} for flying ad-hoc networks (FANETs) to provide massive connectivity to  devices and mobile users. With the objective of sum-rate maximization, the authors propose a NOMA-based multiple-access mechanism and a relay selection algorithm based on VNF migration. The performance of the solution is evaluated in terms of the delay and the sum-rate. Moreover, the Virtualized Environment for multi-UAV network emulation (VENUE) is designed in \cite{AT.C.NFV.9} to offer an ecosystem to implement, prototype and validate the development of multi-UAV services. The framework is based on Linux containers and the NS3 simulator taking into account the specific features of UAV-based networks. Furthermore, the authors introduce an NFV/MEC-based UAV architecture with a security management framework in \cite{AT.C.NFV.3} to investigate network security in NFV-enabled integrated aerial-terrestrial networks. They, also, develop a security VNF placement algorithm optimizing security orchestration and resource utilization and validate it in terms of service deployment time, RAM, and battery consumption.

\begin{table*}
\centering
\begin{tabularx}{\textwidth}{ p{0.0375\textwidth} p{0.145\textwidth} p{0.13\textwidth} p{0.15\textwidth} X }
\hline 
Ref.  & Network architecture & Use case scenario & Implementation Tools & Comments \\ \hline
\cite{AT.C.NFV.1} &
NFV-based ad-hoc UAV networks  &
Multiple UAV applications &
Open Source MANO, OpenStack Ocata &
Study the feasibility of NFV-based ad-hoc UAV system through prototype tests
 \\ \hline

\cite{AT.C.NFV.2} &
MEC/SDN-enabled flying ad-hoc networks &
  Massive users/dev- ices connectivity &
 N/A  &
 Propose a MEC/SDN/NFV-enabled architecture for FANETs with NOMA-mechanism VNF migration-based relay selection algorithm
 \\ \hline

\cite{AT.C.NFV.3} &
MEC/SDN-enabled IoT UAV networks  &
 IoT emergency applications &
 OpenStack Stein, OpenFlow, Open vSwitch  &
 Propose a MEC/SDN and NFV-enabled UAV architecture with a security management framework and a VNF placement scheme
 \\ \hline

\cite{AT.C.NFV.9} &
 NFV-based UAV flying ad-hoc networks &
 5G connectivity &
  Linux containers, NS3 simulator &
 Design a virtualized environment emulation framework (VENUE) to facilitate the development of multi-UAV networks
 \\ \hline
 
\end{tabularx}
\caption{NFV-enabled integrated Aerial-Terrestrial networks: Architectures and experimental implementations.}
\label{tab:NFV_AT_architectures_exp}
\end{table*}

\textbf{\textit{b) Traditional approaches:}} Challenges related to SFC deployment are addressed in \cite{AT.C.NFV.4, AT.C.NFV.5, AT.C.NFV.6}. On the one hand, the SFC migration problem defined as the re-mapping of the ordered VNFs to the network resources under the SFC constraints, is studied in \cite{AT.C.NFV.4} for dynamic MEC-based networks. The authors formulate the problem as an integer programming problem with the objective of long-term cost and latency minimization. Using Lyapunov optimization, they propose a dynamic topology-aware min-latency SFC migration algorithm offering a balanced cost-latency trade-off. Compared to existing SFC migration algorithms, the proposed scheme demonstrates a balance between the migration cost and the SFC latency. On the other hand, the SFC deployment is optimized in \cite{AT.C.NFV.5} for UAV edge computing networks. A heuristic two-stage SFC deployment strategy (ToRu) is designed to simultaneously maximize the revenue and minimize the task completion time. The simulation results show that ToRu presents improved overall revenue, task execution success ratio, and completion time compared to the greedy and random methods. Additionally, the SFC planning problem is formulated as a joint VNF placement and traffic routing problem in \cite{AT.C.NFV.6}. The authors employ the INLP formulation and propose a heuristic approach to solve the problem maximizing the revenue while minimizing the costs for vehicular integrated networks. They also introduce a novel metric, aggregation ratio, to capture the trade-off between communication and computing resource costs. Based on the simulations, the algorithm outperforms the benchmark methods in \cite{li2018virtual, beck2015coordinated} in terms of resources consumption and costs. Additionally, network resilience is examined in \cite{AT.C.NFV.7} where the authors study the resilience of service chains, composed of multiple VNFs, in UAV-based NFV/MEC-enabled networks by designing a quantitative modeling approach to observe the system's behavior and identify potential resilience bottlenecks.

Tables \ref{tab:NFV_AT_architectures_exp} and \ref{tab:NFV_AT_C} give a summary of research efforts on the adaptation of NFV technology in integrated Aerial-Terrestrial networks specifying the research focus, the objective, and the proposed solution.

\begin{table*}[]
\centering
\begin{tabularx}{\textwidth}{p{0.1\textwidth} p{0.0375\textwidth}
p{0.15\textwidth} p{0.16\textwidth} X p{0.2\textwidth}}
\hline
Research focus & Ref. & Network architecture & Optimization objective & Proposed solution & Evaluation metrics \\ \hline

 &
\cite{AT.C.NFV.4} &
MEC/SDN-enabled UAV networks &
Long-term cost and latency minimization  &
 Dynamic topology-aware min-latency SFC migration algorithm  &
 Latency and migration cost
 \\ \cline{2-6}

 \multirow{-2}{*}{SFC embedding}  &
\cite{AT.C.NFV.5} &
Edge computing multi-UAV networks &
Task completion time and revenue optimization  &
 Heuristic two-stage SFC deployment strategy  &
 Overall revenue, task execution time and success ratio 
 \\ \hline


VNF placement &
\cite{AT.C.NFV.6} &
HAPS-based vehicular networks&
Revenue and costs optimization  &
Heuristic VNF placement and routing algorithm   &
Aggregation ratio and resources consumption 
 \\ \hline

Network resiliency &
\cite{AT.C.NFV.7} &
NFV/MEC-enabled UAV networks&
Potential resilience bottlenecks identification  &
Quantitative modeling-based service chains analysis   &
Network resilience
 \\ \hline

\end{tabularx}
\caption{NFV-enabled integrated Aerial-Terrestrial networks: Traditional approaches.}
\label{tab:NFV_AT_C}
\end{table*}

\subsubsection{Network Slicing-based Networks} \hfill\\
Network slicing in the Aerial-Terrestrial segment is still in its infancy focusing on networks employing UAVs or drones as aerial platforms for the terrestrial network extension to provide 5G slices (eMBB, URLLC, mMTC) to end users \cite{AT.AI.NS.1,AT.C.NS.8}. Integrated UAV-Terrestrial networks can benefit from network slicing technologies to increase its reliability, enhance its security and improve its energy efficiency \cite{AT.C.NS.3}. A network slicing framework named AirSlice is proposed in \cite{AT.C.NS.1} for 5G UAV communications. Following the 3GPP standardization, AirSlice is designed to support traffic differentiation based on QoS requirements and a proof of concept implementation is validated offering URLLC services in a realistic setup. The major issues of network slicing in UAV-based networks include mainly RAN resource management, UAV deployment and slicing, which can be addressed through conventional \cite{AT.C.NS.2,AT.C.NS.5,AT.C.NS.7} or AI-based approaches \cite{AT.AI.NS.1,AT.AI.NS.2,AT.AI.NS.3,AT.AI.NS.4}.

    \textbf{\textit{a) Traditional approaches:}} To optimally customize network slices sharing the same infrastructure, the resource management problem is usually considered jointly with UAV deployment and slicing in integrated UAV-Terrestrial networks. For example, the authors of \cite{AT.C.NS.5} propose a RAN resource orchestration algorithm, the repeatedly energy-efficient and fair service coverage (RE$^2$FS) scheme, that jointly optimizes the UAV trajectory, its transmission power, and the slice access requests acceptance, to physically configure the UAV eMBB slices. Based on the successive convex approximation (SCA) method, the RE$^2$FS aims to minimize the UAVs transmit power and maximize the data rates of the payload eMBB slice ground users. Compared to benchmark schemes, the RE$^2$FS shows increased energy efficiency and user fairness. Moreover, the joint RAN resource reservation and UAV deployment is considered in \cite{AT.C.NS.7} where the UAV is optimally deployed to serve eMBB slice users and mMTC slice devices. A binary-search based RAN resource reservation and UAV deployment algorithm is proposed with the goal of BB users average rate maximization. Using an SDN/NFV architecture, the authors of \cite{AT.C.NS.2} study RAN slicing, including RAN inter-and intra-slice resource management, jointly with UAV placement and UAV-device association in multi-drone-small-cells (DSCs) networks. The integrated DSC-terrestrial network provides connectivity to two types of devices; mobile user and IoT machine-type devices having different QoS requirements. The authors design a clique-based joint UAV deployment and resource slicing algorithm that minimizes the radio resource consumption with two-level partitioning. They validate the performance of the proposed scheme in terms of cost and resource utilization. 

    \textbf{\textit{b) AI-based approaches:}} Using AI models, the authors of \cite{AT.C.NS.5} extend their work on UAV slicing in \cite{AT.AI.NS.4}, where they address the problem of inter-slice RAN resource management taking into account the URLLC and MBB slices dedicated for UAV control and UAV payload, respectively. With the objective of optimizing UAV energy consumption and service coverage fairness, they introduce an updated version of the RE$^2$FS algorithm, for RAN resource reservation, employing an echo state network (ESN) based approach and a deep neural network for user location prediction and channel estimation, respectively. Moreover, in \cite{AT.AI.NS.2, AT.AI.NS.3}, AI techniques are utilized for RAN inter-slice resource management in the aerial-terrestrial segment to achieve different objectives. On the one hand, the management and slicing of radio resources is examined in \cite{AT.AI.NS.3} for UAV-aided vehicular communications. To maximize bandwidth efficiency, the authors develop an LSTM-based resource allocation algorithm where the ML model is employed for the prediction of the vehicles and UAVs mobility. Compared to other RL-based methods, the proposed solution shows improved average bandwidth efficiency. On the other hand, the researchers in \cite{AT.AI.NS.2} consider the slicing of three types of resources, i.e. computing, networking, and storage, in a multi-dimensional manner. They investigate the scenario of autonomous vehicles supported by SDN/MEC-enabled networks, where the UAVs are required to meet the QoS of URLLC and eMBB slices for driving services and passengers eMBB services, respectively. With the goal of slice embedding energy consumption minimization, the authors propose an LSTM-based survivable resource slice embedding algorithm. Simulation results demonstrate that the proposed technique offers improved slice request acceptance and recovery ratios and reduced energy consumption compared to existing slice embedding schemes. Meanwhile, in \cite{AT.AI.NS.1}, a system controller, implemented as a VNF on the UAVs, is utilized to manage the computing resources and enable computation offloading in MEC-enabled UAV-terrestrial networks supporting 5G URLLC slices. With the goal of optimizing power consumption, delay and loss probability, a computing resource management scheme is designed leveraging the superiority of RL approaches in decision-making process. 

    Table \ref{tab:NS_AT} summarises the research efforts in network slicing based integrated aerial-terrestrial networks employing conventional and AI-based approaches.

\begin{table*}[]
\centering
\begin{tabularx}{\textwidth}{p{0.12\textwidth} p{0.0375\textwidth} p{0.23\textwidth} X p{0.22\textwidth}}
\hline
 Research focus & Ref. &  Objective & Proposed solution & Evaluation metrics \\ \hline

\multicolumn{5}{c}{Traditional approaches} \\ \hline

    &
 \cite{AT.C.NS.2} & 
  Resource consumption minimization &
  Clique-based joint UAV deployment and resource slicing algorithm  &
 Cost and resource utilization 
\\ \cline{2-5}

    \multirow{-3}{*}{\begin{tabular}[c]{@{}l@{}} Joint RAN resource \\ management and \\ UAV deployment \end{tabular}} &
\cite{AT.C.NS.7} & 
  Users average rate maximization &
   Binary-search based RAN resource reservation and UAV deployment algorithm &
   Average rate increase 
   \\ \hline

 RAN resource management & 
 \cite{AT.C.NS.5}&
  Optimization of UAV transmit power and UE data rate  &
   Repeatedly energy-efficient and fair service coverage resource  orchestration scheme based on SCA  &
   Jain’s fairness index and energy efficiency
   \\ \hline

   \multicolumn{5}{c}{AI-based approaches} \\ \hline

   Traffic scheduling and offloading &
 \cite{AT.AI.NS.1} &
  Optimization of power consumption, delay and loss probability  &
   RL-based computing resource management scheme &
   Delay, loss probability and power consumption \\ \hline 

   &
  \cite{AT.AI.NS.2} & 
  Slice embedding energy consumption minimization  &
   LSTM-based survivable resource slice embedding algorithm  &
   Slice request acceptance and recovery ratios, energy consumption 
   \\ \cline{2-5} 

  \multirow{-2}{*}{\begin{tabular}[c]{@{}l@{}} RAN resource \\ management \end{tabular}} &
  \cite{AT.AI.NS.3}& 
  Bandwidth efficiency maximization  &
   LSTM-based resource allocation algorithm   &
   Average bandwidth efficiency
   \\ \hline

  UAV slicing & 
\cite{AT.AI.NS.4} &
  Optimization of energy consumption and service coverage fairness  &
   Repeatedly energy-efficient and fair service coverage resource reservation scheme based on ESN and DNN  &
   Jain’s fairness index and energy efficiency 
   \\ \hline
\end{tabularx}

\caption{Network Slicing-based integrated Aerial-Terrestrial networks: Traditional and AI-based approaches.}
\label{tab:NS_AT}
\end{table*}

\subsection{Virtualization in the Satellite-Aerial-Terrestrial Segment}

\subsubsection{SDN-enabled Networks} \hfill\\
The data/control planes separation provided by the SDN paradigm facilitates the integration of the satellite, aerial and terrestrial segments producing a three-layered network architecture as introduced in \cite{SAT.C.SDN.1,SAT.C.SDN.3,SAT.AI.SDN.3}. Nonetheless, the large-scale, dynamic and heterogeneous characteristics of these networks result into more complex SDN-related problems compared to terrestrial and other integrated networks. Few works have been reported in the literature addressing such issues including the controller placement problem, the routing optimization and the resource management in the S-A-T segment employing conventional methods \cite{SAT.C.SDN.6,SAT.C.SDN.2} and AI-based techniques \cite{SAT.AI.SDN.1,SAT.AI.SDN.2,SAT.AI.SDN.4}.

\textbf{\textit{a) Architectures and experimental implementations:}} In \cite{SAT.C.SDN.3}, the authors propose a hybrid SDN-based architecture for QoS and security-aware routing, where both SDN and traditional network protocols are adopted. The control structure is composed of hierarchical multi-controllers deployed at different segments including the ground station, and the satellite and aerial platforms. Using this controller configuration, they introduce the routing service composition layer which composes end-to-end paths with the aim of route reliability maximization while satisfying QoS and security requirements. They also validate their proposed architecture through the use case scenario of Vehicle-to-Everything communication. Vehicular communications is also considered in the design of an SDN-based SAGIN architecture in \cite{SAT.C.SDN.8_B.2.2}. Using hierarchical multi-controllers configuration, the proposed framework adopts SDN and network slicing technologies to support both vehicular and legacy services in isolated network slices. Another multi-layered SDN-based architecture is presented in \cite{SAT.C.SDN.1} following a cross-domain design. Employing also hierarchical multi-controllers, the main controller at the ground station performs cross-domain orchestration to improve configuration updating and network efficiency. Meanwhile, researchers in \cite{SAT.AI.SDN.3} adopt not only SDN paradigm, but also AI and MEC technologies to build an integrated aeronautical federation framework. Deploying the SDN controller on HAPS, the proposed framework enables aeronautical applications such as aeronautical edge computing and aircraft in-cabin connectivity and sensing. In \cite{SAT.C.SDN.7}, the authors propose a software-defined space-air-ground integrated moving cells (SAGECELL) framework for ultra-dense networks supporting multiple applications. The architecture is validated through a case study of eMBB services and simulation results show improved throughput performance.

\begin{table*}
\centering
\begin{tabularx}{\textwidth}{ p{0.0375\textwidth} p{0.15\textwidth} p{0.14\textwidth} p{0.14\textwidth} X }
\hline 
Ref.                 & Controller placement & Use case scenario & Implementation Tools & Comments \\ \hline

\cite{SAT.C.SDN.3} &
Ground station, satellite and aerial platforms &
 Vehicle-to-Everything communication &
N/A &
Propose a hybrid SDN-based architecture with hierarchical multi-controllers for QoS and security-aware routing
 \\ \hline

\cite{SAT.C.SDN.8_B.2.2} &
Ground station, satellite and aerial platforms &
Vehicular communications &
N/A &
Design a SAGIN architecture based on SDN and network slicing technologies for vehicular communications
 \\ \hline

\cite{SAT.C.SDN.1} &
Ground station, GEO satellite, and HAPS &
N/A &
OpenFlow protocol &
Propose a cross-domain SDN-based architecture with hierarchical multi-controllers to improve configuration updating
\\ \hline

\cite{SAT.AI.SDN.3} &
 HAPS &
 Aeronautical applications &
 STK toolkit &
Propose AI/SDN-based integrated aeronautical federation framework to enable aeronautical applications
 \\ \hline

\cite{SAT.C.SDN.7} &
Ground station, satellite and aerial platforms & 
Moving cells for ultra-dense networks &
N/A &
Propose a software-defined space-air-ground integrated moving cells framework for ultra-dense networks supporting multiple applications
 \\ \hline

\end{tabularx}
\caption{SDN-enabled integrated Satellite-Aerial-Terrestrial networks: Architectures and experimental implementations.}
\label{tab:SDN_SAT_architectures_exp}
\end{table*}

 \textbf{\textit{b) Traditional approaches:}} Routing optimization is one the major challenges which becomes more complicated in SDN-enabled integrated S-A-T networks. The issue is studied in \cite{SAT.C.SDN.6} and \cite{SAT.C.SDN.2} using a single controller configuration. On the one hand, an intelligent flow forwarding scheme combining multi-path routing and multi-protocol mechanism is proposed in \cite{SAT.C.SDN.6}. With the goal of path reliability maximization, the algorithm offers enhanced resilience and throughput, endogenous security, and reduced delay, compared to conventional routing strategies. On the other hand, the authors of \cite{SAT.C.SDN.2} design a dynamic transmission control technique for SDN-enabled S-A-T networks. The proposed method is based on queueing game theory with the objective of system social welfare maximization, and it presents improved performance in terms of throughput and service value delay as shown by the simulations.

  \textbf{\textit{c) AI-based approaches:}} Since SDN-related issues including controller placement and resource management become more complicated in the Satellite-Aerial-Terrestrial segment, solutions based on conventional techniques become inefficient. Hence, researchers turn to AI models which proved its superiority in solving complex problems. First, a controller deployment scheme with hierarchical multi-controllers structure is designed in \cite{SAT.AI.SDN.4}. Adopting K-means clustering, the authors divide the network into multiple sub-networks, having each a local secondary controller. Then, they formulate the multi-objective optimization of delay and controller load balance and solve it using the Genetic algorithm to determine the optimal controllers deployment scheme. Second, the resource management in SDN-enabled integrated S-A-T networks is investigated in \cite{SAT.AI.SDN.2}. Leveraging the hierarchical multi-controllers configuration, a distributed hierarchical hybrid DRL algorithm is proposed where the RL agents are deployed on the controllers. The DRL-based resource allocation scheme is designed with the goal of the joint optimization of user request acceptance rate and long-term revenue rate. The proposed solution outperforms the conventional and centralized approaches in terms of average revenue and service success rate. Third, the problem of traffic scheduling in the S-A-T segment is examined in \cite{SAT.AI.SDN.1}. With the objective of flow maximization, the authors develop a Q-learning based traffic scheduling algorithm for single controller SAGIN. The RL-based strategy is used to optimized the scheduling decision-making process and the results demonstrate its superiority over existing algorithms in terms of load balancing and network capacity utilization.

Table \ref{tab:SDN_SAT_architectures_exp} and \ref{tab:SDN_SAT} summarize the contributions examining architectural considerations and addressing several challenges in SDN-enabled integrated S-A-T networks.

\begin{table*}[]
\centering
\begin{tabularx}{\textwidth}{p{0.07\textwidth} p{0.0375\textwidth}
p{0.11\textwidth} p{0.08\textwidth} p{0.16\textwidth}  X p{0.17\textwidth}}
\hline
  Research focus &
  Ref. &
  Controller configuration &
  Type of problem &
  Objective &
  Proposed solution &
  Evaluation metrics \\ \hline
  \multicolumn{7}{c}{Traditional approaches} \\ \hline
  
  & \cite{SAT.C.SDN.6} &
  Single controller &
  Multi-path routing &
  Path reliability maximization &
  Intelligent multi-path multi-protocol hybrid flow forwarding scheme &
   Average path reliability, packet delay, throughput
  \\ \cline{2-7}

 \multirow{-2}{*}{\begin{tabular}[c]{@{}l@{}}Routing \\ optimization \end{tabular}} & 
 \cite{SAT.C.SDN.2} &
 Single controller &
 Single-path routing & 
 System social welfare maximization &
   Queueing game-based dynamic transmission control technique &
   Throughput and service value delay 
   \\ \hline
   
\multicolumn{7}{c}{AI-based approaches} \\ \hline

  Controller placement  &
  \cite{SAT.AI.SDN.4} &
   Hierarchical multi-controllers &
   Dynamic CPP &
   Optimization of delay and controller load &
  K-means based controller deployment scheme &
   Latency and load balancing
   \\ \hline

  Resource management & 
 \cite{SAT.AI.SDN.2} &
   Hierarchical multi-controllers &
  Data plane resources &
   Joint optimization of user request acceptance rate and long-term revenue &
   Hierarchical hybrid DRL algorithm &
   Average revenue and service success rate 
  \\ \hline

 Traffic scheduling & 
 \cite{SAT.AI.SDN.1} &
   Single controller &
   N/A &
   Flow maximization &
   Q-learning based SAGIN traffic scheduling algorithm &
  Load balancing and network capacity utilization 
   \\ \hline
   
\end{tabularx}
\caption{SDN-enabled integrated Satellite-Aerial-Terrestrial networks: Traditional and AI-based approaches.}
\label{tab:SDN_SAT}
\end{table*}

\subsubsection{NFV-enabled Networks} \hfill\\
Because of the unique characteristics of these next-generation networks, the adoption of NFV technology in integrated satellite-aerial-terrestrial networks is still in its early stages. Only a handful of studies have been dedicated to the subject matter employing traditional approaches and focusing on two major NFV challenges, including the VNF placement \cite{SAT.C.NFV.2,SAT.C.NFV.3} and the SFC deployment \cite{SAT.C.NFV.1,SAT.C.NFV.4,SAT.C.NFV.5}.

\textbf{\textit{Traditional approaches:}} The VNF placement problem is studied in \cite{SAT.C.NFV.2} and \cite{SAT.C.NFV.3} with different optimization objectives. On the one hand, the authors of \cite{SAT.C.NFV.2} aim to maximize the total profit of the service provider. Taking into account the delay and cost of VNF migration, they jointly formulate the VNF placement and the VNF scheduling problems as a MILP problem and propose two dynamic Tabu search-based VNF mapping and scheduling schemes. The proposed dynamic algorithms (TS-PSCH and TS-MAPSCH) outperform the static and greedy approaches in terms of service acceptance ratio, total profit, and QoS satisfaction for Internet of Vehicles services. On the other hand, the resource utilization is maximized in \cite{SAT.C.NFV.3} while meeting the SFC requests delay constraint. A resource-efficient and delay-aware VNF placement scheme is designed based on Linear Programming and graph matching theory. The simulation results show that the proposed solution demonstrates improved resource utilization efficiency and reduced running time compared to benchmark algorithms.
    
The SFC deployment issue is investigated in \cite{SAT.C.NFV.1,SAT.C.NFV.4,SAT.C.NFV.5} with the goal of minimizing the deployment delay and maximizing the number of completed tasks. In \cite{SAT.C.NFV.5}, the deployment delay is minimized taking into account the constraints on the resources utilization and a delay-aware SFC mapping scheme is designed based on the famous k-shortest path algorithm for delay-sensitive applications. The results show that the proposed SFC deployment algorithm outperforms other techniques with respect to deployment delay, resource consumption, and service acceptance rate. Meanwhile, the authors of \cite{SAT.C.NFV.4} aim to maximize the number of completed tasks while satisfying the deployment, flow and resources constraints. Employing the reconfigurable time expansion graph representation and the MILP formulation, they design an SFC deployment algorithm based on matching game theory. Compared to existing methods, the developed algorithm presents increased number of completed tasks and improved resource utilization efficiency. Moreover, the number of completed missions is jointly optimized with the cost of computing and bandwidth resources in \cite{SAT.C.NFV.1}. The authors propose a bidirectional mission offloading (BDO) framework based on NFV to enhance the flexibility and agility of SAGIN. To validate their framework, they design an SFC embedding scheme for computation-intensive and delay-sensitive applications. The designed BDO-based SFC deployment strategy shows enhanced reliability and resource utilization efficiency, based on simulation results.

Table \ref{tab:NFV_SAT} summarizes the research efforts on NFV-enabled integrated S-A-T networks highlighting the objective and proposed solution of each work.

\begin{table*}[]
\centering
\begin{tabularx}{\textwidth}{p{0.1\textwidth} p{0.0375\textwidth}
p{0.131\textwidth} p{0.165\textwidth} X p{0.19\textwidth}}
\hline
Research focus & Ref. & Network architecture & Objective & Proposed solution & Evaluation metrics \\ \hline

&
\cite{SAT.C.NFV.2} &
SDN/NFV-enabled vehicular networks &
Total profit maximization &
 Two dynamic Tabu search-based VNF mapping and
scheduling algorithms &
Service acceptance ratio, total profit, and QoS satisfaction
 \\ \cline{2-6}

 \multirow{-2}{*}{VNF placement} &
\cite{SAT.C.NFV.3} &
NFV-enabled networks &
Resource utilization maximization &
 Resource-efficient and delay-aware VNF placement scheme  &
 Resource utilization efficiency and running time
 \\ \hline

   &
\cite{SAT.C.NFV.1} &
SDN/NFV-enabled networks &
Optimization of resources cost and number of completed missions &
SFC embedding based on bidirectional mission offloading framework  &
 Reliability and resource utilization efficiency 
 \\ \cline{2-6}

 \multirow{-3}{*}{SFC embedding } &
\cite{SAT.C.NFV.4} &
NFV-enabled networks &
Number of completed tasks maximization  &
Matching game-based SFC deployment algorithm  &
Completed tasks number and resource utilization efficiency 
 \\ \cline{2-6}

&
\cite{SAT.C.NFV.5} &
NFV-enabled delay-sensitive networks  &
 Delay minimization  &
 Delay-aware SFC mapping scheme &
 Delay, resource consumption, and service acceptance rate
 \\ \hline

\end{tabularx}
\caption{NFV-enabled integrated Satellite-Aerial-Terrestrial networks: Traditional approaches.}
\label{tab:NFV_SAT}
\end{table*}

\subsubsection{Network Slicing-based Networks} \hfill\\
Due to the dynamic, large-scale and heterogeneous nature of integrated S-A-T networks, employing network slicing paradigm to improve resource management efficiency and overall network performance is a challenging task \cite{SAT.C.NS.2, SAT.AI.NS.2, SAT.AI.NS.4, SAT.AI.NS.5}. The research works in this area is limited with only a handful of studies that can be found in the literature investigating mainly resource slicing and management issues adopting traditional \cite{SAT.C.NS.1, SAT.C.NS.3} and AI-based methods \cite{SAT.AI.NS.3, SAT.AI.NS.6}.

    \textbf{\textit{a) Traditional approaches:}} Dynamic RAN resource management is examined in \cite{SAT.C.NS.1} and \cite{SAT.C.NS.3} targeting different objectives while considering different use case scenarios. On the one hand, the authors of \cite{SAT.C.NS.1} focus on user association jointly with intra-slice RAN resource allocation for edge computing and SDN-based networks with the goal of maximizing the aggregate transport capacity capturing the overall network performance. To adapt to its dynamics, they study the scaling law describing the behavior of the network performance when increasing its size, under SISO, MISO, and MIMO channels. Then, they design a dynamic resource orchestration and user selection algorithm based on the derived scale laws. On the other hand, joint spectrum resource reservation and UAV deployment is considered in the context of SAGIN vehicular communications in \cite{SAT.C.NS.3}. A service-aware dynamic resource slicing scheme based on Lyapunov optimization is proposed with the objective of long-term revenue and system stability maximization. The algorithm carries out the service request admission and scheduling, the UAV deployment as well as the resource slicing to serve the different network slices. The results show that it outperforms benchmark schemes in terms of time-averaged throughput and queue size.

    \textbf{\textit{b) AI-based approaches:}} The inter-slice RAN resource management problem is solved using AI techniques in \cite{SAT.AI.NS.3} and \cite{SAT.AI.NS.6} with the objective of network utility maximization, and joint throughput, service delay and coverage area optimization, respectively. In \cite{SAT.AI.NS.3}, a distributed dynamic resource slicing (DDRS) scheme is proposed to reserve the processing and transmission resource to the different network slices. The DDRS combines a message passing graph neural network MPNN-based DL model and an online ADMM decomposition technique to obtain optimal resource slicing in SAGIN dynamic environment. Compared to existing dynamic slicing algorithms, the DDRS presents improved user service completion time, network utility, and reliability. Meanwhile, the authors of \cite{SAT.AI.NS.6} develop a dynamic RAN slicing algorithm that can conduct not only dynamic inter- and intra-slice power resources allocation, but also dynamic user association and optimal virtual UAV positioning. To achieve the Pareto optimality of the formulated multi-objective optimization problem, their proposed algorithm is based on a joint central and distributed multi-agent deep deterministic policy gradient (MADDPG) approach. Compared to benchmarks, the proposed solution show increased throughput and SINR as well as reduced average delay. Furthermore, network security and resiliency is studied in \cite{SAT.AI.NS.2, SAT.AI.NS.5} where while the authors of \cite{SAT.AI.NS.2} examine the role of deep learning in privacy preservation of sliced integrated networks, the researchers in \cite{SAT.AI.NS.5} look into the resilience of network slicing in the S-A-T segment and propose a resilient multi-domain slicing framework for S-A-T Edge Computing IoT networks.
    
Table \ref{tab:NS_SAT} gives a summary on the works dedicated for network slicing-based integrated S-A-T networks, indicating the research focus and the proposed traditional or AI-based solution.

\begin{table*}[]
\centering
\begin{tabularx}{\textwidth}{p{0.13\textwidth} p{0.0375\textwidth} p{0.2\textwidth} X p{0.2\textwidth}}
\hline
 {Research focus} &
  Ref. &
  {Objective} &
  {Proposed solution} &
  {Evaluation metrics} \\ \hline

  \multicolumn{5}{c}{Traditional approaches} \\ \hline

  Device/user association & 
   \cite{SAT.C.NS.1} &
   Aggregate transport capacity maximization  &
    Dynamic resource orchestration and user selection algorithm based on scaling laws &
    Aggregate transport capacity
  \\ \hline
  
 Joint RAN resource management and UAV deployment &  
 \cite{SAT.C.NS.3} &
 Long-term revenue and system  stability maximization &
    Service-aware dynamic resource slicing scheme &
    Time-averaged throughput and queue size 
   \\ \hline

\multicolumn{5}{c}{AI-based approaches} \\ \hline
 
   RAN resource management & 
   \cite{SAT.AI.NS.3} &
   Network utility maximization &
   Distributed dynamic resource slicing scheme based on MPNN-based DL and online ADMM methods  &
    User service completion time, network utility, and reliability 
   \\ \hline
  
   Joint RAN resource management and UAV deployment &  
   \cite{SAT.AI.NS.6} &
    Joint optimization of throughput, service delay and coverage area  &
    Dynamic RAN slicing and UAV deployment algorithm based on a joint central and distributed MADDPG approach  &
    Throughput, average delay and SINR 
   \\ \hline

   Network resiliency &
   \cite{SAT.AI.NS.5} &
   Network failure mitigation &
   Resilient multi-domain network slicing framework &
   Network resilience
   \\ \hline
   
\end{tabularx}
\caption{Network Slicing-based integrated Satellite-Aerial-Terrestrial networks: Traditional and AI-based approaches.}
\label{tab:NS_SAT}
\end{table*}

\subsection{Summary $\&$ Learnt Lessons}

In this section, we provided a taxonomy of integrated TN-NTNs virtualization where we categorized the relevant contributions reported in the literature using a four-level classification. From the prospective of the degree of TN-NTN integration, varying degrees of research interest are shown for the implementation of virtualization technologies in the three integrated TN-NTN segments. On the one hand, the satellite-terrestrial segment receives significant attention, compared to the other two segments, owing to the satellites large coverage area, their broadcast/multicast capabilities, and the recent technological advancements in the satellite industry. On the other hand, documented studies investigating virtualized integrated aerial-terrestrial networks are comparatively fewer, primarily due to the low reliability and limited capacity of the UAVs and the immaturity of HAPS technology. In particular, while the virtualization of HAPS-based networks is still in its infancy due to the nascent state of the associated technologies, there is an emphasis on the adaptation of virtualization techniques in UAV-assisted networks because of their flexible deployment and diverse application scenarios. Meanwhile, incorporating network virtualization enablers in the Satellite-Aerial-Terrestrial segment is still in its early stages. This is because the related problems are significantly more complicated compared to terrestrial and other integrated networks, since their complexity increases as the networks become larger and more dynamic. In addition, the deployment of UAVs is typically taken into account in conjunction with other issues in integrated A-T and S-A-T networks, which further complicates the implementation of virtualization approaches in these segments.

In terms of the application of virtualization technologies, researchers largely focus on SDN-enabled networks in the three segments. They typically investigate key architectural considerations by proposing multiple designs of integrated networks based the concepts of the SDN paradigm. Following the SDN data/control planes separation and centralization of the network logic, they examine various use case scenarios, consider different controller structures, and provide several experimental implementations. They, also attempt to solve a number of issues that have emerged as a result of the introduction of SDN in next-generation networks. The controller placement problem, the routing optimization, and the satellite handover management are the main problems that has been studied in SDN-enabled satellite-terrestrial networks. Conversely, in SDN-based aerial-terrestrial networks, resource management, traffic offloading, and routing optimization are primarily researched, whereas the controller placement problem is seldom examined. On the other hand, contributions employing SDN in the satellite-aerial-terrestrial segment are limited and mostly concentrate on architectural perspectives. 

Compared to SDN-enabled networks, the research on the adaptation of NFV is restricted in the three network segments. However, unlike the works on SDN-based networks, the studied NFV-enabled integrated networks can be based on architectures where both SDN and NFV paradigms are adopted. The core challenges that have been addressed include VNF placement, SFC embedding, and virtual resource management. Additionally, few insights on architectures and experimental implementations are provided in the Aerial-Terrestrial segment. As for the adaptation of network slicing, the contributions are considerably scarce compared to SDN and NFV. Taking into account the three segments, the primary issue that has been tackled is RAN resource management. Device/user association, traffic scheduling, and offloading are also considered. Besides, UAV slicing is explored as a special case in network slicing-based integrated Aerial-Terrestrial networks.

To solve the various problems associated with the introduction of SDN, NFV and network slicing in integrated TN-NTNs, researchers tend to utilise traditional methods mainly based on conventional optimization techniques. Nonetheless, a few efforts have been dedicated to the application of AI algorithms, especially in SDN and network slicing-based networks, whereas AI-powered solutions in NFV-enabled networks are lacking. 

\section{Simulation and Implementation Tools}\label{Sim_tools}

Simulation plays a crucial role in developing and implementing virtualization technologies in 6G integrated terrestrial and non-terrestrial networks. There are several reasons why simulation is essential. First, the integrated TN-NTNs are highly complex. Simulation provides a controlled environment for modeling and understanding the interactions and behaviors of these complex networks. Second, most researchers cannot access or afford the high expense of deploying physical testbeds for integrated TN-NTNs. Developing and testing virtualized networks in a simulated environment is cost-effective without affecting any live network. Furthermore, proposing new algorithms and architectures for integrated TN-NTNs often requires multiple explorations on various deployment scenarios, including combinations of vehicles or elements and network configurations. Simulation allows for exploring different settings, which is essential for understanding the trade-offs and performance implications. It is necessary for validating and refining algorithms before actual deployment. Moreover, the simulated environment enables researchers to conduct scalability tests, assess the designed network's fault tolerance and resilience and evaluate system latency, throughput, and reliability in diverse scenarios. In this section, we introduce the simulation and implementation tools that has been used in the literature. The tools are categorized into network simulators and network virtualization tools. Additionally, we highlight a few testbeds that emulate integrated TN-NTNs.

\subsection{Network Simulators}

Multiple tools are built for network simulation and can be classified into two classes; generalized simulators including MATLAB, NS-3 and its extensions, EXata, and OMNeT++, and satellite-oriented simulators including Systems Tool Kit (STK) and OpenSAND.

\begin{itemize} 
\item \textbf{MATLAB} - MATLAB is a well-known and powerful commercial tool for simulation and modeling in many scientific and engineering fields. It is one of the most popular simulation tools utilized by the researchers in integrated TN-NTNs virtualization. Several essential toolboxes contribute to this field. Simulink is a graphical simulation environment for modeling the entire communication systems, including network architecture, base stations, user equipment, and satellite components. 5G Toolbox and LTE Toolbox support mobile network systems such as waveforms, channel models, and link-level simulations. Satellite Communications Toolbox provides functions for modeling and visualizing satellite communications, and performing link analysis and calculations. Other traditional toolboxes, such as the Communications Toolbox, Antenna Toolbox, and Phased Array System Toolbox provide most essential functionalities to simulate wireless communication environments. Furthermore, MATLAB is well utilized in channel modeling, propagation modeling, channel coding, and modulation at the physical level of simulation. Integrating MATLAB with other tools provides a comprehensive environment for research on integrated TN-NTNs. MATLAB has often been jointly used with Systems Tool Kit to simulate satellite and integrated TN-NTN systems. Several advanced features support integration. MATLAB can be used to automate and script actions in STK. The data generated from STK could be imported to MATLAB for further analysis, and MATLAB-generated data could be exported to STK for visualization in dynamic scenarios. MATLAB’s scripting capabilities can be used to automate Monte Carlo simulations in STK, while the Optimization Toolbox can to optimize the parameters.

\item \textbf{NS-3} - NS-3 is an extensible, open-source, discrete-event network simulation platform that is designed for research and educational purposes. NS-3 provides a detailed and comprehensive simulation environment. It has good scalability to model realistic network sizes and conditions. In the context of virtualization, NS-3 can simulate networks that are based on SDN, NFV, and network slicing, while supporting a wide range of networking scenarios, such as cellular, ad-hoc, mesh, and IoT networks. 

\item \textbf{NS-3 extensions for integrated TN-NTNs} - Extensions for NS-3 are developed to support non-terrestrial networks including satellite communications. For example, NS-3 Satellite Mobility Model is designed based on the North American Aerospace Defense Command’s SGP4/SDP4 models to predict near-space and deep-space satellite orbits \cite{Sim.SMM_ns3}. The authors in \cite{Sim.ns3leo} present NS-3-leo for modeling the LEO satellite’s mobility and link characteristics for mega-constellations. In addition, the Satellite Network Simulator 3 (SNS3) is a satellite network extension to NS-3 developed by Magister Solutions \cite{Sim.SNS3}. It simulates a geostationary satellite and transparent star payload in a fully dynamic multi-spot beam satellite network. The reference satellite system is made up of five gateways, 72 spot-beams covering Europe, and Ka-band frequencies. The DVB-RCS2 and DVB-S2 standards are implemented. A newer version of system level simulator that extend NS-3 to support 5G NTN is introduced by the same company \cite{Sim.5G_NTN_SLS}. Another system-level simulator is proposed to integrate TN-NTNs by extending NS-3 \cite{ Sim.NS3_5G_sim}. It supports simulating the NTN handover decisions, selecting bandwidth parts, and configuring component carriers.

\item \textbf{EXata} - EXata is a commercial network modeling and emulation tool published by Keysight (previously developed by Scalable Network Technologies, Inc.) for network digital twins, allowing users to replicate network behavior in simulation and emulation \cite{Sim.EXata}. The Joint Network Emulator library enables the waveforms and data link for satellite communication networks.

\item \textbf{OMNeT++} -
OMNeT++ is a C++-based discrete event network simulator. It is free for non-commercial usage. It has been used for simulating queuing models, internet protocols, wireless networks, ad-hoc networks, vehicular networks, and so on. It provides a simulation IDE for developing and evaluating models. 
\item \textbf{Systems Tool Kit (STK)} - The Ansys Systems Tool Kit (STK) \cite{Sim.STK} (previously Satellite Tool Kit) is a paid software application designed by Analytical Graphics, Inc. for modeling, analyzing, and visualizing satellite systems and their interactions with the Earth. It is widely used in the research discussed in this survey for simulating satellite trajectory and assessing communication links for integrated TN-NTNs. The 3D graphical user interface provides an easy-to-use feature for researchers to create realistic real-time simulation scenarios. The radio frequency propagation models and realistic satellite trajectories create a reliable modeling environment for designing software-defined strategies. Moreover, the flexible development options support STK to work with other programming languages and frameworks such as Python, .NET, and C/C++. We found that STK has been highly used with MATLAB and other tools in the literature. 
For example, in \cite{NTN.C.SDN.5.16}, the authors built the MIRSAT testbed with a substrate network emulated by Mininet, having OpenFlow switches, with RYU SDN controller. They use MATLAB to execute algorithms and integrate it with STK to simulate satellite scenario. The substrate network topology generated in the STK is translated into Mininet for evaluation. Additionally, in \cite{NTN.C.SDN.2.4}, the experiments are implemented in MATLAB, with STK for creating LEO satellite network scenarios and adopting EXata for simulations.

\item \textbf{OpenSAND} - OpenSAND is an open source satellite communication emulation platform developed by Thales Alenia Space and promoted by the French Space Center \cite{Sim.OpenSAND}. Focusing on the DVB-RCS standard, OpenSAND allows users to emulate communication scenarios involving satellite links, ground stations, and numerous network protocols. It facilitates not only the performance evaluation of such systems but also their experimentation and validation by enabling efficient and user-friendly demonstration tools.

\end{itemize}

\subsection{Network Virtualization Tools}

The major tools used in the simulation and implementation of virtualization-based networks are the Mininet environment, the OpenFlow protocol, the different SDN controllers, as well as the Open vSwitch and OpenStack platforms.

\begin{itemize}

\item \textbf{Mininet} - Mininet is the popular network emulator testbed, especially for SDN and NFV research. It provides a flexible and cost-effective way to create and experiment with complex network scenarios in a controlled and reproducible environment. Researchers create the entire virtual SDN topology, including routers, switches, hosts, and links on a single laptop or computer.  It has been widely used for testing SDN applications in integrated TN-NTNs scenarios. Mininet supports the SDN protocols and languages such as OpenFlow and P4. Also, Mininet is versatile enough to emulate NFV and Network slicing by creating multiple virtual networks within a Mininet instance, with each slice having specific characteristics and requirements.

\item \textbf{OpenFlow} - OpenFlow is a well-known communication protocol that enables the SDN paradigm \cite{mckeown2008openflow}. Being a southbound APIs, OpenFlow promotes the data/control planes decoupling and allows the SDN controller to communicate with the forwarding devices, including switches and routers. The protocol is based on the concept of flow tables implemented in data plane components where the flow entries are used to define how incoming packets should be processed. Thus, the SDN controller, acting as the network brain, utilizes OpenFlow to communicate instructions to the forwarding devices, and configure their flow tables according to network policies and conditions. In integrated TN-NTNs, researchers attempt to extend the OpenFlow in order to cope with the unique characteristics of NTN platforms \cite{NTN.C.Comb.10_B.2.3,NTN.C.SDN.1.4}.

\item \textbf{SDN Controllers} - The SDN controller is the central component in SDN-enabled networks offering a centralized intelligence and a global view of the network. Multiple controllers have been developed for terrestrial networks which are also adopted by researchers in their designs of SDN-enabled integrated TN-NTNs. The most frequently used controllers include \textit{OpenDayLight} \cite{Sim.opendaylight}, \textit{Ryu} \cite{Sim.Ryu}, \textit{POX} \cite{Sim.POX}, and \textit{Open Network Operating System (ONOS)} \cite{Sim.ONOS}, and they are open-source projects.

\item \textbf{Open vSwitch} - Open vSwitch, also known as OVS, is an open-source implementation of multilayer virtual network switches, designed to enable virtualized environments, such as SDN/NFV-enabled networks \cite{Sim.openvswitch}. Running on virtual machines and hypervisors, Open vSwitch enables network virtualization through its abstraction of the server architecture and efficient network automation via its programmatic extensions. It also supports various standard protocols and interfaces.

\item \textbf{OpenStack} - OpenStack is an open-source cloud computing platform based on the infrastructure-as-a-service (IaaS) model \cite{Sim.openstack}. The cloud system provides and controls pools of resources in a data center including computing, storage, and networking resources. In the context of NFV-enabled networks, OpenStack is used to facilitate the virtualization of network resources as well as the deployment, orchestration and management of virtual network functions. It also aids network operators in the scaling and management of network services in an efficient and optimized manner through its modular flexible cloud infrastructure.

\end{itemize}

Furthermore, we note that although the aforementioned tools mainly focus on SDN and NFV paradigms, they can also support and enable network slicing since SDN and NFV are key enabling technologies of network slicing.

\subsection{Testbeds and Prototypes}
Despite the difficulty of real-world implementation of testbeds and prototypes in integrated TN-NTNs, few field trials have been documented in the literature that could serve as valuable examples for researchers. For instance, the authors of \cite{AT.C.SDN.1.3} implement SkyCore by building physical prototypes using two UAVs as base stations to demonstrate an Edge-EPC architecture in an SDN-enabled Aerial-Terrestrial network. The SkyCore software switch is built on top of Open vSwitch and it is customized to support the designed flow tables and switch actions. Another variant of SkyCore switch is proposed, operating in the user space and built on top of Lagopus software switch \cite{Sim.Lagopus}. The Lagopus switch and router complied with all the requirements in OpenFlow 1.3. It leverages the open-sourced Data Plane Development Kit \cite{Sim.DPDK} for high-performance packet processing in SDN. The SkyCore prototype is deployed on two DJI Matrice 600 Pro drones. One drone carries a commercial LTE small cell with support LTE UEs, and another drone carries a 4-core server with 8GB RAM and 1.9GHz CPU that executes SkyCore. Moreover, in \cite{NTN.C.SDN.5.7}, a hardware LTE emulator, Amari OTS 100, is employed in an SDN-enabled hybrid satellite-terrestrial network testbed. The machine is developed by Amarisoft, which is a hardware system that emulates a functional private LTE network. The private network includes two PCs, one for the eNB function and another for the EPC function. Commercial smartphones are used to connect this experimental private LTE network. Two OpenFlow switches are used to emulate the satellite and terrestrial paths, while OpenSAND is employed for the implementation of satellite link emulator and the Ryu controller is used in the control plane. Furthermore,
researchers in \cite{NTN.C.SDN.5.8} utilised the Linktropy mini2, developed by Apposite Technologies, to emulate the satellite channel and the open-source Trema framework to control the OpenFlow SDN switches, in their testbed.

In brief, simulation is essential for researchers to develop 6G integrated TN-NTNs technologies. It provides a versatile and controlled environment for testing, validating, and optimizing network resources and components, ensuring the future successful deployment of next-generation networks. We summarize the tools that have been utilized in the literature in Table \ref{tab:Tools_Sum_Up}. Lastly, we refer the reader to reference \cite{Sim.ST.SimSurvey} where the authors provide a comprehensive survey of network simulators for satellite-terrestrial integrated networks.

\begin{table}[]
\centering
\begin{tabularx}{0.49\textwidth}{p{0.09\textwidth} p{0.08\textwidth} X}
\hline
\multicolumn{2}{l}{Simulation Tools}            & Relevant Contributions \\ \hline

\multirow{6}{*}{\begin{tabular}{lll} Network \\ Simulators \end{tabular}}  
& MATLAB  & \cite{NTN.C.SDN.2.3, NTN.C.SDN.2.4, NTN.C.SDN.2.7, NTN.C.SDN.2.8, NTN.C.SDN.2.10, NTN.C.SDN.2.12, NTN.C.SDN.2.13, NTN.C.SDN.3.5, NTN.C.SDN.3.9, NTN.C.SDN.5.2, NTN.C.SDN.5.13, NTN.C.SDN.5.16, NTN.C.NFV.3(1), NTN.C.NFV.5, NTN.C.NFV.18, NTN.C.NFV.19, NTN.C.NS.4, NTN.C.NS.5,NTN.AI.SDN.2,NTN.AI.NS.4, AT.C.SDN.1.8,AT.C.SDN.2.4,AT.C.SDN.2.5,AT.C.SDN.3.2,AT.C.SDN.5.2,AT.C.SDN.6.2, SAT.C.NFV.3,SAT.C.NFV.4,SAT.AI.SDN.4}           \\ \cline{2-3} 
& NS-3 &  \cite{NTN.C.SDN.3.3,AT.C.SDN.1.5,AT.C.SDN.2.3,AT.C.SDN.5.2,AT.C.NFV.9,SAT.AI.SDN.1,NTN.C.SDN.1.4, NTN.C.SDN.2.6}          \\ \cline{2-3} 
& EXata  &  \cite{NTN.C.SDN.2.4, NTN.C.SDN.2.13, NTN.C.SDN.3.2,NTN.AI.SDN.6,	AT.C.SDN.2.6}          \\ \cline{2-3}
& OMNeT++  & \cite{AT.C.SDN.1.4,AT.C.SDN.1.13,SAT.C.SDN.2}           \\ \cline{2-3}
& STK  &   \cite{NTN.C.SDN.2.3, NTN.C.SDN.2.4, NTN.C.SDN.2.10, NTN.C.SDN.2.12, NTN.C.SDN.3.2, NTN.C.SDN.3.5, NTN.C.SDN.4.1,  NTN.C.SDN.4.4, NTN.C.SDN.4.4(1), NTN.C.SDN.5.16, NTN.C.NFV.2, NTN.C.NFV.4, NTN.C.NFV.8, NTN.C.NFV.9, NTN.C.NFV.11, NTN.C.NFV.14, NTN.C.NFV.16, NTN.C.NFV.17, NTN.C.NFV.18, NTN.C.NFV.19, NTN.C.Comb.1_B.2.1, NTN.C.Comb.4,NTN.AI.SDN.1,NTN.AI.SDN.6,NTN.AI.Comb.2, SAT.AI.SDN.3,SAT.C.SDN.4,SAT.AI.SDN.1,SAT.AI.SDN.4}         \\ \cline{2-3}
& OpenSAND & \cite{NTN.C.SDN.5.7, NTN.C.Comb.11, NTN.C.NFV.12, NTN.C.NS.1}           \\ \hline

\multirow{11}{*}{\begin{tabular}{lll} Network \\ Virtualization \\ Tools\end{tabular}} 
& Mininet & \cite{NTN.C.SDN.2.11, NTN.C.SDN.3.4, NTN.C.SDN.5.2, NTN.C.SDN.5.12, NTN.C.SDN.5.16,NTN.AI.SDN.4,NTN.AI.SDN.7, AT.C.SDN.1.2,AT.C.SDN.1.5,AT.C.SDN.1.9,AT.C.SDN.5.4, SAT.AI.SDN.4}           \\ \cline{2-3} 
& OpenFlow & \cite{NTN.C.SDN.1.4, NTN.C.SDN.1.6, NTN.C.SDN.3.4, NTN.C.SDN.5.2, NTN.C.SDN.5.7, NTN.C.SDN.5.8, NTN.C.SDN.5.12, NTN.C.SDN.5.16, NTN.C.Comb.11, NTN.C.NFV.12, NTN.C.Comb.10_B.2.3,NTN.AI.SDN.1,NTN.AI.SDN.7, AT.C.SDN.1.1, AT.C.SDN.1.2, AT.C.SDN.1.3, AT.C.SDN.1.5,AT.C.SDN.1.11,AT.C.SDN.1.14,AT.C.SDN.2.1,AT.C.SDN.2.5,AT.C.SDN.2.6,AT.C.SDN.5.4,AT.C.NFV.3,AT.AI.NS.2,SAT.C.SDN.1,SAT.AI.SDN.4}           \\ \cline{2-3} 

& SDN Controllers &            \\ \cline{2-3} 
& OpenDayLight &  \cite{NTN.C.NFV.3, NTN.C.Comb.10_B.2.3, NTN.C.Comb.12, AT.C.SDN.1.5,AT.C.SDN.1.8,AT.AI.NS.2}          \\ 
& Ryu & \cite{NTN.C.SDN.5.2, NTN.C.SDN.5.7, NTN.C.SDN.5.16,NTN.AI.SDN.7,AT.C.SDN.1.14}           \\ 
& POX &  \cite{NTN.C.SDN.5.12,AT.C.SDN.1.9}          \\ 
& ONOS &  \cite{NTN.C.SDN.2.9, NTN.C.SDN.2.11,	AT.AI.SDN.3}          \\  
\cline{2-3} 

& Open vSwitch & \cite{NTN.C.NFV.3, NTN.C.NFV.12, NTN.C.Comb.10_B.2.3, NTN.C.Comb.12, AT.C.SDN.1.3, AT.C.SDN.1.5,AT.C.NFV.3,AT.AI.NS.2,AT.AI.SDN.3, SAT.AI.SDN.4}           \\ \cline{2-3} 
& OpenStack &   \cite{NTN.C.Comb.12,AT.C.NFV.1,AT.C.NFV.3}	        \\ \hline

\end{tabularx}
\caption{Summary of Most Popular Network Simulators and Virtualization Tools in Integrated TN-NTNs.}
\label{tab:Tools_Sum_Up}
\end{table}

\section{Open issues and research directions}\label{Open_issues}
In this section, we highlight several open issues and discuss potential research directions for the advancement of integrated TN-NTNs virtualization. The primary challenges facing the adaptation of virtualization technologies in next-generation networks involve coping with non-terrestrial networks characteristics, dealing with multi-domain network architecture, and ensuring network security and resiliency. Besides, because of the unique peculiarities of NTN platforms, the development of specialised simulation tools is necessary to design, optimize, and evaluate communication systems in integrated TN-NTNs. Moreover, since AI is expected to play a major role in the establishment of 6G networks, overcoming the obstacles arising from the introduction of AI algorithms becomes another open issue. Additionally, emerging innovations such as digital twin, blockchain, and quantum communications can be leveraged and combined with virtualization technologies to enhance the efficiency and security of next-generation networks.

\subsection{Coping with NTNs characteristics} 

Because of the unique characteristics of NTNs, the implementation of virtualization technologies in next-generation networks faces several difficulties. These features mainly include the dynamic environment, the large-scale topology and the limited resources on-board NTN platforms. On the one hand, the high mobility of network nodes increases the complexity of network management and operation. These mobile nodes can follow either predictable patterns such as satellites moving according to their predefined orbits, or unpredictable patterns such as UAVs which can exhibit varying flying trajectories. This results into unstable connectivity, frequent handovers, and service interruption. Hence, novel mobility and handover management strategies are crucial to guarantee QoS requirements and seamless connectivity \cite{SAT.AI.NS.4,NTN.C.Comb.10_B.2.3,AT.C.SDN.6.1}. In SDN/NFV-enabled networks, continuous flow rules computation, forwarding tables updates, and NFV service reconfiguration are necessary to avoid disruptions and assure service continuity. Besides, the high mobility and frequent handovers yield to variation of network resource availability. This affects the provisioning of network slices where the resources reserved for one slice may no longer be accessible, causing failure to meet QoS constraints. Therefore, adaptive network slicing schemes are needed and developing dynamic resource reservation and orchestration is imperative \cite{NTN.AI.NS.1,NTN.C.Comb.5,SAT.AI.NS.5}. On the other hand, scalability issues emerge as the number of network nodes and end-users grows. Mega-constellations of NGEO satellites and HAPS as well as large UAV swarms can cause network performance degradation. Hence, efficient scalable network management procedures and hierarchical architectures should be designed to alleviate the scalability problem \cite{AT.C.SDN.6.1,B.2.3.2}. In particular, the physically centralized single controller structure is not inadequate for SDN-enabled integrated TN-NTNs. This due not only to the  single controller's restricted computing powers in comparison to the network's scale, but also to the high latency and increased control overhead caused by this type of control structure. As a result, a logically distributed hierarchical control structure is required to satisfy the growing service demands of these large-scale networks \cite{NTN.C.Comb.10_B.2.3, NTN.C.SDN.5.14, NTN.C.NS.1}. Another scalability challenge involves the placement of VNFs and the embedding of SFCs in NFV-based networks where the complexity of these optimization problems escalates because of the large size of the network and the limited resources of NTN platforms \cite{AT.C.NFV.6}. Therefore, designing suitable network architectures and effective network operation and management algorithms is important to overcome the scalability obstacles. Furthermore, the limited resources on-board NTN platforms introduces constraints on the network's ability to cope with its dynamic large-scale topology. The restricted communication, computing, and caching resources can impose limitations on the NTN nodes functionalities such as collecting network information, processing data, and executing complex algorithms. In addition, multiple connectivity interruptions and limited service duration can be caused by the energy depletion of satellite and aerial nodes, relying on batteries and solar power \cite{AT.C.NFV.2,SAT.AI.NS.4}. The energy constraints can result in service discontinuity and network failure especially in UAV-assisted networks, where the energy supplies are used for connectivity and flight purposes. Thus, it is critical to develop energy-efficient lightweight schemes taking into account the limited resources and characteristics of the different nodes in integrated TN-NTNs.

\subsection{Dealing with multi-domain network architecture}

A multi-domain multi-tenant architecture is created by virtualizing integrated terrestrial and non-terrestrial networks using SDN, NFV, and network slicing. This introduces a number of challenges stemming from the essential seamless orchestration and management of multi-dimensional resources across multiple network domains while catering to the needs of diverse tenants. In this multi-domain architecture, network resources are owned by numerous service and infrastructure providers across different administrative domains \cite{SAT.AI.NS.1_B.4.3.6}. For instance, space, airborne and terrestrial platforms are managed and operated by different entities including traditional terrestrial telecommunication companies, and aerospace agencies. Cloud services and edge computing infrastructure providers are also major stakeholders, as next-generation networks rely significantly on technologies requiring unprecedented computational capabilities. Besides, the heterogeneity of the underlying network equipments supported by a variety of communication standards and technologies further complicates the issue and limits the network interoperability \cite{SAT.C.SDN.8_B.2.2,AT.C.SDN.6.1}. Consequently, it becomes necessary not only to provide an unified methodology to abstract the network resources offered by various providers but also to promote for the standardization of the protocols and interfaces to facilitate the exchange of these resources and the seamless integration of different network components in virtualized integrated 6G networks \cite{NTN.C.SDN.1.3,SAT.AI.NS.1_B.4.3.6,SAT.AI.NS.5,NTN.C.SDN.5.3}. The next challenge imposed by such architecture is the design of efficient cross-domain coordination and collaboration mechanisms between the different entities. Developing efficient cost-effective schemes to share and orchestrate resources across various domains while meeting the stringent requirements of diverse services is necessary, to create customized network slices in multi-domain networks. Moreover, the availability of network resources is directly affected by the dynamic topology of 6G networks, necessitating a dynamic SLA decomposition across the different domains \cite{SAT.AI.NS.5}. However, ensuring the SLA in this multi-domain architecture is difficult, which demands the implementation of innovative cross-domain orchestration and coordination approaches capable of adapting to the characteristics of non-terrestrial networks. Furthermore, through network slicing, the multi-domain integrated TN-NTN architectures offer tailored services to multiple tenants by enabling the creation of various network slices on top of the shared infrastructure. This raises a number of obstacles, notably in terms of the properties of slice isolation, elasticity and scalability \cite{NTN.C.Comb.6_B.2.3.4,NTN.C.SDN.1.3}. It is challenging to ensure an isolated, elastic, and scalable allocation of network resources for each tenant due to the large-scale topology, the highly mobile NTN platforms, and constantly changing user demands. Thus, it is essential to design network slicing strategies capable of maintaining high levels of QoS satisfaction for each network slice while dealing with 6G networks features.

\subsection{Ensuring network security and resiliency}

Compared to terrestrial networks, the unique characteristics of integrated TN-NTNs complicates the task of ensuring the network security, resiliency, and data privacy. In fact, the large-scale, dynamic and heterogeneous topology combined with the limited on-board resources imposes numerous challenges. First, a crucial security challenge is the vulnerability of data transmission due to the wireless and broadcast nature of communication links in integrated TN-NTNs. Jamming, eavesdropping, disruption and falsification of data are potential threats in this scenario \cite{NTN.C.SDN.1.8,B.2.3.2,AT.C.SDN.6.1}. Particularly, in SDN-enabled networks, the communications between the data and control planes can be susceptible to such menaces which can compromise the network nodes \cite{NTN.C.SDN.1.8, B.2.3.2}. Additionally, hijacking and unauthorized access to non-terrestrial platforms including satellites, HAPS, and UAVs are other major vulnerabilities \cite{AT.AI.SDN.2, mershad2021cloud, AT.C.SDN.6.1}. Hence, lightweight low-complexity solutions for physical layer security are of paramount importance. Novel techniques for anti-jamming, encryption, authentication, and intrusion detection are necessary to safeguard data transmission in highly dynamic networks. Also, blockchain and quantum communication can be leveraged to protect the data and secure satellites optical links, respectively. Second, since the SDN paradigm offers the centralization of the control logic, the security of SDN controllers is another important concern \cite{NTN.C.SDN.1.6,NTN.C.Comb.7,NTN.C.SDN.5.14}. On the level of the control plane, cyber-attacks and malicious activities, such as controllers unauthorized access and hijacking, DDoS and target selection attacks, and software vulnerabilities can be fatal where the attacker can gain access to the entire network. Thus, it is necessary to design security protocols to ensure the protection of SDN controllers especially if they are deployed on NTN nodes. AI models can be employed for attacks detection and mitigation, while blockchain techniques can be used to ensure the trustworthiness and integrity of the network entities. Besides, the optimal orchestration and placement of security VNFs, such as virtual IDSs, firewalls, and proxies can aid in the mitigation of cyber-attacks on the network \cite{AI_NFV_16,AT.C.NFV.3}. However, the virtualization of network functions as VNFs, in NFV-enabled networks, can increase their vulnerability because of software flaws \cite{NTN.C.SDN.1.8}. Moreover, the slicing of a shared underlying infrastructure introduces other security and data privacy challenges in network slicing-based integrated TN-NTNs. Since multiple tenant can share the same physical network node to deploy their virtual networks, an attacker can exploit one slice to gain access to another slice and exhaust its resources \cite{SAT.AI.NS.1_B.4.3.6,SAT.AI.NS.2}. Another security concern in sliced networks is the data leakage during the communication between end-users and network slices. This type of communication involves the exchange of sensitive user information such as location, device type, and user demands. The interception and tampering of such data can result into the user association to an exposed network slice. Therefore, efficient security policies should be enforced including traditional and AI-enabled authentication and slice access control measures. In addition, the multi-domain architecture of integrated networks requires the development of efficient mechanisms to seamlessly orchestrate security protocols across the different domains in \cite{NTN.C.SDN.5.14}. Furthermore, network resiliency is a major issue in integrated TN-NTNs due to the network characteristics. The large communication ranges of satellites and the high dynamicity of UAV-assisted networks, in particular, render the TN-NTNs more susceptible to failures and interruptions \cite{SAT.AI.NS.5}. For sliced networks, robust network slicing solutions that can countermeasure various types of network failures are necessary to sustain network performance and ensure service continuity during the slices life cycles.

\subsection{Designing dedicated simulations tools}

The network performance evaluation phase is mandatory before deploying new network architectures and implementing novel protocols in a real-world environment. As a result, it is vital to test and validate communication systems using simulation tools and experimental implementations. However, this can be a challenging task in integrated TN-NTNs because real-life experimental evaluation of NTN platforms is difficult, and existing network simulators lack the adaptability to NTNs characteristics \cite{NTN.C.SDN.1.8,AT.C.NFV.9,B.2.3.2}. Field trials using NTN platforms such as satellites, HAPS, and UAVs can involve significant expenses, safety risks, and regulatory constraints, which limits their scale and frequency. In addition, existing simulation tools are not suitable for these networks since they are built for terrestrial networks and do not capture the specificities of non-terrestrial nodes. In particular, the current simulation tools that incorporate virtualization technologies and protocols need to be extended and novel tools need to be built to include the constraints imposed by the use of satellites, HAPS, and UAVs. For example, the well known OpenFlow protocol used in SDN-enabled networks should be extended and the development of novel extensions capable of dealing with the NTN features is required \cite{NTN.C.SDN.5.14, NTN.C.Comb.5}. Nonetheless, few efforts have recently been directed to the design of specialised simulation tools for next-generation networks. For instance, virtualized environment emulation framework (VENUE) is introduced in \cite{AT.C.NFV.9} to facilitate the validation and prototyping of NFV-enabled UAV-assisted networks. Additionally, extensions for the network simulator NS3 and the OpenFlow protocol are proposed in \cite{NTN.C.SDN.1.4} and \cite{NTN.C.Comb.10_B.2.3} to implement SDN-based satellite-terrestrial networks architectures and evaluate routing algorithms and network management strategies. However, such studies are still in their early stages and further research is necessary. Meanwhile, theoretical modeling can be utilized to understand the network behavior and evaluate the performance of the proposed architectures and algorithms \cite{NTN.C.SDN.5.14}.	

\subsection{Applying AI algorithms}

AI will play a critical role in the development of 6G networks. Particularly, it can not only solve multiple complex problems in network virtualization as discussed before but also enhance network performance through prediction and pattern recognition, as well as enable autonomous network planning and operation \cite{B.2.3.1}. Nonetheless, using AI algorithms in next-generation networks raises a number of issues that can be observed from two aspects. On the one hand, issues caused by the inherent characteristics of AI models can complicate its implementation in 6G networks. While supervised and unsupervised learning, used for prediction and classification problems, require large realistic training datasets causing data collection and analysis challenges, RL algorithms, used for decision-making tasks, struggle to solve complex optimization problems with numerous constraints \cite{NTN.AI.NS.1}. Another concern with using AI in 6G networks is algorithm selection, as there is no one-size-fits-all solution. Different factors should be considered in choosing a suitable AI technique to address a particular network problem \cite{SAT.AI.NS.5}. This includes the type of the problem, the needed resources to execute the algorithm, and the desired level of performance. Thus, it is necessary to conduct an analysis that examines the cost-benefit trade-off between the selected AI model and its anticipated performance. On the other hand, applying AI in integrated TN-NTNs is a challenging task due to the unique features of SAGIN including the highly dynamic environment, the large-scale topology and the limited on-board resources. The high mobility of non-terrestrial platforms introduces increased dynamicity to the network topology. This results into the need for designing adaptive algorithms with continuous updating capable of obtaining optimized strategies at different time slots for resource allocation, device/user association, routing, and controller placement etc. Supervised and unsupervised ML techniques lack resilience to adapt to such dynamic environment \cite{NTN.AI.NS.1,SAT.AI.NS.5,B.2.3.2}. This is mainly due to their dependence on the training dataset where the real dataset may be statistically different and constantly changing causing degradation in the ML algorithm performance. Reinforcement learning can be a solution for this issue since the RL agents can continuously learn new optimal policies adapting the dynamic environment in a dataset-free fashion \cite{B.4.2.1}. However, the large-scale topology of 6G networks increases the dimensionality of the state space for RL agents imposing another challenge on the learning and optimization process of these AI models. This network characteristic brings additional obstacles in the application of AI methods regarding algorithmic complexity, feature extraction, and massive data collection and analysis \cite{SAT.AI.NS.5}. Additionally, AI models are expected to deliver high performance in order to satisfy the needs of this expanding network with increased demands, diverse services, and stringent QoS requirements. Nevertheless, the limited resources on-board non-terrestrial nodes further complicates this task where the satellites, HAPS, and UAVs may not have sufficient resources in terms of energy, computing and storage, necessary for the implementation of powerful AI solutions \cite{NTN.AI.NS.1, B.2.3.2}. Therefore, the development of low-complexity, lightweight, and energy-efficient AI algorithms is required in 6G networks.

\subsection{Leveraging other emerging technologies}
Virtualization technologies can be combined with other emerging innovations such as digital twins, blockchain and quantum communications to improve the performance and the security of next-generation networks.

Multiple definitions can be found in the literature describing the digital twin (DT) paradigm. One way to characterize DTs is by viewing them as replications of physical entities (objects, people, environments, etc) where virtual representations of the physical assets are accurately created and uni/bi-directional communication links are established enabling the interaction between the two sides \cite{digital_twin_survey}. Powered by AI, digital twins can optimize and enhance the performance of next-generation networks by monitoring their status, analyzing their operation, and predicting failures in a real-time manner using a closed loop between the physical and digital versions of the network \cite{digital_twin_survey,khan2022digital}. In the context of SDN/NFV-enabled integrated TN-NTNs, DTs can further enable network virtualization, improve the adaptability to highly dynamic topologies, and provide network operators with real-time insights into their network performance. They can be built in the SDN controller to enable proactive dynamic and intelligent network control \cite{zhao2020intelligentDT}. Moreover, DTs can be used to create simulation and emulation environments, especially for networks incorporating non-terrestrial platforms to test and validate different applications instead of relying on the physical infrastructure which can either be costly and/or dangerous \cite{B.2.3.2,AT.C.SDN.2.2}. For instance, using physical satellites to design, optimize, and test satellite-assisted networks can be very expensive and require interactions with satellite infrastructure providers, while digital twins of such networks can be built allowing researchers to flexibly and easily conduct their experiments and apply their modifications. Similarly, deploying actual UAVs during the development and optimization stages of UAV-based networks can be both dangerous and costly, hence DTs can aid in designing, validating, and ensuring the safety of UAV-assisted networks. End-user virtualization is another approach for implementing digital twins in virtualized networks where it can be utilized to describe the state and service requirements of end-users \cite{B.3.1}. While technologies such as SDN, NFV, and network slicing focus on the virtualization of network infrastructure and resources, DTs of end-users enable the virtualization of end-users providing significant real-time end-user data that can boost the network's decision-making, management, and simulation capabilities. Furthermore, network slicing and digital twin technologies can enable service-centric and user-centric networking, respectively. While network slicing creates customized slices for different services enabling service-centric management in 6G networks, digital twins characterize end-users allowing user-centric management in 6G networks  \cite{B.3.1,khan2022digital}. In fact, after the creation of service-tailored slices, the data provided by the DT of individual end-users in each slice can be exploited to enable user-oriented decision-making and improve intra-slice network management, thereby increasing the granularity and adaptability of network management, particularly in highly dynamic environments with diverse end-users. 

Blockchain is a groundbreaking innovation that has revolutionized data storage, sharing, and verification. Originally developed for crypto-currencies, it is defined as a distributed and transparent ledger that ensures secure recording of transactions and assets \cite{B.2.3.2}. Blockchain can play a pivotal role in improving the security, privacy, and reliability of next-generation networks. In particular, for integrated TN-NTNs that use non-terrestrial platforms including satellites, HAPS, and UAVs, it is crucial to ensure the security and privacy of the exchanged critical data between network nodes, especially that it is wirelessly transmitted. In addition, the decentralized consensus mechanism of blockchain can enhance the trustworthiness of the network entities across different domains, verify the integrity of network data and nodes access control, and aid in cyber-attacks and malicious activities detection \cite{B.2.3.1,B.2.3.2,mershad2021cloud}. Moreover, SDN-enabled integrated TN-NTNs can benefit from blockchain by securing distributed SDN controllers and verifying OpenFlow tables \cite{wang2021blockchain}. Sliced networks also can employ blockchain to support authenticated slice brokering and trustworthy infrastructure sharing between the MNOs by offering traceable and transparent slice ledgers that can autonomously track the slice sharing and leasing behaviors \cite{wang2021blockchain,wijethilaka2021survey}.

Quantum technologies including communication, computing and sensing are reshaping multiple fields such as cyber-security, high-performance computing, and networking. In particular, quantum communication is transforming the way information is transmitted. While classical communications rely on the classical zero and one bits, quantum communications leverage the principle of quantum physics to transmit quantum bits, known as qubits \cite{A.3}. This would inherently result in secure and efficient data transmission where cyber-attacks and malicious activities can be effortlessly detected and mitigated rendering it appropriate for integrated terrestrial and non-terrestrial networks \cite{A.2,B.2.3.1}. Moreover, the SDN paradigm can be combined with quantum communications in future networks to enhance quantum resource management and task administration \cite{A.3}. The SDN controller can continuously monitor the quantum parameters including the secret key rate of the quantum key distribution (QKD) protocol and the quantum bit error rate.

\section{Conclusion}\label{Conclusion}

To support the large variety of applications and satisfy the target KPIs of 6G networks, integrated TN-NTNs are envisioned as 6G key enabling technologies. However, the TN-NTNs integration faces several issues that can be solved using network virtualization technologies such as SDN, NFV and network slicing. This survey provides a comprehensive review on the adaptation of these networking paradigms in next-generation networks. We commenced by covering the fundamentals of NTNs and virtualization techniques. Then, we brought attention to the intersection of AI and network virtualization, summarizing the major research areas where AI models play a pivotal role in enhancing SDN, NFV, and network slicing. Building on this foundation, the survey highlighted the prevalent problems emerging from the adaptation of these techniques in integrated TN-NTNs and proposed a taxonomy of integrated TN-NTNs virtualization based on a four-level classification. This taxonomy offers a structured and comprehensive review of relevant contributions synthesising and clarifying the different facets of the subject. In addition, we include a summary on the simulation and implementation tools utilized in the performance evaluation and testing of such networks, offering guidance on the assessment methodologies. Lastly, this survey discusses open issues and gives insights on future research directions for the advancement of integrated TN-NTNs virtualization in the 6G era. In conclusion, while adopting network virtualization technologies in 6G integrated TN-NTNs offers efficient network management and improved network performance, numerous research gaps should be addressed and further investigations are required to realise the full potential of these technologies.

\bibliographystyle{IEEEtran}
\bibliography{IEEEabrv,citations}

\begin{thebibliography}{100}
\providecommand{\url}[1]{#1}
\csname url@samestyle\endcsname
\providecommand{\newblock}{\relax}
\providecommand{\bibinfo}[2]{#2}
\providecommand{\BIBentrySTDinterwordspacing}{\spaceskip=0pt\relax}
\providecommand{\BIBentryALTinterwordstretchfactor}{4}
\providecommand{\BIBentryALTinterwordspacing}{\spaceskip=\fontdimen2\font plus
\BIBentryALTinterwordstretchfactor\fontdimen3\font minus
  \fontdimen4\font\relax}
\providecommand{\BIBforeignlanguage}[2]{{%
\expandafter\ifx\csname l@#1\endcsname\relax
\typeout{** WARNING: IEEEtran.bst: No hyphenation pattern has been}%
\typeout{** loaded for the language `#1'. Using the pattern for}%
\typeout{** the default language instead.}%
\else
\language=\csname l@#1\endcsname
\fi
#2}}
\providecommand{\BIBdecl}{\relax}
\BIBdecl

\bibitem{Dang2020}
S.~Dang, O.~Amin, B.~Shihada, and M.~S. Alouini, ``What should 6g be?''
  \emph{Nature Electronics}, vol.~3, pp. 20--29, 1 2020.

\bibitem{tong_zhu_2021}
W.~Tong and P.~Zhu, \emph{6G, the Next Horizon: From Connected People and
  Things to Connected Intelligence}.\hskip 1em plus 0.5em minus 0.4em\relax
  Cambridge University Press, 2021.

\bibitem{series2015imt}
M.~Series, ``Imt vision--framework and overall objectives of the future
  development of imt for 2020 and beyond,'' \emph{Recommendation ITU}, vol.
  2083, p.~0, 2015.

\bibitem{wu20216g}
Y.~Wu, S.~Singh, T.~Taleb, A.~Roy, H.~S. Dhillon, M.~R. Kanagarathinam, and
  A.~De, \emph{6G mobile wireless networks}.\hskip 1em plus 0.5em minus
  0.4em\relax Springer, 2021.

\bibitem{B.3.1}
X.~Shen, J.~Gao, W.~Wu, M.~Li, C.~Zhou, and W.~Zhuang, ``Holistic network
  virtualization and pervasive network intelligence for 6g,'' \emph{IEEE
  Communications Surveys \& Tutorials}, vol.~24, no.~1, pp. 1--30, 2021.

\bibitem{SDN2014survey}
B.~A.~A. Nunes, M.~Mendonca, X.-N. Nguyen, K.~Obraczka, and T.~Turletti, ``A
  survey of software-defined networking: Past, present, and future of
  programmable networks,'' \emph{IEEE Communications surveys \& tutorials},
  vol.~16, no.~3, pp. 1617--1634, 2014.

\bibitem{B.1.9}
B.~Yi, X.~Wang, K.~Li, M.~Huang \emph{et~al.}, ``A comprehensive survey of
  network function virtualization,'' \emph{Computer Networks}, vol. 133, pp.
  212--262, 2018.

\bibitem{B.1.2}
I.~Afolabi, T.~Taleb, K.~Samdanis, A.~Ksentini, and H.~Flinck, ``Network
  slicing and softwarization: A survey on principles, enabling technologies,
  and solutions,'' \emph{IEEE Communications Surveys \& Tutorials}, vol.~20,
  no.~3, pp. 2429--2453, 2018.

\bibitem{rinaldi2020non}
F.~Rinaldi, H.-L. Maattanen, J.~Torsner, S.~Pizzi, S.~Andreev, A.~Iera,
  Y.~Koucheryavy, and G.~Araniti, ``Non-terrestrial networks in 5g \& beyond: A
  survey,'' \emph{IEEE access}, vol.~8, pp. 165\,178--165\,200, 2020.

\bibitem{Liu2018}
J.~Liu, Y.~Shi, Z.~M. Fadlullah, and N.~Kato, ``Space-air-ground integrated
  network: A survey,'' \emph{IEEE Communications Surveys and Tutorials},
  vol.~20, pp. 2714--2741, 10 2018.

\bibitem{A.2}
M.~M. Azari, S.~Solanki, S.~Chatzinotas, O.~Kodheli, H.~Sallouha, A.~Colpaert,
  J.~F. Mendoza~Montoya, S.~Pollin, A.~Haqiqatnejad, A.~Mostaani, E.~Lagunas,
  and B.~Ottersten, ``Evolution of non-terrestrial networks from 5g to 6g: A
  survey,'' \emph{IEEE Communications Surveys \& Tutorials}, vol.~24, no.~4,
  pp. 2633--2672, 2022.

\bibitem{B.1.1}
D.~Kreutz, F.~M. Ramos, P.~E. Verissimo, C.~E. Rothenberg, S.~Azodolmolky, and
  S.~Uhlig, ``Software-defined networking: A comprehensive survey,''
  \emph{Proceedings of the IEEE}, vol. 103, no.~1, pp. 14--76, 2014.

\bibitem{B.1.5}
R.~Mijumbi, J.~Serrat, J.-L. Gorricho, N.~Bouten, F.~De~Turck, and R.~Boutaba,
  ``Network function virtualization: State-of-the-art and research
  challenges,'' \emph{IEEE Communications surveys \& tutorials}, vol.~18,
  no.~1, pp. 236--262, 2015.

\bibitem{B.1.8}
A.~A. Barakabitze, A.~Ahmad, R.~Mijumbi, and A.~Hines, ``5g network slicing
  using sdn and nfv: A survey of taxonomy, architectures and future
  challenges,'' \emph{Computer Networks}, vol. 167, p. 106984, 2020.

\bibitem{B.2.3.1}
O.~S. Oubbati, M.~Atiquzzaman, T.~A. Ahanger, and A.~Ibrahim, ``Softwarization
  of uav networks: A survey of applications and future trends,'' \emph{IEEE
  Access}, vol.~8, pp. 98\,073--98\,125, 2020.

\bibitem{B.2.3.2}
W.~Jiang, ``Software defined satellite networks: A survey,'' \emph{Digital
  Communications and Networks}, 2023.

\bibitem{B.2.3.3}
T.~Bouzid, N.~Chaib, M.~L. Bensaad, and O.~S. Oubbati, ``5g network slicing
  with unmanned aerial vehicles: Taxonomy, survey, and future directions,''
  \emph{Transactions on Emerging Telecommunications Technologies}, 2022.

\bibitem{kurt2021vision}
G.~K. Kurt, M.~G. Khoshkholgh, S.~Alfattani, A.~Ibrahim, T.~S. Darwish, M.~S.
  Alam, H.~Yanikomeroglu, and A.~Yongacoglu, ``A vision and framework for the
  high altitude platform station (haps) networks of the future,'' \emph{IEEE
  Communications Surveys \& Tutorials}, vol.~23, no.~2, pp. 729--779, 2021.

\bibitem{wang2021blockchain}
Y.~Wang, Z.~Su, J.~Ni, N.~Zhang, and X.~Shen, ``Blockchain-empowered
  space-air-ground integrated networks: Opportunities, challenges, and
  solutions,'' \emph{IEEE Communications Surveys \& Tutorials}, vol.~24, no.~1,
  pp. 160--209, 2021.

\bibitem{zhang2020survey}
S.~Zhang, D.~Zhu, and Y.~Wang, ``A survey on space-aerial-terrestrial
  integrated 5g networks,'' \emph{Computer Networks}, vol. 174, p. 107212,
  2020.

\bibitem{SAT.C.SDN.8_B.2.2}
N.~Zhang, S.~Zhang, P.~Yang, O.~Alhussein, W.~Zhuang, and X.~S. Shen,
  ``Software defined space-air-ground integrated vehicular networks: Challenges
  and solutions,'' \emph{IEEE Communications Magazine}, vol.~55, no.~7, pp.
  101--109, 2017.

\bibitem{geraci2022integrating}
G.~Geraci, D.~Lopez-Perez, M.~Benzaghta, and S.~Chatzinotas, ``Integrating
  terrestrial and non-terrestrial networks: 3d opportunities and challenges,''
  \emph{IEEE Communications Magazine}, 2022.

\bibitem{A.3}
H.~Al-Hraishawi, M.~Razavi, S.~Chatzinotas \emph{et~al.}, ``Characterizing and
  utilizing the interplay between quantum technologies and non-terrestrial
  networks,'' \emph{arXiv preprint arXiv:2211.08508}, 2022.

\bibitem{B.2.3}
S.~Xu, X.-W. Wang, and M.~Huang, ``Software-defined next-generation satellite
  networks: Architecture, challenges, and solutions,'' \emph{IEEE Access},
  vol.~6, pp. 4027--4041, 2018.

\bibitem{B.1.4}
M.~Condoluci and T.~Mahmoodi, ``Softwarization and virtualization in 5g mobile
  networks: Benefits, trends and challenges,'' \emph{Computer Networks}, vol.
  146, pp. 65--84, 2018.

\bibitem{B.1.3}
N.~M.~K. Chowdhury and R.~Boutaba, ``A survey of network virtualization,''
  \emph{Computer Networks}, vol.~54, no.~5, pp. 862--876, 2010.

\bibitem{SDNsurvey2019application}
R.~Sahay, W.~Meng, and C.~D. Jensen, ``The application of software defined
  networking on securing computer networks: A survey,'' \emph{Journal of
  Network and Computer Applications}, vol. 131, pp. 89--108, 2019.

\bibitem{kafetzis2022SDN}
D.~Kafetzis, S.~Vassilaras, G.~Vardoulias, and I.~Koutsopoulos,
  ``Software-defined networking meets software-defined radio in mobile ad hoc
  networks: State of the art and future directions,'' \emph{IEEE Access},
  vol.~10, pp. 9989--10\,014, 2022.

\bibitem{mckeown2008openflow}
N.~McKeown, T.~Anderson, H.~Balakrishnan, G.~Parulkar, L.~Peterson, J.~Rexford,
  S.~Shenker, and J.~Turner, ``Openflow: enabling innovation in campus
  networks,'' \emph{ACM SIGCOMM computer communication review}, vol.~38, no.~2,
  pp. 69--74, 2008.

\bibitem{ETSI_NFV}
\BIBentryALTinterwordspacing
ETSI. Nfv: Architectural framework. [Online]. Available: \url{http://www.etsi.
  org/deliver/etsi_gs/nfv/001_099/002/01.02.01_60/gs_nfv002v010201p.pdf}
\BIBentrySTDinterwordspacing

\bibitem{el2015ngmn}
R.~El~Hattachi and J.~Erfanian, ``Ngmn 5g white paper ngmn alliance,''
  \emph{NGNM}, 2015.

\bibitem{3gpp2016study}
3GPP, ``Study on architecture for next generation system (release 14): Tr23.
  799 v14. 0.0,'' 2016.

\bibitem{khan2020network}
L.~U. Khan, I.~Yaqoob, N.~H. Tran, Z.~Han, and C.~S. Hong, ``Network slicing:
  Recent advances, taxonomy, requirements, and open research challenges,''
  \emph{IEEE Access}, vol.~8, pp. 36\,009--36\,028, 2020.

\bibitem{alliance2016description}
N.~Alliance, ``Description of network slicing concept,'' \emph{NGMN 5G P},
  vol.~1, no.~1, pp. 1--11, 2016.

\bibitem{SAT.AI.NS.1_B.4.3.6}
J.~Wang, J.~Liu, J.~Li, and N.~Kato, ``Artificial intelligence-assisted network
  slicing: Network assurance and service provisioning in 6g,'' \emph{IEEE
  Vehicular Technology Magazine}, vol.~18, pp. 49--58, 3 2023.

\bibitem{B.4.2}
C.~Ssengonzi, O.~P. Kogeda, and T.~O. Olwal, ``A survey of deep reinforcement
  learning application in 5g and beyond network slicing and virtualization,''
  \emph{Array}, p. 100142, 2022.

\bibitem{AI_Comb_1}
T.~Rakkiannan, G.~Ekambaram, N.~Palanisamy, R.~R. Ramasamy, S.~Muthusamy, A.~K.
  Loganathan, H.~Panchal, K.~Thangaraj, and A.~Ravindaran, ``An automated
  network slicing at edge with software defined networking and network function
  virtualization: A federated learning approach,'' \emph{Wireless Personal
  Communications}, 2023.

\bibitem{B.1.7}
Y.~Wu, H.-N. Dai, H.~Wang, Z.~Xiong, and S.~Guo, ``A survey of intelligent
  network slicing management for industrial iot: integrated approaches for
  smart transportation, smart energy, and smart factory,'' \emph{IEEE
  Communications Surveys \& Tutorials}, vol.~24, no.~2, pp. 1175--1211, 2022.

\bibitem{B.4.3.4_B.1.6}
X.~Shen, J.~Gao, W.~Wu, K.~Lyu, M.~Li, W.~Zhuang, X.~Li, and J.~Rao,
  ``Ai-assisted network-slicing based next-generation wireless networks,''
  \emph{IEEE Open Journal of Vehicular Technology}, vol.~1, pp. 45--66, 2020.

\bibitem{B.4.3.2}
C.~Ssengonzi, O.~P. Kogeda, and T.~O. Olwal, ``A survey of deep reinforcement
  learning application in 5g and beyond network slicing and virtualization,''
  \emph{Array}, vol.~14, 7 2022.

\bibitem{B.4.3.7}
N.~Saha, M.~Zangooei, M.~Golkarifard, and R.~Boutaba, ``Deep reinforcement
  learning approaches to network slice scaling and placement: A survey,''
  \emph{IEEE Communications Magazine}, vol.~61, pp. 82--87, 2 2023.

\bibitem{MLmeetNMOinEdge}
A.~Shahraki, T.~Ohlenforst, and F.~Krey{\ss}, ``When machine learning meets
  network management and orchestration in edge-based networking paradigms,''
  \emph{Journal of Network and Computer Applications}, vol. 212, p. 103558,
  2023.

\bibitem{B.4.2.2}
S.~Lange, N.~V. Tu, S.~Y. Jeong, D.~Y. Lee, H.~G. Kim, J.~Hong, J.~H. Yoo, and
  J.~W.~K. Hong, ``A network intelligence architecture for efficient vnf
  lifecycle management,'' \emph{IEEE Transactions on Network and Service
  Management}, vol.~18, pp. 1476--1490, 6 2021.

\bibitem{B.4.4.3}
A.~A. Gebremariam, M.~Usman, and M.~Qaraqe, ``Applications of artificial
  intelligence and machine learning in the area of sdn and nfv: A survey,''
  \emph{SSD'19 : the 16th International Multiconference on Systems, Signals \&
  Devices}, 2019.

\bibitem{B.4.2.1}
D.~M. Manias and A.~Shami, ``The need for advanced intelligence in nfv
  management and orchestration,'' \emph{IEEE Network}, vol.~35, pp. 365--371, 3
  2021.

\bibitem{B.4.9}
J.~Xie, F.~R. Yu, T.~Huang, R.~Xie, J.~Liu, C.~Wang, and Y.~Liu, ``A survey of
  machine learning techniques applied to software defined networking (sdn):
  Research issues and challenges,'' \emph{IEEE Communications Surveys and
  Tutorials}, vol.~21, pp. 393--430, 1 2019.

\bibitem{DL_book}
F.~o. Chollet, \emph{\BIBforeignlanguage{English}{Deep learning with
  Python}}.\hskip 1em plus 0.5em minus 0.4em\relax Shelter Island, New York:
  Manning Publications Co, 2018.

\bibitem{AI_book}
S.~J. Russell and P.~Norvig, \emph{\BIBforeignlanguage{English}{Artificial
  intelligence: a modern approach}}, 4th~ed.\hskip 1em plus 0.5em minus
  0.4em\relax Hoboken, NJ: Pearson, 2021.

\bibitem{AI_book2}
M.~Negnevitsky, \emph{\BIBforeignlanguage{English}{Artificial intelligence: a
  guide to intelligent systems}}, 2nd~ed.\hskip 1em plus 0.5em minus
  0.4em\relax New York;Harlow, England;: Addison-Wesley, 2005.

\bibitem{Bishop_book}
C.~M. Bishop, \emph{\BIBforeignlanguage{English}{Pattern Recognition and
  Machine Learning}}.\hskip 1em plus 0.5em minus 0.4em\relax New York:
  Springer, 2006.

\bibitem{ML_book}
A.~Géron, \emph{\BIBforeignlanguage{English}{Hands-On Machine Learning with
  Scikit-Learn, Keras, and TensorFlow, 2nd Edition}}.\hskip 1em plus 0.5em
  minus 0.4em\relax S.l.: O'Reilly Media, Inc, 2019.

\bibitem{RL_book}
R.~S. Sutton and A.~G. Barto, \emph{Reinforcement learning: an
  introduction}.\hskip 1em plus 0.5em minus 0.4em\relax Cambridge, Mass: MIT
  Press, 1998.

\bibitem{bengio2017deep}
Y.~Bengio, I.~Goodfellow, and A.~Courville, \emph{Deep learning}.\hskip 1em
  plus 0.5em minus 0.4em\relax MIT press Cambridge, MA, USA, 2017, vol.~1.

\bibitem{fourati2021artificial}
F.~Fourati and M.-S. Alouini, ``Artificial intelligence for satellite
  communication: A review,'' \emph{Intelligent and Converged Networks}, vol.~2,
  no.~3, pp. 213--243, 2021.

\bibitem{jiang2016machine}
C.~Jiang, H.~Zhang, Y.~Ren, Z.~Han, K.-C. Chen, and L.~Hanzo, ``Machine
  learning paradigms for next-generation wireless networks,'' \emph{IEEE
  Wireless Communications}, vol.~24, no.~2, pp. 98--105, 2016.

\bibitem{chen2019artificial}
M.~Chen, U.~Challita, W.~Saad, C.~Yin, and M.~Debbah, ``Artificial neural
  networks-based machine learning for wireless networks: A tutorial,''
  \emph{IEEE Communications Surveys \& Tutorials}, vol.~21, no.~4, pp.
  3039--3071, 2019.

\bibitem{sheth2020taxonomy}
K.~Sheth, K.~Patel, H.~Shah, S.~Tanwar, R.~Gupta, and N.~Kumar, ``A taxonomy of
  ai techniques for 6g communication networks,'' \emph{Computer
  communications}, vol. 161, pp. 279--303, 2020.

\bibitem{bhattacharyyamachine}
A.~Bhattacharyya, S.~M. Nambiar, R.~Ojha, A.~Gyaneshwar, U.~Chadha, and
  K.~Srinivasan, ``Machine learning and deep learning powered satellite
  communications: Enabling technologies, applications, open challenges, and
  future research directions,'' \emph{International Journal of Satellite
  Communications and Networking}, 2023.

\bibitem{B.4.7}
Y.~Qian, J.~Wu, R.~Wang, F.~Zhu, and W.~Zhang, ``Survey on reinforcement
  learning applications in communication networks,'' \emph{Journal of
  Communications and Information Networks}, vol.~4, 2019.

\bibitem{B.4.4}
Y.~Zhao, Y.~Li, X.~Zhang, G.~Geng, W.~Zhang, and Y.~Sun, ``A survey of
  networking applications applying the software defined networking concept
  based on machine learning,'' \emph{IEEE Access}, vol.~7, pp.
  95\,397--95\,417, 2019.

\bibitem{B.4.5}
M.~Latah and L.~Toker, ``Artificial intelligence enabled software-defined
  networking: A comprehensive overview,'' \emph{IET Networks}, vol.~8, pp.
  79--99, 3 2019.

\bibitem{B.4.3.5_B.4.10}
R.~Dangi, A.~Jadhav, G.~Choudhary, N.~Dragoni, M.~K. Mishra, and P.~Lalwani,
  ``Ml-based 5g network slicing security: A comprehensive survey,''
  \emph{Future Internet}, vol.~14, 4 2022.

\bibitem{B.4.3.3_B.4.11}
J.~A.~H. Sánchez, K.~Casilimas, and O.~M.~C. Rendon, ``Deep reinforcement
  learning for resource management on network slicing: A survey,''
  \emph{Sensors}, vol.~22, 4 2022.

\bibitem{AI_Comb_11}
G.~Ramya and R.~Manoharan, ``Traffic-aware dynamic controller placement in sdn
  using nfv,'' \emph{Journal of Supercomputing}, vol.~79, pp. 2082--2107, 2
  2023.

\bibitem{AI_SDN_2}
R.~Amin, E.~Rojas, A.~Aqdus, S.~Ramzan, D.~Casillas-Perez, and J.~M. Arco, ``A
  survey on machine learning techniques for routing optimization in sdn,''
  \emph{IEEE Access}, vol.~9, pp. 104\,582--104\,611, 2021.

\bibitem{AI_SDN_10}
M.~Cicioğlu and A.~Çalhan, ``Mlar: machine-learning-assisted centralized
  link-state routing in software-defined-based wireless networks,''
  \emph{Neural Computing and Applications}, vol.~35, pp. 5409--5420, 3 2023.

\bibitem{AI_NFV_10}
R.~Zhu, P.~Wang, Z.~Geng, Y.~Zhao, and S.~Yu, ``Double-agent reinforced vnfc
  deployment in eons for cloud-edge computing,'' \emph{Journal of Lightwave
  Technology}, 2023.

\bibitem{AI_NFV_11}
D.~Basu, S.~Kal, U.~Ghosh, and R.~Datta, ``Drive: Dynamic resource
  introspection and vnf embedding for 5g using machine learning,'' \emph{IEEE
  Internet of Things Journal}, pp. 1--1, 1 2023.

\bibitem{AI_NFV_12}
X.~Fu, F.~R. Yu, J.~Wang, Q.~Qi, and J.~Liao, ``Dynamic service function chain
  embedding for nfv-enabled iot: A deep reinforcement learning approach,''
  \emph{IEEE Transactions on Wireless Communications}, vol.~19, pp. 507--519, 1
  2020.

\bibitem{AI_NFV_13}
M.~A. Khoshkholghi and T.~Mahmoodi, ``Edge intelligence for service function
  chain deployment in nfv-enabled networks,'' \emph{Computer Networks}, vol.
  219, 12 2022.

\bibitem{AI_NFV_21}
H.~Huang, C.~Zeng, Y.~Zhao, G.~Min, Y.~Zhu, W.~Miao, and J.~Hu, ``Scalable
  orchestration of service function chains in nfv-enabled networks: A federated
  reinforcement learning approach,'' \emph{IEEE Journal on Selected Areas in
  Communications}, vol.~39, pp. 2558--2571, 8 2021.

\bibitem{AI_NFV_22}
X.~Fu, F.~R. Yu, J.~Wang, Q.~Qi, and J.~Liao, ``Service function chain
  embedding for nfv-enabled iot based on deep reinforcement learning,''
  \emph{IEEE Communications Magazine}, vol.~57, pp. 102--108, 11 2019.

\bibitem{AI_Comb_6}
J.~Zhang, Y.~Liu, Z.~Li, and Y.~Lu, ``Forecast-assisted service function chain
  dynamic deployment for sdn $\&$ nfv-enabled cloud management systems,''
  \emph{IEEE Systems Journal}, 2023.

\bibitem{AI_Comb_8}
H.~Xuan, Y.~Zhou, X.~Zhao, and Z.~Liu, ``Multi-agent deep reinforcement
  learning algorithm with self-adaption division strategy for vnf-sc deployment
  in sdn/nfv-enabled networks,'' \emph{Applied Soft Computing}, p. 110189, 5
  2023.

\bibitem{AI_NFV_7}
Z.~Ning, N.~Wang, and R.~Tafazolli, ``Deep reinforcement learning for nfv-based
  service function chaining in multi-service networks,'' \emph{IEEE 21st
  International Conference on High Performance Switching and Routing (HPSR)},
  2020.

\bibitem{B.4.3.1_B.4.8}
W.~Wu, C.~Zhou, M.~Li, H.~Wu, H.~Zhou, N.~Zhang, X.~S. Shen, and W.~Zhuang,
  ``Ai-native network slicing for 6g networks,'' \emph{IEEE Wireless
  Communications}, vol.~29, pp. 96--103, 2 2022.

\bibitem{AI_NS_11}
S.~Khan, S.~Khan, Y.~Ali, M.~Khalid, Z.~Ullah, and S.~Mumtaz, ``Highly accurate
  and reliable wireless network slicing in 5th generation networks: A hybrid
  deep learning approach,'' \emph{Journal of Network and Systems Management},
  vol.~30, 4 2022.

\bibitem{AI_NS_2}
E.~Thomatos, A.~Sgora, and P.~Chatzimisios, ``A survey on ai based network
  slicing standards,'' \emph{2021 IEEE Conference on Standards for
  Communications and Networking, CSCN 2021}, pp. 136--141, 2021.

\bibitem{AI_NS_6}
Q.~Liu, N.~Choi, and T.~Han, ``Deep reinforcement learning for end-to-end
  network slicing: Challenges and solutions,'' \emph{IEEE Network}, 2022.

\bibitem{AI_NS_29}
M.~Iannelli, M.~R. Rahman, N.~Choi, and L.~Wang, ``Applying machine learning to
  end-to-end slice sla decomposition,'' \emph{NetSoft}, 2020.

\bibitem{AI_SDN_4}
Y.~Cao, R.~Wang, M.~Chen, and A.~Barnawi, ``Ai agent in software-defined
  network: Agent-based network service prediction and wireless resource
  scheduling optimization,'' \emph{IEEE Internet of Things Journal}, vol.~7,
  pp. 5816--5826, 7 2020.

\bibitem{AI_SDN_7}
P.~Wang, F.~Ye, X.~Chen, and Y.~Qian, ``Datanet: Deep learning based encrypted
  network traffic classification in sdn home gateway,'' \emph{IEEE Access},
  vol.~6, pp. 55\,380--55\,391, 2018.

\bibitem{AI_NS_8}
D.~Bega, M.~Gramaglia, M.~Fiore, A.~Banchs, and X.~Costa-Perez, ``Deepcog:
  Cognitive network management in sliced 5g networks with deep learning,''
  \emph{IEEE INFOCOM 2019 - IEEE Conference on Computer Communications}, 2019.

\bibitem{AI_SDN_5}
G.~Manogaran, T.~Baabdullah, D.~B. Rawat, and P.~M. Shakeel, ``Ai-assisted
  service virtualization and flow management framework for 6g-enabled
  cloud-software-defined network-based iot,'' \emph{IEEE Internet of Things
  Journal}, vol.~9, pp. 14\,644--14\,654, 8 2022.

\bibitem{AI_SDN_6}
J.~Singh, P.~Singh, M.~Hedabou, and N.~Kumar, ``An efficient machine
  learning-based resource allocation scheme for sdn-enabled fog computing
  environment,'' \emph{IEEE Transactions on Vehicular Technology}, 2023.

\bibitem{AI_NS_3}
A.~Thantharate and C.~Beard, ``Adaptive6g: Adaptive resource management for
  network slicing architectures in current 5g and future 6g systems,''
  \emph{Journal of Network and Systems Management}, vol.~31, 3 2023.

\bibitem{AI_NS_4}
\BIBentryALTinterwordspacing
A.~Filali, B.~Nour, S.~Cherkaoui, and A.~Kobbane, ``Communication and
  computation o-ran resource slicing for urllc services using deep
  reinforcement learning,'' \emph{arXiv preprint arXiv:2301.04696}, 2 2022.
  [Online]. Available: \url{http://arxiv.org/abs/2202.06439}
\BIBentrySTDinterwordspacing

\bibitem{AI_NS_5}
S.~Messaoud, A.~Bradai, O.~B. Ahmed, P.~T.~A. Quang, M.~Atri, and M.~S.
  Hossain, ``Deep federated q-learning-based network slicing for industrial
  iot,'' \emph{IEEE Transactions on Industrial Informatics}, vol.~17, pp.
  5572--5582, 8 2021.

\bibitem{AI_NS_7}
K.~Suh, S.~Kim, Y.~Ahn, S.~Kim, H.~Ju, and B.~Shim, ``Deep reinforcement
  learning-based network slicing for beyond 5g,'' \emph{IEEE Access}, vol.~10,
  pp. 7384--7395, 2022.

\bibitem{AI_NS_9}
W.~Wu, N.~Chen, C.~Zhou, M.~Li, X.~Shen, W.~Zhuang, and X.~Li, ``Dynamic ran
  slicing for service-oriented vehicular networks via constrained learning,''
  \emph{IEEE Journal on Selected Areas in Communications}, vol.~39, pp.
  2076--2089, 7 2021.

\bibitem{AI_NS_12}
J.~Wu, G.~Member, Y.~Gao, S.~Member, L.~Wang, J.~Zhang, and D.~O. Wu, ``How to
  allocate resources in cloud native networks towards 6g,'' \emph{IEEE
  Network}, 2023.

\bibitem{AI_NS_17}
X.~Chen, Z.~Zhao, C.~Wu, M.~Bennis, H.~Liu, Y.~Ji, and H.~Zhang, ``Multi-tenant
  cross-slice resource orchestration: A deep reinforcement learning approach,''
  \emph{IEEE Journal on Selected Areas in Communications}, vol.~37, pp.
  2377--2392, 2019.

\bibitem{AI_NS_19}
D.~Bega, M.~Gramaglia, A.~Garcia-Saavedra, M.~Fiore, A.~Banchs, and
  X.~Costa-Perez, ``Network slicing meets artificial intelligence: An ai-based
  framework for slice management,'' \emph{IEEE Communications Magazine},
  vol.~58, pp. 32--38, 6 2020.

\bibitem{AI_NS_21}
E.~S. Xavier, N.~Agoulmine, and J.~S.~B. Martins, ``On modeling network slicing
  communication resources with sarsa optimization,'' \emph{arXiv preprint
  arXiv:2301.04696}, 1 2023.

\bibitem{AI_NFV_1}
X.~Chen, Z.~Han, H.~Zhang, G.~Xue, Y.~Xiao, and M.~Bennis, ``Wireless resource
  scheduling in virtualized radio access networks using stochastic learning,''
  \emph{IEEE Transactions on Mobile Computing}, vol.~17, no.~4, pp. 961--974,
  2017.

\bibitem{AI_NFV_3}
O.~Houidi, O.~Soualah, W.~Louati, and D.~Zeghlache, ``An enhanced reinforcement
  learning approach for dynamic placement of virtual network functions,''
  \emph{IEEE 31st Annual International Symposium on Personal, Indoor and Mobile
  Radio Communications}, 2020.

\bibitem{AI_NFV_6}
H.~R. Khezri, P.~A. Moghadam, and M.~K. Farshbafan, ``Deep reinforcement
  learning for dynamic reliability aware nfv-based service provisioning,''
  \emph{IEEE Global Communications Conference (GLOBECOM)}, 2019.

\bibitem{AI_NFV_5}
M.~Moradi, M.~Ahmadi, and R.~Nikbazm, ``Comparison of machine learning
  techniques for vnf resource requirements prediction in nfv,'' \emph{Journal
  of Network and Systems Management}, vol.~30, 1 2022.

\bibitem{AI_NFV_18}
Z.~Li, L.~Wu, X.~Zeng, X.~Yue, Y.~Jing, W.~Wu, and K.~Su, ``Online coordinated
  nfv resource allocation via novel machine learning techniques,'' \emph{IEEE
  Transactions on Network and Service Management}, 3 2022.

\bibitem{AI_NFV_19}
A.~Nouruzi, A.~Zakeri, M.~R. Javan, N.~Mokari, R.~Hussain, and S.~M. Kazmi,
  ``Online service provisioning in nfv-enabled networks using deep
  reinforcement learning,'' \emph{IEEE Transactions on Network and Service
  Management}, vol.~19, pp. 3276--3289, 9 2022.

\bibitem{AI_NFV_20}
J.~Pei, P.~Hong, M.~Pan, J.~Liu, and J.~Zhou, ``Optimal vnf placement via deep
  reinforcement learning in sdn/nfv-enabled networks,'' \emph{IEEE Journal on
  Selected Areas in Communications}, vol.~38, pp. 263--278, 2 2020.

\bibitem{AI_NFV_23}
R.~Mijumbi, S.~Hasija, S.~Davy, A.~Davy, B.~Jennings, and R.~Boutaba,
  ``Topology-aware prediction of virtual network function resource
  requirements,'' \emph{IEEE Transactions on Network and Service Management},
  vol.~14, pp. 106--120, 3 2017.

\bibitem{AI_Comb_12}
J.~Pei, P.~Hong, and D.~Li, ``Virtual network function selection and chaining
  based on deep learning in sdn and nfv-enabled networks,'' \emph{2018 IEEE
  International Conference on Communications Workshops, ICC Workshops 2018 -
  Proceedings}, pp. 1--6, 7 2018.

\bibitem{AI_SDN_1}
N.~M. Yungaicela-Naula, C.~Vargas-Rosales, J.~A. Pérez-Díaz, and D.~F.
  Carrera, ``A flexible sdn-based framework for slow-rate ddos attack
  mitigation by using deep reinforcement learning,'' \emph{Journal of Network
  and Computer Applications}, vol. 205, 9 2022.

\bibitem{AI_SDN_9}
L.~M. Halman and M.~J. Alenazi, ``Mcad: A machine learning based cyberattacks
  detector in software-defined networking (sdn) for healthcare systems,''
  \emph{IEEE Access}, 2023.

\bibitem{AI_Comb_5}
F.~Naeem, M.~Ali, and G.~Kaddoum, ``Federated-learning-empowered
  semi-supervised active learning framework for intrusion detection in zsm,''
  \emph{IEEE Communications Magazine}, vol.~61, pp. 88--94, 2 2023.

\bibitem{AI_NS_1}
G.~Rahmanian, H.~S. Shahhoseini, and A.~H.~J. Pozveh, ``A review of network
  slicing in 5g and beyond: Intelligent approaches and challenges,'' \emph{2021
  ITU Kaleidoscope: Connecting Physical and Virtual Worlds, ITU K 2021}, 2021.

\bibitem{AI_NS_27}
J.~Wang and J.~Liu, ``Secure and reliable slicing in 5g and beyond vehicular
  networks,'' \emph{IEEE Wireless Communications}, vol.~29, pp. 126--133, 2
  2022.

\bibitem{AI_NS_13}
Y.~Shi, Y.~E. Sagduyu, T.~Erpek, and M.~C. Gursoy, ``How to attack and defend
  nextg radio access network slicing with reinforcement learning,'' \emph{IEEE
  Open Journal of Vehicular Technology}, vol.~4, pp. 181--192, 2023.

\bibitem{AI_NFV_16}
G.~W. de~Oliveira, M.~Nogueira, A.~L. dos Santos, and D.~M. Batista,
  ``Intelligent vnf placement to mitigate ddos attacks on industrial iot,''
  \emph{IEEE Transactions on Network and Service Management}, 2023.

\bibitem{das2019survey}
T.~Das, V.~Sridharan, and M.~Gurusamy, ``A survey on controller placement in
  sdn,'' \emph{IEEE communications surveys \& tutorials}, vol.~22, no.~1, pp.
  472--503, 2019.

\bibitem{NTN.C.SDN.2.4}
S.~Wu, X.~Chen, L.~Yang, C.~Fan, and Y.~Zhao, ``Dynamic and static controller
  placement in software-defined satellite networking,'' \emph{Acta
  Astronautica}, vol. 152, pp. 49--58, 11 2018.

\bibitem{chowdhury2006handover}
P.~K. Chowdhury, M.~Atiquzzaman, and W.~Ivancic, ``Handover schemes in
  satellite networks: State-of-the-art and future research directions,''
  \emph{IEEE Communications Surveys \& Tutorials}, vol.~8, no.~4, pp. 2--14,
  2006.

\bibitem{NTN.C.SDN.4.1}
W.~Yang, H.~Guyu, J.~Fenglin, and Z.~Jiachen, ``A satellite handover strategy
  based on the potential game in leo satellite networks,'' \emph{IEEE Access},
  vol.~7, pp. 133\,641--133\,652, 2019.

\bibitem{AT.C.SDN.3.1}
C.~Pan, J.~Yi, C.~Yin, J.~Yu, and X.~Li, ``Joint 3d uav placement and resource
  allocation in software-defined cellular networks with wireless backhaul,''
  \emph{IEEE Access}, vol.~7, pp. 104\,279--104\,293, 2019.

\bibitem{bari2016orchestrating}
F.~Bari, S.~R. Chowdhury, R.~Ahmed, R.~Boutaba, and O.~C. M.~B. Duarte,
  ``Orchestrating virtualized network functions,'' \emph{IEEE Transactions on
  Network and Service Management}, vol.~13, no.~4, pp. 725--739, 2016.

\bibitem{NTN.C.NFV.2}
X.~Gao, R.~Liu, A.~Kaushik, and H.~Zhang, ``Dynamic resource allocation for
  virtual network function placement in satellite edge clouds,'' \emph{IEEE
  Transactions on Network Science and Engineering}, vol.~9, pp. 2252--2265,
  2022.

\bibitem{hantouti2020service}
H.~Hantouti, N.~Benamar, and T.~Taleb, ``Service function chaining in 5g \&
  beyond networks: Challenges and open research issues,'' \emph{IEEE Network},
  vol.~34, no.~4, pp. 320--327, 2020.

\bibitem{NTN.C.NFV.7}
H.~Zhang, J.~Xu, X.~Liu, K.~Long, and V.~C. Leung, ``Joint optimization of
  caching placement and power allocation in virtualized satellite-terrestrial
  network,'' \emph{IEEE Transactions on Wireless Communications}, 2023.

\bibitem{AT.C.NS.2}
H.~Shen, Q.~Ye, W.~Zhuang, W.~Shi, G.~Bai, and G.~Yang,
  ``Drone-small-cell-assisted resource slicing for 5g uplink radio access
  networks,'' \emph{IEEE Transactions on Vehicular Technology}, vol.~70, pp.
  7071--7086, 7 2021.

\bibitem{AT.C.NS.4}
A.~E. Garcıa, S.~Hofmann, C.~Sous, L.~Garcia, A.~Baltaci, C.~Bach, R.~Wellens,
  D.~Gera, D.~Schupke, and H.~E. Gonzalez, ``Performance evaluation of network
  slicing for aerial vehicle communications,'' \emph{IEEE International
  Conference on Communications Workshops (ICC Workshops)}, 2019.

\bibitem{AT.C.NS.7}
I.~Donevski, J.~J. Nielsen, and P.~Popovski, ``Standalone deployment of a
  dynamic drone cell for wireless connectivity of two services,'' vol.
  2021-March.\hskip 1em plus 0.5em minus 0.4em\relax Institute of Electrical
  and Electronics Engineers Inc., 2021.

\bibitem{AT.C.SDN.5.4}
S.~Vashisht and S.~Jain, ``Software-defined network-enabled opportunistic
  offloading and charging scheme in multi-unmanned aerial vehicle ecosystem,''
  \emph{International Journal of Communication Systems}, vol.~32, 5 2019.

\bibitem{SAT.AI.NS.5}
H.~H. Esmat, B.~Lorenzo, and W.~Shi, ``Toward resilient network slicing for
  satellite-terrestrial edge computing iot,'' \emph{IEEE Internet of Things
  Journal}, vol.~10, pp. 14\,621--14\,645, 8 2023.

\bibitem{NTN.C.SDN.1.1}
\BIBentryALTinterwordspacing
J.~Feng, L.~Jiang, Y.~Shen, W.~Ma, and M.~Yin, ``A scheme for software defined
  ors satellite networking,'' \emph{2014 IEEE Fourth International Conference
  on Big Data and Cloud Computing}, pp. 716--721, 12 2014. [Online]. Available:
  \url{http://ieeexplore.ieee.org/document/7034865/}
\BIBentrySTDinterwordspacing

\bibitem{NTN.C.SDN.1.2}
J.~Li, K.~Xue, J.~Liu, Y.~Zhang, and Y.~Fang, ``An icn/sdn-based network
  architecture and efficient content retrieval for future satellite-terrestrial
  integrated networks,'' \emph{IEEE Network}, 2020.

\bibitem{NTN.C.SDN.1.3}
B.~Feng, H.~Zhou, H.~Zhang, G.~Li, H.~Li, S.~Yu, and H.~C. Chao, ``Hetnet: A
  flexible architecture for heterogeneous satellite-terrestrial networks,''
  \emph{IEEE Network}, vol.~31, pp. 86--92, 11 2017.

\bibitem{NTN.C.SDN.1.4}
P.~Dong, M.~Gao, F.~Tang, L.~Cao, X.~Zhang, P.~Han, Y.~Yang, W.~Xu, and
  X.~Zhang, ``Multi-layer and heterogeneous resource management in sdn-based
  space-terrestrial integrated networks,'' \emph{Proceedings - 2020 IEEE 22nd
  International Conference on High Performance Computing and Communications,
  IEEE 18th International Conference on Smart City and IEEE 6th International
  Conference on Data Science and Systems, HPCC-SmartCity-DSS 2020}, pp.
  377--384, 12 2020.

\bibitem{NTN.C.SDN.1.5}
J.~Bao, B.~Zhao, W.~Yu, Z.~Feng, C.~Wu, and Z.~Gong, ``Opensan: A
  software-defined satellite network architecture,'' \emph{Computer
  Communication Review}, vol.~44, pp. 347--348, 2 2015.

\bibitem{NTN.C.SDN.1.6}
T.~Li, H.~Zhou, H.~Luo, and S.~Yu, ``Service: A software defined framework for
  integrated space-terrestrial satellite communication,'' \emph{IEEE
  Transactions on Mobile Computing}, vol.~17, pp. 703--716, 3 2018.

\bibitem{NTN.C.SDN.1.8}
Y.~Bi, G.~Han, S.~Xu, X.~Wang, C.~Lin, Z.~Yu, and P.~Sun, ``Software defined
  space-terrestrial integrated networks: Architecture, challenges, and
  solutions,'' \emph{IEEE Network}, vol.~33, pp. 22--28, 1 2019.

\bibitem{NTN.C.SDN.1.9}
M.~Sheng, Y.~Wang, J.~Li, R.~Liu, D.~Zhou, and L.~He, ``Toward a flexible and
  reconfigurable broadband satellite network: Resource management architecture
  and strategies,'' \emph{IEEE Wireless Communications}, vol.~24, pp. 127--133,
  6 2017.

\bibitem{NTN.C.SDN.5.2}
F.~Mendoza, M.~Minardi, S.~Chatzinotas, L.~Lei, and T.~X. Vu, ``An sdn based
  testbed for dynamic network slicing in satellite-terrestrial networks,''
  \emph{2021 IEEE International Mediterranean Conference on Communications and
  Networking, MeditCom 2021}, pp. 36--41, 2021.

\bibitem{NTN.C.SDN.5.16}
M.~Minardi, T.~X. Vu, L.~Lei, C.~Politis, and S.~Chatzinotas, ``Virtual network
  embedding for ngso systems: Algorithmic solution and sdn-testbed
  validation,'' \emph{IEEE Transactions on Network and Service Management},
  2022.

\bibitem{NTN.C.Comb.4}
B.~Liu, T.~Zhang, L.~Zhang, and Z.~Ma, ``Online virtual network embedding for
  both the delay sensitive and tolerant services in sdn-enabled
  satellite-terrestrial networks,'' vol. 2023-March.\hskip 1em plus 0.5em minus
  0.4em\relax Institute of Electrical and Electronics Engineers Inc., 2023.

\bibitem{NTN.C.SDN.5.7}
F.~Mendoza, R.~Ferrus, and O.~Sallent, ``Experimental proof of concept of an
  sdn-based traffic engineering solution for hybrid satellite-terrestrial
  mobile backhauling,'' \emph{International Journal of Satellite Communications
  and Networking}, vol.~37, pp. 630--645, 11 2019.

\bibitem{NTN.C.SDN.5.8}
N.~Yoshino, H.~Oguma, S.~Kamedm, and N.~Suematsu, ``Feasibility study of
  expansion of openflow network using satellite communication to wide area,''
  \emph{International Conference on Ubiquitous and Future Networks, ICUFN}, pp.
  647--651, 7 2017.

\bibitem{NTN.C.SDN.5.12}
A.~Mudonhi, C.~Sacchi, and F.~Granelli, ``Sdn-based multimedia content delivery
  in 5gmmwave hybrid satellite-terrestrial networks,'' \emph{IEEE 29th Annual
  International Symposium on Personal, Indoor, and Mobile Radio Communications
  (PIMRC)}, 2018.

\bibitem{NTN.C.Comb.10_B.2.3}
S.~Xu, X.~W. Wang, and M.~Huang, ``Software-defined next-generation satellite
  networks: Architecture, challenges, and solutions,'' \emph{IEEE Access},
  vol.~6, pp. 4027--4041, 1 2018.

\bibitem{NTN.C.Comb.11}
T.~Ahmed, E.~Dubois, J.~B. Dupé, R.~Ferrús, P.~Gélard, and N.~Kuhn,
  ``Software-defined satellite cloud ran,'' \emph{International Journal of
  Satellite Communications and Networking}, vol.~36, pp. 108--133, 1 2018.

\bibitem{NTN.C.SDN.2.3}
S.~Xu, X.~Wang, B.~Gao, M.~Zhang, and M.~Huang, ``Controller placement in
  software-defined satellite networks,'' \emph{Proceedings - 14th International
  Conference on Mobile Ad-Hoc and Sensor Networks, MSN 2018}, pp. 146--151, 7
  2018.

\bibitem{NTN.C.SDN.2.6}
X.~Zhang, F.~Tang, L.~Cao, L.~Chen, J.~Yu, W.~Xu, X.~Zhang, J.~Lei, and
  Z.~Wang, ``Dynamical controller placement among sdn space-terrestrial
  integrated networks,'' \emph{Proceedings - 2020 IEEE 22nd International
  Conference on High Performance Computing and Communications, IEEE 18th
  International Conference on Smart City and IEEE 6th International Conference
  on Data Science and Systems, HPCC-SmartCity-DSS 2020}, pp. 352--359, 12 2020.

\bibitem{NTN.C.SDN.2.9}
L.~Chen, F.~Tang, and X.~Li, ``Mobility- and load-adaptive controller placement
  and assignment in leo satellite networks,'' \emph{Proceedings - IEEE
  INFOCOM}, vol. 2021-May, 5 2021.

\bibitem{NTN.C.SDN.2.10}
Z.~Han, C.~Xu, Z.~Xiong, G.~Zhao, and S.~Yu, ``On-demand dynamic controller
  placement in software defined satellite-terrestrial networking,'' \emph{IEEE
  Transactions on Network and Service Management}, vol.~18, pp. 2915--2928, 9
  2021.

\bibitem{NTN.C.SDN.2.2}
N.~Torkzaban and J.~S. Baras, ``Controller placement in sdn-enabled 5g
  satellite-terrestrial networks,'' in \emph{2021 IEEE Global Communications
  Conference (GLOBECOM)}.\hskip 1em plus 0.5em minus 0.4em\relax IEEE, 2021,
  pp. 1--6.

\bibitem{NTN.C.SDN.2.2(1)}
------, ``Joint satellite gateway deployment \& controller placement in
  software-defined 5g-satellite integrated networks,'' \emph{arXiv preprint
  arXiv:2103.08735}, 2021.

\bibitem{NTN.C.SDN.2.5}
A.~Papa, T.~De~Cola, P.~Vizarreta, M.~He, C.~M. Machuca, and W.~Kellerer,
  ``Dynamic sdn controller placement in a leo constellation satellite
  network,'' in \emph{2018 IEEE Global Communications Conference
  (GLOBECOM)}.\hskip 1em plus 0.5em minus 0.4em\relax IEEE, 2018, pp. 206--212.

\bibitem{NTN.C.SDN.2.7}
J.~Liu, Y.~Shi, L.~Zhao, Y.~Cao, W.~Sun, and N.~Kato, ``Joint placement of
  controllers and gateways in sdn-enabled 5g-satellite integrated network,''
  \emph{IEEE Journal on Selected Areas in Communications}, vol.~36, pp.
  221--232, 2 2018.

\bibitem{NTN.C.SDN.2.8}
D.~K. Luong, Y.-F. Hu, J.-P. Li, and M.~Ali, ``Metaheuristic approaches to the
  joint controller and gateway placement in 5g-satellite sdn networks,'' in
  \emph{ICC 2020-2020 IEEE international conference on communications
  (ICC)}.\hskip 1em plus 0.5em minus 0.4em\relax IEEE, 2020, pp. 1--6.

\bibitem{NTN.C.SDN.2.11}
X.~Li, F.~Tang, L.~Fu, J.~Yu, L.~Chen, J.~Liu, Y.~Zhu, and L.~T. Yang,
  ``Optimized controller provisioning in software-defined leo satellite
  networks,'' \emph{IEEE Transactions on Mobile Computing}, 2022.

\bibitem{NTN.C.SDN.2.12}
J.~Guo, L.~Yang, D.~Rinc{\'o}n, S.~Sallent, C.~Fan, Q.~Chen, and X.~Li, ``Sdn
  controller placement in leo satellite networks based on dynamic topology,''
  in \emph{2021 IEEE/CIC International Conference on Communications in China
  (ICCC)}.\hskip 1em plus 0.5em minus 0.4em\relax IEEE, 2021, pp. 1083--1088.

\bibitem{NTN.C.SDN.2.13}
S.~Wu, X.~Liu, Q.~Chen, J.~Guo, L.~Yang, Y.~Zhao, and C.~Fan, ``Update method
  for controller placement problem in software-defined satellite networking,''
  in \emph{2019 28th International Conference on Computer Communication and
  Networks (ICCCN)}.\hskip 1em plus 0.5em minus 0.4em\relax IEEE, 2019, pp.
  1--7.

\bibitem{NTN.C.SDN.3.1}
X.~Tao, K.~Ota, M.~Dong, H.~Qi, and K.~Li, ``Congestion-aware scheduling for
  software-defined sag networks,'' \emph{IEEE Transactions on Network Science
  and Engineering}, vol.~8, pp. 2861--2871, 2021.

\bibitem{NTN.C.SDN.3.4}
Z.~Ma, X.~Di, J.~Li, L.~Cong, and P.~Li, ``Mptcp based load balancing mechanism
  in software defined satellite networks,'' \emph{International Conference on
  Wireless and Satellite Systems}, vol. 280, pp. 294--302, 2019.

\bibitem{NTN.C.SDN.3.5}
M.~Ouyang, X.~Duan, J.~Liu, R.~Zhang, T.~Huang, and H.~Lu, ``Multi-path
  transmission scheme based on segment control in low-earth-orbit satellite
  network,'' \emph{IEEE International Conference on High Performance Switching
  and Routing, HPSR}, vol. 2021-June, 6 2021.

\bibitem{NTN.C.SDN.3.6}
A.~Kak and I.~F. Akyildiz, ``Online intra-domain segment routing for
  software-defined cubesat networks,'' \emph{2020 IEEE Global Communications
  Conference, GLOBECOM 2020 - Proceedings}, 12 2020.

\bibitem{NTN.C.SDN.3.7}
M.~Jia, S.~Zhu, L.~Wang, Q.~Guo, H.~Wang, and Z.~Liu, ``Routing algorithm with
  virtual topology toward to huge numbers of leo mobile satellite network based
  on sdn,'' \emph{Mobile Networks and Applications}, vol.~23, pp. 285--300, 4
  2018.

\bibitem{NTN.C.SDN.3.8}
Z.~Jiang, Q.~Wu, H.~Li, and J.~Wu, ``Scmptcp: Sdn cooperated multipath transfer
  for satellite network with load awareness,'' \emph{IEEE Access}, vol.~6, pp.
  19\,823--19\,832, 3 2018.

\bibitem{NTN.C.SDN.5.5}
Y.~Xiao, H.~Zhang, Q.~Ji, Y.~Zhang, and J.~Wang, ``Efficient transmission
  protocols for the satellite-terrestrial integrated networks,''
  \emph{Lecture Notes of the Institute for Computer Sciences,
  Social-Informatics and Telecommunications Engineering, LNICST}, vol. 481
  LNICST, pp. 86--96, 2023.

\bibitem{NTN.C.SDN.5.15}
M.~Hu, J.~Li, C.~Cai, T.~Deng, W.~Xu, and Y.~Dong, ``Software defined multicast
  for large-scale multi-layer leo satellite networks,'' \emph{IEEE Transactions
  on Network and Service Management}, vol.~19, pp. 2119--2130, 9 2022.

\bibitem{NTN.C.SDN.3.2}
Z.~Han, G.~Zhao, Y.~Xing, N.~Sun, C.~Xu, and S.~Yu, ``Dynamic routing for
  software-defined leo satellite networks based on isl attributes,'' \emph{2021
  IEEE Global Communications Conference, GLOBECOM 2021 - Proceedings}, 2021.

\bibitem{NTN.C.SDN.3.3}
P.~Kumar, S.~Bhushan, D.~Halder, and A.~M. Baswade, ``Fybrrlink: Efficient
  qos-aware routing in sdn enabled future satellite networks,'' \emph{IEEE
  Transactions on Network and Service Management}, vol.~19, pp. 2107--2118, 9
  2022.

\bibitem{NTN.C.SDN.3.9}
Q.~Guo, R.~Gu, T.~Dong, J.~Yin, Z.~Liu, L.~Bai, and Y.~Ji, ``Sdn-based
  end-to-end fragment-aware routing for elastic data flows in leo
  satellite-terrestrial network,'' \emph{IEEE Access}, vol.~7, pp. 396--410,
  2019.

\bibitem{NTN.C.SDN.5.6}
C.~Bu, X.~Wang, H.~Cheng, M.~Huang, K.~Li, and S.~K. Das, ``Enabling adaptive
  routing service customization via the integration of sdn and nfv,''
  \emph{Journal of Network and Computer Applications}, vol.~93, pp. 123--136, 9
  2017.

\bibitem{NTN.C.SDN.5.9}
M.~Ouyang, J.~Liu, R.~Zhang, B.~Wang, L.~Liu, N.~Xin, and J.~Tong, ``Flow
  granularity multi-path transmission optimization design for satellite
  networks,'' \emph{IEEE Wireless Communications and Networking Conference,
  WCNC}, vol. 2023-March, 2023.

\bibitem{NTN.C.SDN.4.2}
M.~M. Aurizzi, T.~Rossi, E.~Raso, L.~Funari, and E.~Cianca, ``An sdn-based
  traffic handover control procedure and sgd management logic for ehf satellite
  networks,'' \emph{Computer Networks}, vol. 196, 9 2021.

\bibitem{NTN.C.SDN.4.3}
B.~Yang, Y.~Wu, X.~Chu, and G.~Song, ``Seamless handover in software-defined
  satellite networking,'' \emph{IEEE Communications Letters}, vol.~20, pp.
  1768--1771, 9 2016.

\bibitem{NTN.C.SDN.4.4}
T.~Li, H.~Zhou, H.~Luo, W.~Quan, Q.~Xu, G.~Li, and G.~Li, ``Timeout
  strategy-based mobility management for software defined satellite networks,''
  \emph{IEEE Conference on Computer Communications Workshops (INFOCOM WKSHPS)},
  2017.

\bibitem{NTN.C.SDN.4.4(1)}
T.~Li, H.~Zhou, H.~Luo, I.~You, and Q.~Xu, ``Sat-flow: Multi-strategy flow
  table management for software defined satellite networks,'' \emph{IEEE
  Access}, vol.~5, pp. 14\,952--14\,965, 7 2017.

\bibitem{NTN.C.SDN.5.11}
\BIBentryALTinterwordspacing
D.~Yan, M.~Gu, L.~Wang, and X.~He, ``Sada: Sdn architecture based secure
  dynamic access scheme for satellite network,'' \emph{International Conference
  on Mobile Wireless Middleware, Operating Systems, and Applications}, 2022.
  [Online]. Available:
  \url{https://link.springer.com/10.1007/978-3-031-34497-8_15}
\BIBentrySTDinterwordspacing

\bibitem{NTN.C.Comb.13_NTN.C.SDN.4.1}
R.~Ferrus, O.~Sallent, T.~Ahmed, and R.~Fedrizzi, ``Towards sdn/nfv-enabled
  satellite ground segment systems: End-to-end traffic engineering use
  case.''\hskip 1em plus 0.5em minus 0.4em\relax Institute of Electrical and
  Electronics Engineers Inc., 6 2017, pp. 888--893.

\bibitem{NTN.C.SDN.5.13}
F.~Mendoza, R.~Ferrús, and O.~Sallent, ``Sdn-based traffic engineering for
  improved resilience in integrated satellite-terrestrial backhaul networks,''
  \emph{4th International Conference on Information and Communication
  Technologies for Disaster Management (ICT-DM)}, 2017.

\bibitem{NTN.C.SDN.5.4}
J.~Du, C.~Jiang, H.~Zhang, Y.~Ren, and M.~Guizani, ``Auction design and
  analysis for sdn-based traffic offloading in hybrid satellite-terrestrial
  networks,'' \emph{IEEE Journal on Selected Areas in Communications}, vol.~36,
  pp. 2202--2217, 10 2018.

\bibitem{NTN.AI.SDN.2}
K.~Yang, B.~Zhang, and D.~Guo, ``Controller and gateway partition placement
  insdn-enabled integrated satellite-terrestrialnetwork,'' \emph{IEEE
  International Conference on Communications Workshops (ICC Workshops)}, 2019.

\bibitem{NTN.AI.SDN.5}
A.~Wei, H.~Yu, X.~Lang, and B.~Yang, ``Dynamic controller placement for
  software-defined leo network using deep reinforcement learning,'' \emph{2021
  7th International Conference on Computer and Communications, ICCC 2021}, pp.
  1314--1320, 2021.

\bibitem{NTN.AI.SDN.4}
Z.~Xing, H.~Qi, X.~Di, J.~Liu, and L.~Cong, ``Deep reinforcement learning based
  congestion control mechanism for sdn and ndn in satellite networks,'' in
  \emph{International Conference on Mobile Wireless Middleware, Operating
  Systems, and Applications}.\hskip 1em plus 0.5em minus 0.4em\relax Springer,
  2022, pp. 13--29.

\bibitem{NTN.AI.SDN.6}
S.~Wu, L.~Yang, J.~Guo, Q.~Chen, X.~Liu, and C.~Fan, ``Intelligent quality of
  service routing in software-defined satellite networking,'' \emph{IEEE
  Access}, vol.~7, pp. 155\,281--155\,298, 2019.

\bibitem{NTN.AI.SDN.3}
C.~Qiu, H.~Yao, F.~R. Yu, F.~Xu, and C.~Zhao, ``Deep q-learning aided
  networking, caching, and computing resources allocation in software-defined
  satellite-terrestrial networks,'' \emph{IEEE Transactions on Vehicular
  Technology}, vol.~68, pp. 5871--5883, 6 2019.

\bibitem{NTN.AI.SDN.1}
D.~Vickramasingam and S.~Bangar, ``A link planning and ddos attack detection in
  sdn based integrated space-terrestrial networks,'' \emph{Journal of
  Communications}, vol.~18, pp. 267--273, 4 2023.

\bibitem{NTN.AI.SDN.7}
\BIBentryALTinterwordspacing
R.~Uddin and S.~Kumar, ``Sdn-based federated learning approach for
  satellite-iot framework to enhance data security and privacy in space
  communication,'' \emph{IEEE Journal of Radio Frequency Identification}, pp.
  1--1, 2023. [Online]. Available:
  \url{https://ieeexplore.ieee.org/document/10138184/}
\BIBentrySTDinterwordspacing

\bibitem{NTN.C.NFV.8}
Z.~Jiayz, M.~Shengy, J.~Liy, R.~Liuy, K.~Guoy, Y.~Wangy, D.~Chenz, and R.~Ding,
  ``Joint optimization of vnf deployment and routing in software defined
  satellite networks,'' \emph{IEEE 88th Vehicular Technology Conference
  (VTC-Fall)}, 2018.

\bibitem{NTN.C.NFV.4}
H.~Yang, W.~Liu, X.~Wang, and J.~Li, ``Group sparse space information network
  with joint virtual network function deployment and maximum flow routing
  strategy,'' \emph{IEEE Transactions on Wireless Communications}, pp. 1--1, 1
  2023.

\bibitem{NTN.C.NFV.10}
A.~Petrosino, G.~Piro, L.~A. Grieco, and G.~Boggia, ``On the optimal deployment
  of virtual network functions in non-terrestrial segments,'' \emph{IEEE
  Transactions on Network and Service Management}, 2023.

\bibitem{NTN.C.NFV.14}
X.~Gao, R.~Liu, and A.~Kaushik, ``Service chaining placement based on satellite
  mission planning in ground station networks,'' \emph{IEEE Transactions on
  Network and Service Management}, vol.~18, pp. 3049--3063, 9 2021.

\bibitem{NTN.C.NFV.16}
\BIBentryALTinterwordspacing
H.~Yang, W.~Liu, J.~Li, and T.~Q. Quek, ``Space information network with joint
  virtual network function deployment and flow routing strategy with qos
  constraints,'' \emph{IEEE Journal on Selected Areas in Communications}, pp.
  1--1, 2023. [Online]. Available:
  \url{https://ieeexplore.ieee.org/document/10121438/}
\BIBentrySTDinterwordspacing

\bibitem{NTN.C.NFV.1}
A.~Petrosino, G.~Piro, L.~A. Grieco, and G.~Boggia, ``An optimal allocation
  framework of security virtual network functions in 6g satellite
  deployments,'' \emph{Proceedings - IEEE Consumer Communications and
  Networking Conference, CCNC}, pp. 917--920, 2022.

\bibitem{NTN.C.NFV.17}
X.~Gao, R.~Liu, and A.~Kaushik, ``Virtual network function placement in
  satellite edge computing with a potential game approach,'' \emph{IEEE
  Transactions on Network and Service Management}, vol.~19, pp. 1243--1259, 6
  2022.

\bibitem{NTN.C.Comb.1_B.2.1}
I.~Maity, T.~X. Vu, S.~Chatzinotas, and M.~Minardi, ``D-vine: Dynamic virtual
  network embedding in non-terrestrial networks,'' in \emph{IEEE Wireless
  Communications and Networking Conference, WCNC}, vol. 2022-April, 2022, pp.
  166--171.

\bibitem{NTN.C.NFV.13}
C.~Pan, J.~Shi, L.~Yang, and Z.~Kong, ``Satellite network load balancing
  strategy for sdn/nfv collaborative deployment,'' \emph{Proceedings - 2019
  IEEE SmartWorld, Ubiquitous Intelligence and Computing, Advanced and Trusted
  Computing, Scalable Computing and Communications, Internet of People and
  Smart City Innovation, SmartWorld/UIC/ATC/SCALCOM/IOP/SCI 2019}, pp.
  1406--1411, 8 2019.

\bibitem{NTN.C.NFV.5}
G.~Li, H.~Zhou, B.~Feng, G.~Li, and Q.~Xu, ``Horizontal based orchestration for
  multi-domain sfc in sdn nfv-enabled satellite terrestrial networks,''
  \emph{China Communications}, 2018.

\bibitem{NTN.C.NFV.15}
X.~Qin, T.~Ma, Z.~Tang, X.~Zhang, H.~Zhou, and L.~Zhao, ``Service-aware
  resource orchestration in ultra-dense leo satellite-terrestrial integrated
  6g: A service function chain approach,'' \emph{IEEE Transactions on Wireless
  Communications}, 2023.

\bibitem{NTN.C.NFV.3}
B.~Feng, G.~Li, G.~Li, Y.~Zhang, H.~Zhou, and S.~Yu, ``Enabling efficient
  service function chains at terrestrial-satellite hybrid cloud networks,''
  \emph{IEEE Network}, vol.~33, pp. 94--99, 11 2019.

\bibitem{NTN.C.NFV.3(1)}
B.~Feng, G.~Li, G.~Li, H.~Zhou, H.~Zhang, and S.~Yu, ``Efﬁcient mappings of
  service function chains at terrestrial-satellite hybrid cloud networks,''
  \emph{IEEE Global Communications Conference (GLOBECOM)}, 2018.

\bibitem{NTN.C.NFV.6}
Y.~Ouyang, J.~Lin, T.~Feng, C.~Yang, L.~Zhang, T.~Li, and Z.~Han,
  ``Intent-driven cox resource management for space-terrestrial networks,''
  \emph{IEEE Wireless Communications}, 2023.

\bibitem{NTN.C.NFV.18}
Z.~Jia†, M.~Sheng†, J.~Li†, Y.~Zhu†, W.~Bai†, and Z.~Han, ``Virtual
  network functions orchestration in software defined leo small satellite
  networks,'' \emph{IEEE International Conference on Communications}, 2020.

\bibitem{NTN.C.NFV.19}
Z.~Jia, M.~Sheng, J.~Li, D.~Zhou, and Z.~Han, ``Vnf-based service provision in
  software defined leo satellite networks,'' \emph{IEEE Transactions on
  Wireless Communications}, vol.~20, pp. 6139--6153, 9 2021.

\bibitem{NTN.C.NS.2}
S.~Hendaoui and C.~N. Zangarz, ``Leveraging sdn slicing isolation for improved
  adaptive satellite-5g downlink scheduler,'' \emph{2021 International
  Symposium on Networks, Computers and Communications, ISNCC 2021}, 2021.

\bibitem{NTN.C.NS.4}
T.~Kim, J.~Kwak, and J.~P. Choi, ``Satellite edge computing architecture and
  network slice scheduling for iot support,'' \emph{IEEE Internet of Things
  Journal}, vol.~9, pp. 14\,938--14\,951, 8 2022.

\bibitem{NTN.C.NS.5}
------, ``Satellite network slice planning: Architecture, performance analysis,
  and open issues,'' \emph{IEEE Vehicular Technology Magazine}, 2023.

\bibitem{NTN.C.NS.3}
T.~Ahmed, A.~Alleg, R.~Ferrus, and R.~Riggio, ``On-demand network slicing using
  sdn/nfv-enabled satellite ground segment systems,'' \emph{2018 4th IEEE
  Conference on Network Softwarization and Workshops, NetSoft 2018}, pp.
  378--383, 9 2018.

\bibitem{NTN.AI.NS.1}
L.~Lei, Y.~Yuan, T.~X. Vu, S.~Chatzinotas, M.~Minardi, and J.~F.~M. Montoya,
  ``Dynamic-adaptive ai solutions for network slicing management in
  satellite-integrated b5g systems,'' \emph{IEEE Network}, vol.~35, pp. 91--97,
  11 2021.

\bibitem{NTN.AI.NS.2}
H.~Wu, J.~Chen, C.~Zhou, J.~Li, X.~Shen, H.~Q. Wu, C.~H. Zhou, X.~M. Shen,
  J.~Y. Chen, and J.~L. Li, ``Learning-based joint resource slicing and
  scheduling in space-terrestrial integrated vehicular networks,'' 2021.

\bibitem{NTN.AI.NS.4}
T.~D. Cola and I.~Bisio, ``Qos optimisation of embb services in converged
  5g-satellite networks,'' \emph{IEEE Transactions on Vehicular Technology},
  vol.~69, pp. 12\,098--12\,110, 10 2020.

\bibitem{NTN.AI.NS.3}
T.~K. Rodrigues and N.~Kato, ``Network slicing with centralized and distributed
  reinforcement learning for combined satellite/ground networks in a 6g
  environment,'' \emph{IEEE Wireless Communications}, vol.~29, pp. 104--110, 2
  2022.

\bibitem{AT.C.SDN.1.1}
B.~Barritt, T.~Kichkaylo, K.~Mandke, A.~Zalcman, and V.~Lin, ``Operating a uav
  mesh \& internet backhaul network using temporospatial sdn.''\hskip 1em plus
  0.5em minus 0.4em\relax IEEE Aerospace Conference Proceedings, 6 2017.

\bibitem{AT.C.SDN.1.2}
V.~Sharma, F.~Song, I.~You, and H.~C. Chao, ``Efficient management and fast
  handovers in software defined wireless networks using uavs,'' \emph{IEEE
  Network}, vol.~31, pp. 78--85, 11 2017.

\bibitem{AT.C.SDN.1.3}
M.~Moradi, K.~Sundaresan, E.~Chai, S.~Rangarajan, and Z.~M. Mao, ``Skycore:
  Moving core to the edge for untethered and reliable uav-based lte
  networks.''\hskip 1em plus 0.5em minus 0.4em\relax Association for Computing
  Machinery, 10 2018, pp. 35--49.

\bibitem{AT.C.SDN.1.4}
Z.~Zhao, P.~Cumino, A.~Souza, D.~Rosário, T.~Braun, E.~Cerqueira, and
  M.~Gerla, ``Software-defined unmanned aerial vehicles networking for video
  dissemination services,'' \emph{Ad Hoc Networks}, vol.~83, pp. 68--77, 2
  2019.

\bibitem{AT.C.SDN.1.5}
H.~Iqbal, J.~M. Kenneth~Stranc, K.~Palmer, and P.~Benbenek, ``A
  software-defined networking architecture for aerial network
  optimization.''\hskip 1em plus 0.5em minus 0.4em\relax IEEE, 2016.

\bibitem{AT.C.SDN.1.6}
M.~Sara, I.~Jawhar, and M.~Nader, ``A softwarization architecture for uavs and
  wsns as part of the cloud environment.''\hskip 1em plus 0.5em minus
  0.4em\relax Institute of Electrical and Electronics Engineers Inc., 8 2016,
  pp. 13--18.

\bibitem{AT.C.SDN.1.7}
O.~S. Oubbati, M.~Atiquzzaman, P.~Lorenz, A.~Baz, and H.~Alhakami, ``Search: An
  sdn-enabled approach for vehicle path-planning,'' \emph{IEEE Transactions on
  Vehicular Technology}, vol.~69, pp. 14\,523--14\,536, 12 2020.

\bibitem{AT.C.SDN.1.8}
M.~A. B.~S. Abir, M.~Z. Chowdhury, and Y.~M. Jang, ``A software-defined uav
  network using queueing model,'' \emph{IEEE Access}, 2023.

\bibitem{AT.C.SDN.1.9}
H.~Wang and H.~Z.~X. Zhang, ``An sdn framework for uav backbone network towards
  knowledge centric networking,'' 2018.

\bibitem{AT.C.SDN.1.10}
A.~Alioua, S.~M. Senouci, S.~Moussaoui, H.~Sedjelmaci, and M.~A. Messous,
  ``Efficient data processing in software-defined uav-assisted vehicular
  networks: A sequential game approach,'' \emph{Wireless Personal
  Communications}, vol. 101, pp. 2255--2286, 8 2018.

\bibitem{AT.C.SDN.1.11}
Z.~Latif, C.~Lee, K.~Sharif, F.~Li, and S.~P. Mohanty, ``An sdn-based framework
  for load balancing and flight control in uav networks,'' \emph{IEEE Consumer
  Electronics Magazine}, vol.~12, pp. 43--51, 1 2023.

\bibitem{AT.C.SDN.1.12}
F.~Xiong, A.~Li, H.~Wang, and L.~Tang, ``An sdn-mqtt based communication system
  for battlefield uav swarms,'' \emph{IEEE Communications Magazine}, vol.~57,
  pp. 41--47, 8 2019.

\bibitem{AT.C.SDN.1.13}
T.~D.~E. Silva, C.~F. E.~D. Melo, P.~Cumino, D.~Rosario, E.~Cerqueira, and
  E.~P.~D. Freitas, ``Stfanet: Sdn-based topology management for flying ad hoc
  network,'' \emph{IEEE Access}, vol.~7, pp. 173\,499--173\,514, 2019.

\bibitem{AT.C.SDN.1.14}
C.~GUERBER, N.~LARRIEU, and M.~ROYER, ``Software deﬁned network based
  architecture to improve security in a swarm of drones,'' 2019.

\bibitem{AT.C.SDN.6.2}
D.~K. Luong, Y.-F. Hu, J.-P. Li, F.~Benamrane, M.~Ali, and K.~Abdo,
  ``Traffic-aware dynamic controller placement using ai techniques in sdn-based
  aeronautical networks,'' 2019.

\bibitem{AT.C.SDN.2.1}
W.~Qi, Q.~Song, X.~Kong, and L.~Guo, ``A traffic-differentiated routing
  algorithm in flying ad hoc sensor networks with sdn cluster controllers,''
  \emph{Journal of the Franklin Institute}, vol. 356, pp. 766--790, 1 2019.

\bibitem{AT.C.SDN.2.3}
A.~Ramaprasath, A.~Srinivasan, C.~H. Lung, and M.~St-Hilaire, ``Intelligent
  wireless ad hoc routing protocol and controller for uav networks,'' in
  \emph{Lecture Notes of the Institute for Computer Sciences,
  Social-Informatics and Telecommunications Engineering, LNICST}, vol. 184
  LNICST.\hskip 1em plus 0.5em minus 0.4em\relax Springer Verlag, 2017, pp.
  92--104.

\bibitem{AT.C.SDN.2.4}
G.~Sec¸inti, P.~B. Darian, B.~Canberk, and K.~R. Chowdhury, ``Resilient
  end-to-end connectivity for software deﬁned unmanned aerial vehicular
  networks,'' in \emph{IEEE 28th Annual International Symposium on Personal,
  Indoor, and Mobile Radio Communications (PIMRC)}, 2017.

\bibitem{AT.C.SDN.2.5}
G.~Secinti, P.~B. Darian, B.~Canberk, and K.~R. Chowdhury, ``Sdns in the sky:
  Robust end-to-end connectivity for aerial vehicular networks,'' \emph{IEEE
  Communications Magazine}, vol.~56, pp. 16--21, 1 2018.

\bibitem{AT.C.SDN.2.6}
K.~Chen, S.~Zhao, N.~Lv, W.~Gao, X.~Wang, and X.~Zou, ``Segment routing based
  traffic scheduling for the software-defined airborne backbone network,''
  \emph{IEEE Access}, vol.~7, pp. 106\,162--106\,178, 2019.

\bibitem{AT.C.SDN.3.3}
R.~M. Shukla, S.~Sengupta, and A.~N. Patra, ``Software-deﬁned network based
  resource allocation in distributed servers for unmanned aerial vehicles,''
  2018.

\bibitem{AT.C.SDN.5.1}
L.~Zhao, K.~Yang, Z.~Tan, X.~Li, S.~Sharma, and Z.~Liu, ``A novel cost
  optimization strategy for sdn-enabled uav-assisted vehicular computation
  offloading,'' \emph{IEEE Transactions on Intelligent Transportation Systems},
  vol.~22, pp. 3664--3674, 6 2021.

\bibitem{AT.C.SDN.5.3}
M.~A. Ali, Y.~Zeng, and A.~Jamalipour, ``Software-defined coexisting uav and
  wifi: Delay-oriented traffic offloading and uav placement,'' \emph{IEEE
  Journal on Selected Areas in Communications}, vol.~38, pp. 988--998, 6 2020.

\bibitem{AT.C.SDN.4.1}
Y.~Tan, J.~Liu, and J.~Wang, ``How to protect key drones in unmanned aerial
  vehicle networks? an sdn-based topology deception scheme,'' \emph{IEEE
  Transactions on Vehicular Technology}, vol.~71, pp. 13\,320--13\,331, 12
  2022.

\bibitem{AT.AI.SDN.4}
``Deploying sdn control in internet of uavs: Q-learning-based edge
  scheduling,'' \emph{IEEE Transactions on Network and Service Management},
  vol.~18, pp. 526--537, 3 2021.

\bibitem{AT.AI.SDN.3}
M.~Ariman, M.~Akkoc, T.~Sari, M.~R. Erol, G.~Secinti, and B.~Canberk,
  ``Energy-efficient rl-based aerial network deployment testbed for disaster
  areas,'' \emph{Journal of Communications and Networks}, pp. 1--10, 1 2023.

\bibitem{AT.AI.SDN.2}
R.~Gupta, M.~M. Patel, S.~Tanwar, N.~Kumar, and S.~Zeadally, ``Blockchain-based
  data dissemination scheme for 5g-enabled softwarized uav networks,''
  \emph{IEEE Transactions on Green Communications and Networking}, vol.~5, pp.
  1712--1721, 12 2021.

\bibitem{AT.C.NFV.1}
B.~Nogales, V.~Sanchez-Aguero, I.~Vidal, F.~Valera, and J.~Garcia-Reinoso, ``A
  nfv system to support configurable and automated multi-uav service
  deployments,'' \emph{Proceedings of the 2018 ACM International Conference on
  Mobile Systems, Applications and Services}, pp. 39--44, 6 2018.

\bibitem{AT.C.NFV.2}
N.~Nomikos, E.~T. Michailidis, P.~Trakadas, D.~Vouyioukas, H.~Karl, J.~Martrat,
  T.~Zahariadis, K.~Papadopoulos, and S.~Voliotis, ``A uav-based moving 5g ran
  for massive connectivity of mobile users and iot devices,'' \emph{Vehicular
  Communications}, vol.~25, 10 2020.

\bibitem{AT.C.NFV.3}
A.~Hermosilla, A.~M. Zarca, J.~B. Bernabe, J.~Ortiz, and A.~Skarmeta,
  ``Security orchestration and enforcement in nfv/sdn-aware uav deployments,''
  \emph{IEEE Access}, vol.~8, pp. 131\,779--131\,795, 2020.

\bibitem{AT.C.NFV.9}
V.~Sanchez-Aguero, F.~Valera, B.~Nogales, L.~F. Gonzalez, and I.~Vidal,
  ``Venue: Virtualized environment for multi-uav network emulation,''
  \emph{IEEE Access}, vol.~7, pp. 154\,659--154\,671, 2019.

\bibitem{AT.C.NFV.6}
G.~Wang, S.~Zhou, S.~Zhang, Z.~Niu, and X.~Shen, ``Sfc-based service
  provisioning for reconfigurable space-air-ground integrated networks,''
  \emph{IEEE Journal on Selected Areas in Communications}, vol.~38, pp.
  1478--1489, 7 2020.

\bibitem{AT.C.NFV.4}
Y.~Qin, D.~Guo, L.~Luo, J.~Zhang, and M.~Xu, ``Service function chain migration
  with the long-term budget in dynamic networks,'' \emph{Computer Networks},
  vol. 223, 3 2023.

\bibitem{AT.C.NFV.5}
Y.~Wang, H.~Wang, X.~Wei, K.~Zhao, J.~Fan, J.~Chen, Y.~Hu, and R.~Jia,
  ``Service function chain scheduling in heterogeneous multi-uav edge
  computing,'' \emph{Drones}, vol.~7, 2 2023.

\bibitem{AT.C.NFV.7}
J.~Bai, X.~Chang, R.~J. Rodriguez, K.~S. Trivedi, and S.~Li, ``Towards
  uav-based mec service chain resilience evaluation: A quantitative modeling
  approach,'' \emph{IEEE Transactions on Vehicular Technology}, vol.~72, pp.
  5181--5194, 4 2023.

\bibitem{AT.C.NS.5}
P.~Yang, X.~Xi, T.~Q. Quek, J.~Chen, X.~Cao, and D.~Wu, ``Repeatedly
  energy-efficient and fair service coverage: Uav slicing,'' \emph{IEEE Global
  Communications Conference, GLOBECOM 2020 - Proceedings}, 12 2020.

\bibitem{AT.AI.NS.4}
P.~Yang, X.~Xi, K.~Guo, T.~Q. Quek, J.~Chen, and X.~Cao, ``Proactive uav
  network slicing for urllc and mobile broadband service multiplexing,''
  \emph{IEEE Journal on Selected Areas in Communications}, vol.~39, pp.
  3225--3244, 10 2021.

\bibitem{AT.AI.NS.1}
G.~Faraci, C.~Grasso, and G.~Schembra, ``Design of a 5g network slice extension
  with mec uavs managed with reinforcement learning,'' \emph{IEEE Journal on
  Selected Areas in Communications}, vol.~38, pp. 2356--2371, 10 2020.

\bibitem{AT.AI.NS.2}
G.~Wu, B.~Zhang, and Y.~Li, ``Intelligent and survivable resource slicing for
  6g-oriented uav-assisted edge computing networks,'' \emph{Computer
  Communications}, vol. 202, pp. 154--165, 3 2023.

\bibitem{AT.AI.NS.3}
Y.~H. Xu, J.~H. Li, W.~Zhou, and C.~Chen, ``Learning-empowered resource
  allocation for air slicing in uav-assisted cellular v2x communications,''
  \emph{IEEE Systems Journal}, vol.~17, pp. 1008--1011, 3 2023.

\bibitem{SAT.C.SDN.1}
Y.~Shi, Y.~Cao, J.~Liu, and N.~Kato, ``A cross-domain sdn architecture for
  multi-layered space-terrestrial integrated networks,'' \emph{IEEE Network},
  vol.~33, pp. 29--35, 1 2019.

\bibitem{SAT.C.SDN.3}
M.~H. Eiza and A.~Raschellà, ``A hybrid sdn-based architecture for secure and
  qos aware routing in space-air-ground integrated networks (sagins),''
  \emph{IEEE Wireless Communications and Networking Conference, WCNC}, vol.
  2023-March, 2023.

\bibitem{SAT.AI.SDN.3}
A.~Papa, J.~V. Mankowski, H.~Vijayaraghavan, B.~Mafakheri, L.~Goratti, and
  W.~Kellerer, ``Enabling 6g applications in the sky: Aeronautical federation
  framework,'' \emph{IEEE Network}, 2023.

\bibitem{SAT.C.SDN.7}
Z.~Zhou, J.~Feng, C.~Zhang, Z.~Chang, Y.~Zhang, and K.~M.~S. Huq, ``Sagecell:
  Software-defined space-air-ground integrated moving cells,'' \emph{IEEE
  Communications Magazine}, vol.~56, no.~8, pp. 92--99, 2018.

\bibitem{SAT.C.SDN.2}
C.~Guo, C.~Gong, H.~Xu, L.~Zhang, and Z.~Han, ``A dynamic handover
  software-defined transmission control scheme in space-air-ground integrated
  networks,'' \emph{IEEE Transactions on Wireless Communications}, vol.~21, pp.
  6110--6124, 8 2022.

\bibitem{SAT.C.SDN.6}
Z.~Li, Y.~Hu, D.~Zhu, J.~Wu, and Y.~Gu, ``Esmd-flow: An intelligent flow
  forwarding scheme with endogenous security based on mimic defense in
  space-air-ground integrated network,'' \emph{China Communications Magazine},
  2022.

\bibitem{SAT.AI.SDN.4}
C.~Chen, Z.~Liao, Y.~Ju, C.~He, K.~Yu, and S.~Wan, ``Hierarchical domain-based
  multicontroller deployment strategy in sdn-enabled space-air-ground
  integrated network,'' \emph{IEEE Transactions on Aerospace and Electronic
  Systems}, vol.~58, pp. 4864--4879, 12 2022.

\bibitem{SAT.AI.SDN.2}
P.~Zhang, N.~Chen, S.~Shen, S.~Yu, N.~Kumar, and C.-H. Hsu, ``Ai-enabled
  space-air-ground integrated networks: Management and optimization,''
  \emph{IEEE Network}, 2023.

\bibitem{SAT.AI.SDN.1}
J.~Tao, S.~Liu, and C.~Liu, ``A traffic scheduling scheme for load balancing in
  sdn-based space-air-ground integrated networks,'' \emph{IEEE International
  Conference on High Performance Switching and Routing, HPSR}, vol. 2022-June,
  pp. 95--100, 2022.

\bibitem{SAT.C.NFV.2}
J.~Li, W.~Shi, H.~Wu, S.~Zhang, and X.~Shen, ``Cost-aware dynamic sfc mapping
  and scheduling in sdn/nfv-enabled space-air-ground-integrated networks for
  internet of vehicles,'' \emph{IEEE Internet of Things Journal}, vol.~9, pp.
  5824--5838, 4 2022.

\bibitem{SAT.C.NFV.3}
Y.~Yue, X.~Tang, W.~Yang, X.~Zhang, Z.~Zhang, C.~Gao, and L.~Xu, ``Delay-aware
  and resource-efficient vnf placement in 6g non-terrestrial networks,''
  \emph{IEEE Wireless Communications and Networking Conference, WCNC}, vol.
  2023-March, 2023.

\bibitem{SAT.C.NFV.1}
S.~Zhou, G.~Wang, S.~Zhang, Z.~Niu, and X.~S. Shen, ``Bidirectional mission
  offloading for agile space-air-ground integrated networks,'' \emph{IEEE
  Wireless Communications}, vol.~26, pp. 38--45, 4 2019.

\bibitem{SAT.C.NFV.4}
\BIBentryALTinterwordspacing
Y.~Cao, Z.~Jia, C.~Dong, Y.~Wang, J.~You, and Q.~Wu, ``Sfc deployment in
  space-air-ground integrated networks based on matching game,'' \emph{arXiv
  preprint arXiv:2303.01020}, 3 2023. [Online]. Available:
  \url{http://arxiv.org/abs/2303.01020}
\BIBentrySTDinterwordspacing

\bibitem{SAT.C.NFV.5}
P.~Zhang, P.~Yang, N.~Kumar, and M.~Guizani, ``Space-air-ground integrated
  network resource allocation based on service function chain,'' \emph{IEEE
  Transactions on Vehicular Technology}, vol.~71, pp. 7730--7738, 7 2022.

\bibitem{SAT.C.NS.1}
X.~Zhang, Q.~Zhu, and H.~V. Poor, ``Heterogeneous statistical qos provisioning
  for scalable software-defined 6g mobile networks.''\hskip 1em plus 0.5em
  minus 0.4em\relax Institute of Electrical and Electronics Engineers Inc.,
  2023.

\bibitem{SAT.C.NS.3}
F.~Lyu, P.~Yang, H.~Wu, C.~Zhou, J.~Ren, Y.~Zhang, and X.~Shen,
  ``Service-oriented dynamic resource slicing and optimization for
  space-air-ground integrated vehicular networks,'' \emph{IEEE Transactions on
  Intelligent Transportation Systems}, vol.~23, pp. 7469--7483, 7 2022.

\bibitem{SAT.AI.NS.3}
A.~Asheralieva, D.~Niyato, and Y.~Miyanaga, ``Efficient dynamic distributed
  resource slicing in 6g multi-access edge computing networks with online admm
  and message passing graph neural networks,'' \emph{IEEE Transactions on
  Mobile Computing}, 2023.

\bibitem{SAT.AI.NS.6}
G.~Zhou, L.~Zhao, G.~Zheng, S.~Song, J.~Zhang, and L.~Hanzo, ``Multi-objective
  optimization of space-air-ground integrated network slicing relying on a pair
  of central and distributed learning algorithms,'' \emph{IEEE Internet of
  Things Journal}, 2023.

\bibitem{SAT.AI.NS.2}
Y.~Wu, Y.~Ma, H.~N. Dai, and H.~Wang, ``Deep learning for privacy preservation
  in autonomous moving platforms enhanced 5g heterogeneous networks,''
  \emph{Computer Networks}, vol. 185, 2 2021.

\bibitem{NTN.C.SDN.5.14}
\BIBentryALTinterwordspacing
Y.~Miao, Z.~Cheng, W.~Li, H.~Ma, X.~Liu, and Z.~Cui, ``Software defined
  integrated satellite-terrestrial network: A survey,'' \emph{International
  conference on space information network}, vol. 688, 2017. [Online].
  Available: \url{http://link.springer.com/10.1007/978-981-10-4403-8}
\BIBentrySTDinterwordspacing

\bibitem{NTN.C.SDN.5.3}
\BIBentryALTinterwordspacing
X.~Huang, Z.~Zhao, X.~Meng, and H.~Zhang, ``Architecture and application of
  sdn/nfv-enabled space-terrestrial integrated network,'' \emph{International
  Conference on Space Information Network}, vol. 688, 2017. [Online].
  Available: \url{http://link.springer.com/10.1007/978-981-10-4403-8}
\BIBentrySTDinterwordspacing

\bibitem{NTN.C.Comb.12}
G.~Gardikis, H.~Koumaras, C.~Sakkas, and V.~Koumaras, ``Towards sdn/nfv-enabled
  satellite networks,'' \emph{Telecommunication Systems}, vol.~66, pp.
  615--628, 12 2017.

\bibitem{yan2016sadr}
D.~Yan, J.~Guo, L.~Wang, and P.~Zhan, ``Sadr: Network status adaptive qos
  dynamic routing for satellite networks,'' in \emph{2016 IEEE 13th
  International Conference on Signal Processing (ICSP)}.\hskip 1em plus 0.5em
  minus 0.4em\relax IEEE, 2016, pp. 1186--1190.

\bibitem{gkizeli2001hybrid}
M.~Gkizeli, R.~Tafazolli, and B.~G. Evans, ``Hybrid channel adaptive handover
  scheme for non-geo satellite diversity based systems,'' \emph{IEEE
  Communications Letters}, vol.~5, no.~7, pp. 284--286, 2001.

\bibitem{kaur2022review}
K.~Kaur, V.~Mangat, and K.~Kumar, ``A review on virtualized infrastructure
  managers with management and orchestration features in nfv architecture,''
  \emph{Computer Networks}, vol. 217, p. 109281, 2022.

\bibitem{NTN.C.Comb.2}
T.~Rossi, M.~D. Sanctis, E.~Cianca, C.~Fragale, M.~Ruggieri, and H.~Fenech,
  ``Future space-based communications infrastructures based on high throughput
  satellites and software defined networking.''\hskip 1em plus 0.5em minus
  0.4em\relax Institute of Electrical and Electronics Engineers Inc., 10 2015,
  pp. 332--337.

\bibitem{he2019game}
Q.~He, G.~Cui, X.~Zhang, F.~Chen, S.~Deng, H.~Jin, Y.~Li, and Y.~Yang, ``A
  game-theoretical approach for user allocation in edge computing
  environment,'' \emph{IEEE Transactions on Parallel and Distributed Systems},
  vol.~31, no.~3, pp. 515--529, 2019.

\bibitem{greenberg2009vl2}
A.~Greenberg, J.~R. Hamilton, N.~Jain, S.~Kandula, C.~Kim, P.~Lahiri, D.~A.
  Maltz, P.~Patel, and S.~Sengupta, ``Vl2: A scalable and flexible data center
  network,'' in \emph{Proceedings of the ACM SIGCOMM 2009 conference on Data
  communication}, 2009, pp. 51--62.

\bibitem{guo2009bcube}
C.~Guo, G.~Lu, D.~Li, H.~Wu, X.~Zhang, Y.~Shi, C.~Tian, Y.~Zhang, and S.~Lu,
  ``Bcube: a high performance, server-centric network architecture for modular
  data centers,'' in \emph{Proceedings of the ACM SIGCOMM 2009 conference on
  Data communication}, 2009, pp. 63--74.

\bibitem{leiserson1985fat}
C.~E. Leiserson, ``Fat-trees: Universal networks for hardware-efficient
  supercomputing,'' \emph{IEEE transactions on Computers}, vol. 100, no.~10,
  pp. 892--901, 1985.

\bibitem{zhang2017network}
N.~Zhang, Y.-F. Liu, H.~Farmanbar, T.-H. Chang, M.~Hong, and Z.-Q. Luo,
  ``Network slicing for service-oriented networks under resource constraints,''
  \emph{IEEE Journal on Selected Areas in Communications}, vol.~35, no.~11, pp.
  2512--2521, 2017.

\bibitem{wang2017multi}
T.~Wang, H.~Xu, and F.~Liu, ``Multi-resource load balancing for virtual network
  functions,'' in \emph{2017 IEEE 37th International Conference on Distributed
  Computing Systems (ICDCS)}.\hskip 1em plus 0.5em minus 0.4em\relax IEEE,
  2017, pp. 1322--1332.

\bibitem{riggio2015virtual}
R.~Riggio, A.~Bradai, T.~Rasheed, J.~Schulz-Zander, S.~Kuklinski, and T.~Ahmed,
  ``Virtual network functions orchestration in wireless networks,'' in
  \emph{2015 11th International conference on network and service management
  (CNSM)}.\hskip 1em plus 0.5em minus 0.4em\relax IEEE, 2015, pp. 108--116.

\bibitem{zervoudakis2020mayfly}
K.~Zervoudakis and S.~Tsafarakis, ``A mayfly optimization algorithm,''
  \emph{Computers \& Industrial Engineering}, vol. 145, p. 106559, 2020.

\bibitem{NTN.C.NS.1}
Y.~Drif, E.~Chaput, E.~Lavinal, P.~Berthou, B.~T. Jou, O.~Grémillet, and
  F.~Arnal, ``An extensible network slicing framework for satellite integration
  into 5g,'' \emph{International Journal of Satellite Communications and
  Networking}, vol.~39, pp. 339--357, 7 2021.

\bibitem{NTN.C.NS.6}
Y.~Drif, E.~Lavinal, E.~Chaput, P.~Berthou, B.~T. Jou, O.~Gr{\'e}millet, and
  F.~Arnal, ``Slice aware non terrestrial networks,'' in \emph{2021 IEEE 46th
  Conference on Local Computer Networks (LCN)}.\hskip 1em plus 0.5em minus
  0.4em\relax IEEE, 2021, pp. 24--31.

\bibitem{NTN.AI.Comb.1}
J.~Zhang, X.~Zhang, P.~Wang, L.~Liu, and Y.~Wang, ``Double edge intelligent
  integrated satellite terrestrial networks,'' \emph{China Communications},
  2020.

\bibitem{AT.C.SDN.6.1}
M.~A. B.~S. Abir, M.~Z. Chowdhury, and Y.~M. Jang, ``Software-defined uav
  networks for 6g systems: Requirements, opportunities, emerging techniques,
  challenges, and research directions,'' \emph{IEEE Open Journal of the
  Communications Society}, 2023.

\bibitem{AT.C.SDN.3.2}
S.~U. Rahman, G.~H. Kim, Y.~Z. Cho, and A.~Khan, ``Positioning of uavs for
  throughput maximization in software-defined disaster area uav communication
  networks,'' \emph{Journal of Communications and Networks}, vol.~20, pp.
  452--463, 10 2018.

\bibitem{li2018virtual}
D.~Li, P.~Hong, K.~Xue \emph{et~al.}, ``Virtual network function placement
  considering resource optimization and sfc requests in cloud datacenter,''
  \emph{IEEE Transactions on Parallel and Distributed Systems}, vol.~29, no.~7,
  pp. 1664--1677, 2018.

\bibitem{beck2015coordinated}
M.~T. Beck and J.~F. Botero, ``Coordinated allocation of service function
  chains,'' in \emph{2015 IEEE global communications Conference
  (GLOBECOM)}.\hskip 1em plus 0.5em minus 0.4em\relax IEEE, 2015, pp. 1--6.

\bibitem{AT.C.NS.8}
G.~K. Xilouris, M.~C. Batistatos, G.~E. Athanasiadou, G.~Tsoulos, H.~B.
  Pervaiz, and C.~C. Zarakovitis, ``Uav-assisted 5g network architecture with
  slicing and virtualization,'' 2018.

\bibitem{AT.C.NS.3}
P.~M. Payagalage, C.~M. Basnayaka, D.~N. Dushantha, and A.~Kumar, ``Network
  virtualization and slicing in uav-enabled future networks.''\hskip 1em plus
  0.5em minus 0.4em\relax Institute of Electrical and Electronics Engineers
  Inc., 2023, pp. 98--103.

\bibitem{AT.C.NS.1}
Z.~Yuan and G.~M. Muntean, ``Airslice: A network slicing framework for uav
  communications,'' \emph{IEEE Communications Magazine}, vol.~58, pp. 62--68,
  11 2020.

\bibitem{SAT.C.NS.2}
C.~C. Gonzalez, S.~Pizzi, M.~Murroni, and G.~Araniti, ``Multicasting over 6g
  non-terrestrial networks: A softwarization-based approach,'' \emph{IEEE
  Vehicular Technology Magazine}, vol.~18, pp. 91--99, 3 2023.

\bibitem{SAT.AI.NS.4}
H.~Wu, J.~Chen, C.~Zhou, W.~Shi, N.~Cheng, W.~Xu, W.~Zhuang, and X.~S. Shen,
  ``Resource management in space-air-ground integrated vehicular networks: Sdn
  control and ai algorithm design,'' \emph{IEEE Wireless Communications},
  vol.~27, pp. 52--60, 12 2020.

\bibitem{Sim.SMM_ns3}
\BIBentryALTinterwordspacing
P.~Silva, ``ns-3 satellite mobility model,'' 2017. [Online]. Available:
  \url{https://gitlab.inesctec.pt/pmms/ns3-satellite}
\BIBentrySTDinterwordspacing

\bibitem{Sim.ns3leo}
T.~Schubert, L.~Wolf, and U.~Kulau, ``ns-3-leo: Evaluation tool for satellite
  swarm communication protocols,'' \emph{IEEE Access}, vol.~10, pp.
  11\,527--11\,537, 2022.

\bibitem{Sim.SNS3}
J.~Puttonen, B.~Herman, S.~Rantanen, F.~Laakso, and J.~Kurjenniemi, ``Satellite
  network simulator 3,'' 2015.

\bibitem{Sim.5G_NTN_SLS}
J.~Puttonen, L.~Sormunen, H.~Martikainen, S.~Rantanen, and J.~Kurjenniemi, ``A
  system simulator for 5g non-terrestrial network evaluations.''\hskip 1em plus
  0.5em minus 0.4em\relax Institute of Electrical and Electronics Engineers
  Inc., 6 2021, pp. 292--297.

\bibitem{Sim.NS3_5G_sim}
``Ns-3-based 5g satellite-terrestrial integrated network simulator.''\hskip 1em
  plus 0.5em minus 0.4em\relax Institute of Electrical and Electronics
  Engineers Inc., 2022, pp. 154--159.

\bibitem{Sim.EXata}
\BIBentryALTinterwordspacing
Keysight, ``{EX}ata.'' [Online]. Available:
  \url{https://www.keysight.com/us/en/product/SN100EXBA/exata-network-modeling.html}
\BIBentrySTDinterwordspacing

\bibitem{Sim.STK}
\BIBentryALTinterwordspacing
Ansys, ``Systems {T}ool {K}it ({STK}).'' [Online]. Available:
  \url{https://www.ansys.com/products/missions/ansys-stk}
\BIBentrySTDinterwordspacing

\bibitem{Sim.OpenSAND}
\BIBentryALTinterwordspacing
``Opensand.'' [Online]. Available: \url{https://www.opensand.org/}
\BIBentrySTDinterwordspacing

\bibitem{Sim.opendaylight}
\BIBentryALTinterwordspacing
T.~L. Foundation, ``Open{D}aylight.'' [Online]. Available:
  \url{https://www.opendaylight.org/}
\BIBentrySTDinterwordspacing

\bibitem{Sim.Ryu}
\BIBentryALTinterwordspacing
R.~S.~F. Community, ``Build {SDN} agilely.'' [Online]. Available:
  \url{https://ryu-sdn.org/}
\BIBentrySTDinterwordspacing

\bibitem{Sim.POX}
\BIBentryALTinterwordspacing
N.~Repo, ``The {POX} network software platform.'' [Online]. Available:
  \url{https://github.com/noxrepo/pox}
\BIBentrySTDinterwordspacing

\bibitem{Sim.ONOS}
\BIBentryALTinterwordspacing
O.~N. Foundation, ``{O}pen {N}etwork {O}perating {S}ystem ({ONOS}).'' [Online].
  Available: \url{https://opennetworking.org/onos/}
\BIBentrySTDinterwordspacing

\bibitem{Sim.openvswitch}
\BIBentryALTinterwordspacing
T.~L. Foundation, ``Open v{S}witch.'' [Online]. Available:
  \url{https://www.openvswitch.org/}
\BIBentrySTDinterwordspacing

\bibitem{Sim.openstack}
\BIBentryALTinterwordspacing
T.~O. Foundation, ``Open{S}tack.'' [Online]. Available:
  \url{https://www.openstack.org/}
\BIBentrySTDinterwordspacing

\bibitem{Sim.Lagopus}
\BIBentryALTinterwordspacing
``Lagopus.'' [Online]. Available: \url{https://www.lagopus.org/}
\BIBentrySTDinterwordspacing

\bibitem{Sim.DPDK}
\BIBentryALTinterwordspacing
``Data plane development kit (dpdk).'' [Online]. Available:
  \url{https://www.dpdk.org/}
\BIBentrySTDinterwordspacing

\bibitem{Sim.ST.SimSurvey}
W.~Jiang, Y.~Zhan, X.~Xiao, and G.~Sha, ``Network simulators for
  satellite-terrestrial integrated networks: A survey,'' \emph{IEEE Access},
  vol.~11, pp. 98\,269--98\,292, 2023.

\bibitem{AT.C.SDN.5.2}
M.~A. Sayeed, R.~Kumar, and V.~Sharma, ``Efficient data management and control
  over wsns using sdn-enabled aerial networks,'' \emph{International Journal of
  Communication Systems}, vol.~33, 1 2020.

\bibitem{NTN.C.NFV.9}
H.~Yang, W.~Liu, H.~Li, and J.~Li, ``Maximum flow routing strategy for space
  information network with service function constraints,'' \emph{IEEE
  Transactions on Wireless Communications}, vol.~21, pp. 2909--2923, 5 2022.

\bibitem{NTN.C.NFV.11}
\BIBentryALTinterwordspacing
B.~Guo, H.~Li, Z.~Zhang, and Y.~Yan, ``Online network slicing for real time
  applications in large-scale satellite networks,'' \emph{IEEE ICC}, 1 2023.
  [Online]. Available: \url{http://arxiv.org/abs/2301.09372}
\BIBentrySTDinterwordspacing

\bibitem{NTN.AI.Comb.2}
\BIBentryALTinterwordspacing
R.~Zhu, G.~Li, Y.~Zhang, Z.~Fang, and J.~Wang, ``Load-balanced virtual network
  embedding based on deep reinforcement learning for 6g regional satellite
  networks,'' \emph{IEEE Transactions on Vehicular Technology}, pp. 1--14,
  2023. [Online]. Available:
  \url{https://ieeexplore.ieee.org/document/10132506/}
\BIBentrySTDinterwordspacing

\bibitem{SAT.C.SDN.4}
Z.~Liao, C.~Chen, Y.~Ju, C.~He, J.~Jiang, and Q.~Pei, ``Multi-controller
  deployment in sdn-enabled 6g space–air–ground integrated network,''
  \emph{Remote Sensing}, vol.~14, 3 2022.

\bibitem{NTN.C.NFV.12}
T.~Ahmed, R.~Ferrus, R.~Fedrizzi, O.~Sallent, N.~Kuhn, E.~Dubois, and
  P.~Gelard, ``Satellite gateway diversity in sdn/nfv-enabled satellite ground
  segment systems,'' \emph{IEEE International Conference on Communications (ICC
  Workshops)}, 2017.

\bibitem{NTN.C.Comb.5}
L.~Boero, R.~Bruschi, F.~Davoli, M.~Marchese, and F.~Patrone, ``Satellite
  networking integration in the 5g ecosystem: Research trends and open
  challenges,'' \emph{IEEE Network}, vol.~32, pp. 9--15, 9 2018.

\bibitem{NTN.C.Comb.6_B.2.3.4}
R.~Ferrús, H.~Koumaras, O.~Sallent, G.~Agapiou, T.~Rasheed, M.~A. Kourtis,
  C.~Boustie, P.~Gélard, and T.~Ahmed, ``Sdn/nfv-enabled satellite
  communications networks: Opportunities, scenarios and challenges,''
  \emph{Physical Communication}, vol.~18, pp. 95--112, 3 2016, this is not a
  survey, it is just a paper describing scenarios where SDN/NFV can improve the
  operation and management of integrated satellite and terrestrial networks.

\bibitem{mershad2021cloud}
K.~Mershad, H.~Dahrouj, H.~Sarieddeen, B.~Shihada, T.~Al-Naffouri, and M.-S.
  Alouini, ``Cloud-enabled high-altitude platform systems: Challenges and
  opportunities,'' \emph{Frontiers in Communications and Networks}, vol.~2, p.
  716265, 2021.

\bibitem{NTN.C.Comb.7}
S.~Yao, J.~Guan, Z.~Yan, and K.~Xu, ``Si-stin: A smart identifier framework for
  space and terrestrial integrated network,'' \emph{IEEE Network}, vol.~33, pp.
  8--14, 1 2019.

\bibitem{digital_twin_survey}
S.~Mihai, M.~Yaqoob, D.~V. Hung, W.~Davis, P.~Towakel, M.~Raza, M.~Karamanoglu,
  B.~Barn, D.~Shetve, R.~V. Prasad \emph{et~al.}, ``Digital twins: A survey on
  enabling technologies, challenges, trends and future prospects,'' \emph{IEEE
  Communications Surveys \& Tutorials}, 2022.

\bibitem{khan2022digital}
L.~U. Khan, W.~Saad, D.~Niyato, Z.~Han, and C.~S. Hong, ``Digital-twin-enabled
  6g: Vision, architectural trends, and future directions,'' \emph{IEEE
  Communications Magazine}, vol.~60, no.~1, pp. 74--80, 2022.

\bibitem{zhao2020intelligentDT}
L.~Zhao, G.~Han, Z.~Li, and L.~Shu, ``Intelligent digital twin-based
  software-defined vehicular networks,'' \emph{IEEE Network}, vol.~34, no.~5,
  pp. 178--184, 2020.

\bibitem{AT.C.SDN.2.2}
M.~A. B.~S. Abir and M.~Z. Chowdhury, ``Digital twin-based software-defined uav
  networks using queuing model,'' in \emph{Proceedings of the 10th
  International Conference on Signal Processing and Integrated Networks, SPIN
  2023}.\hskip 1em plus 0.5em minus 0.4em\relax Institute of Electrical and
  Electronics Engineers Inc., 2023, pp. 479--483.

\bibitem{wijethilaka2021survey}
S.~Wijethilaka and M.~Liyanage, ``Survey on network slicing for internet of
  things realization in 5g networks,'' \emph{IEEE Communications Surveys \&
  Tutorials}, vol.~23, no.~2, pp. 957--994, 2021.

\end{thebibliography}

\end{document}